\documentclass{emulateapj}

\newcommand{\liratio}{$^6$Li/$^7$Li}
\newcommand{\liratioinverse}{$^7$Li/$^6$Li}
\newcommand{\lisix}{$^6$Li}
\newcommand{\liseven}{$^7$Li}
\newcommand{\benine}{$^9$Be}
\newcommand{\lisixbe}{$^6$Li/Be}

\newcommand{\hi}{H\,{\sc i}}
\newcommand{\lii}{Li\,{\sc i}}

\newcommand{\oi}{O\,{\sc i}}
\newcommand{\oifor}{[O\,{\sc i}]}
\newcommand{\mgi}{Mg\,{\sc i}}
\newcommand{\ki}{K\,{\sc i}}
\newcommand{\cai}{Ca\,{\sc i}}
\newcommand{\fei}{Fe\,{\sc i}}
\newcommand{\feii}{Fe\,{\sc ii}}
\newcommand{\logh}{log\,$\epsilon_{\rm H}$}
\newcommand{\loghe}{log\,$\epsilon_{\rm He}$}
\newcommand{\logli}{log\,$\epsilon_{\rm Li}$}
\newcommand{\logliseven}{log\,$\epsilon_{\rm ^7Li}$}
\newcommand{\loglisix}{log\,$\epsilon_{\rm ^6Li}$}

\newcommand{\logo}{log\,$\epsilon_{\rm O}$}
\newcommand{\logfe}{log\,$\epsilon_{\rm Fe}$}

\newcommand{\feh}{[Fe/H]}
\newcommand{\feih}{[\fei /H]}
\newcommand{\feiih}{[\feii /H]}
\newcommand{\beh}{[Be/H]}
\newcommand{\oh}{[O/H]}

\newcommand{\ofe}{[O/Fe]}

\newcommand{\alphafe}{[$\alpha$/Fe]}
\newcommand{\fefesun}{Fe/Fe$_\sun$}
\newcommand{\oosun}{O/O$_\sun$}

\newcommand{\teff}{$T_{\rm eff}$}
\newcommand{\logteff}{log\,$T_{\rm eff}$}
\newcommand{\logg}{log\,$g$}
\newcommand{\micro}{$\xi_{\rm turb}$}
\newcommand{\macro}{$\zeta_{\rm macro}$}
\newcommand{\vsini}{$v_{\rm rot} \sin i$}
\newcommand{\vrot}{\mbox{$v_{\rm rot}$}}
\newcommand{\vrad}{$v_{\rm rad}$}
\newcommand{\chitwo}{$\chi^2$}
\newcommand{\chitwomin}{$\chi^2_{\rm min}$}

\newcommand{\kms}{km\,s$^{-1}$}
\newcommand{\mv}{\mbox{$M_V$}}

\newcommand{\vk}{$(V-K)$}
\newcommand{\by}{$(b-y)$}
\newcommand{\ha}{H$\alpha$}
\newcommand{\hb}{H$\beta$}
\newcommand{\marcs}{{\sc marcs}}

\slugcomment{Accepted by ApJ}

\shorttitle{Lithium isotopic abundances in metal-poor halo stars}
\shortauthors{Asplund et al.}

\begin{document}

\title{Lithium isotopic abundances in metal-poor halo stars
\footnote{Based on observations collected 
at the European Southern Observatory, Paranal, Chile
(observing programs 65.L-0131, 68.D-0091 and 273.D-5043)}
}

\author{Martin Asplund}
\affil{Research School of Astronomy and Astrophysics,
Mt. Stromlo Observatory, Cotter Road, Weston, ACT 2611, Australia
\email{martin@mso.anu.edu.au}
}

\author{David L. Lambert}
\affil{The W.J. McDonald Observatory, The University of Texas at Austin, Austin, TX 78712-1083, USA}

\author{Poul Erik Nissen}
\affil{Department of Physics and Astronomy, Aarhus University, DK-8000, Aarhus C, Denmark}

\author{Francesca Primas}
\affil{European Southern Observatory, Karl-Schwarzschild Str. 2,
D-85748 Garching b. M\"unchen, Germany}

\and

\author{Verne V. Smith}
\affil{National Optical Astronomy Observatory, P.O. Box 26732, Tucson, AZ 85726, USA}

\begin{abstract}

Very high-quality spectra of 24 metal-poor
halo dwarfs and subgiants have been acquired with ESO's VLT/UVES for the purpose
of determining Li isotopic abundances. 
The derived 1D, non-LTE \liseven\ abundances from the \lii\ 670.8\,nm line 
reveal a pronounced dependence on metallicity 
but with negligible scatter around this trend.  
Very good agreement is found between the abundances from the \lii\ 670.8\,nm line and
the \lii\ 610.4\,nm line. 
The estimated primordial \liseven\ abundance is $^7{\rm Li/H} = 1.1-1.5 \cdot 10^{-10}$, 
which is a factor of three to four lower than predicted from standard Big Bang nucleosynthesis
with the baryon density inferred from the cosmic microwave background. 
Interestingly, \lisix\ is detected in nine of our 24 stars at the $\ge 2\sigma$ significance level. 
Our observations suggest the existence of a \lisix\ plateau at the level of \loglisix\,$\approx 0.8$;
however, taking into account predictions for \lisix\ destruction  during the pre-main sequence 
evolution tilts the plateau such that the \lisix\ abundances apparently
increase with metallicity. 
Our most noteworthy result is 
the detection of \lisix\ in the very metal-poor star LP\,815-43. 
Such a high \lisix\ abundance during these early Galactic epochs is very difficult to achieve 
by Galactic cosmic ray spallation and $\alpha$-fusion reactions.
It is concluded that both Li isotopes have a pre-Galactic origin. 
Possible \lisix\ production channels include proto-galactic shocks and
late-decaying or annihilating supersymmetric particles 
during the era of Big Bang nucleosynthesis.  
The presence of \lisix\ limits the possible degree of stellar \liseven\ depletion
and thus sharpens the discrepancy with 
standard Big Bang nucleosynthesis. 


\end{abstract}

\section{Introduction}
\label{s:introduction}

Lithium's two stable isotopes -- \lisix\ and \liseven\ -- continue to
pose intriguing  questions
for astrophysicists concerned with
understanding the origins of this light element. 
These questions
concern the abundance of lithium in cool interstellar
gas and every type of star in
which lines of neutral lithium are either detected or are
potentially detectable.
In this paper, we address by new observations questions
about the origin of the lithium seen in the
atmospheres of metal-poor stars. In particular, we  
provide an adequate set of determinations of the isotopic
abundance ratio \liratio\ with which to constrain some theoretical
ideas on the origins of the rare \lisix\ isotope  and
of the majority \liseven\ isotope for which the
Big Bang is widely identified as the initial and major origin.

An observational link of \liseven\ with the Big Bang was promoted first
by Spite \& Spite (1982) who showed that the lithium abundance in
the warmest metal-poor dwarfs was independent of metallicity
for \feh\,$< -1.5$. Their
derived abundance of lithium in the dwarfs was given as \logli\,$= 2.05\pm0.15$ 
on the usual scale where \logh\,$=12.00$.
The constant lithium abundance defining what is commonly called `the
Spite plateau'  suggested that this may be the lithium abundance
in pre-Galactic gas provided by the Big Bang.
In the succeeding years, many observers have reinvestigated the
lithium abundance of the Spite plateau
but the lithium
(\liseven ) abundance has strayed little from the Spites' original
estimate.  Here, we also present a new determination of the
plateau's lithium abundance as extrapolated to zero metallicity that 
remains faithful to the 1982 estimate. 

Early observations  showed that
the lithium on the plateau was primarily \liseven , as expected
from a standard Big Bang: Maurice et al. (1984) set an
upper limit \liratio\,$<0.1$ for a couple of stars.
The first probable detection of \lisix\ in a very metal-poor star was reported by
Smith et al. (1993) for HD\,84937 with 
\liratio\,$= 0.06\pm0.03$. Confirmation was provided by
Hobbs \& Thorburn (1994, 1997), Smith et al. (1998) 
and subsequently by Cayrel et al. (1999) who
obtained \liratio\,$= 0.052\pm0.019$.  
Smith et al. (1998) observed seven single stars and
reported \lisix\ in one additional very metal-poor star 
(BD\,$+26\arcdeg 3578$, also known as HD\,338529). Cayrel et al.
provided a possible detection of \lisix\  in BD\,$+42\arcdeg 2667$ 
as did Nissen et al. (2000) for G\,271-162. Non-detections
 gave upper
limits variously in the range \liratio\,$<0.03-0.10$.

Two questions now face students of the abundances of \lisix\ and \liseven\
in very metal-poor stars. First, predictions of the
nucleosynthesis by the Big Bang
are  tightly constrained  because analysis of the anisotropies
of the cosmic microwave background determine 
the one free parameter in a standard cosmological models
that sets the abundances
(relative to $^1$H) of the products $^2$H, $^3$He, $^4$He, and
\liseven . The free parameter is the ratio of baryons to photons
$\eta = n_{\rm b}/n_\gamma$ or equivalently the fraction of the
critical density $\Omega_b$ provided by baryons: 
$\Omega_{\rm b} = 3.652 \times 10^7\eta/h^2$ where $h$ is the Hubble
constant in units of 100\,\kms\,Mpc$^{-1}$.
The recent WMAP-based analysis of the cosmic microwave background implies
$\Omega_{\rm b} h^2 = 0.0224 \pm 0.0009$ (Spergel et al. 2003), which
in standard Big Bang nucleosynthesis corresponds to 
an abundance of  \logliseven\,$ = 2.65 \pm 0.10$
(Coc et al. 2004; Cuoco et al. 2004; Cyburt 2004).
This predicted abundance of \liseven\ is
about 0.5 dex or a factor of three larger than the measured abundance
of lithium on the Spite plateau.
Thus, one key question has become -- How does one  bridge the 0.5 dex
gap between  observation and prediction?

The second question arises from the detections of $^6$Li. 
Our results suggest that there may be a \lisix\ plateau
parallel to the Spite plateau for \liseven\ with the  implication
that the major fraction of the
\lisix\ (and presumably some of the \liseven ) may have been synthesised
prior to the onset of star formation in the Galaxy. 
The \lisix\
abundance of the plateau exceeds by a large factor the 
\lisix\ expected from a standard Big Bang.  
Thus, our second key question is -- How does one account for the
high abundance of \lisix\ in some metal-poor stars?

Our principal  concern in undertaking the observations reported here
was to extend high-precision measurements of \lisix\
to more metal-poor stars and to lower metallicities. The observations
and the analysis techniques
are described below in Sections \ref{s:observations} and 
\ref{s:analysis}, respectively.
Section \ref{s:parameters} deals with the determination of stellar
parameters, abundances, and macroturbulent velocities.
The derivation of the Li isotopic 
abundances from the observed line shapes of the \lii\ 670.8\,nm resonance
line is outlined in 
Section \ref{s:li6708}, while the analysis of the subordinate \lii\ 610.4\,nm
line is presented in Section \ref{s:li6104}. 
In Section \ref{s:evolution}, we take up the two questions
just outlined: How can one reconcile the predicted \liseven\ abundance
using the WMAP-based $\Omega_{\rm b} h^2$ with the measurement of the
abundance for the Spite plateau? How does one account for those
\lisix\ abundances that greatly exceed the Big Bang prediction?

\section{Observations and data reduction}
\label{s:observations}

\subsection{Selection of Stars}
\label{s:sample}

Programme stars were selected from the $uvby$-$\beta$ catalogue
of Schuster \& Nissen \cite{schuster88} and from the Li abundance
survey of very metal-poor stars by Ryan et al. \cite{ryan99}.
In order to be able to study the possible evolution of the Li isotopes
in the Galactic halo, we aimed at a group of stars covering the
metallicity range $-3.0 <$ [Fe/H] $< -1.0$ and
situated in the turnoff region of the Hertzsprung-Russell-diagram, where the upper
convection zone of the stars is relatively thin so that proton destruction of
Li is minimal.  Furthermore, the stars had to be brighter than
$V \sim 11$ in order to obtain very high
resolution spectra with sufficiently high $S/N$. Finally, known double-lined 
spectroscopic binaries were avoided. 

\subsection{Spectroscopic Observations}
\label{s:spectroscopy}

Spectra of 23 stars were obtained with UVES (Dekker et al. 2000)
on the ESO VLT 8.2m Kueyen
telescope during three observing runs: July 23-25, 2000; February
5-7, 2002, and August 28-31, 2004 (5.5 hours in service mode).
About 1/3 of the time during the first two periods was lost due to 
high humidity.
The seeing varied from about $0.5\arcsec$
most of the time in July 2000 to occasionally more than $2\arcsec$ 
in February 2002. One of the nights in July 2000 was
interrupted by ``Target of Opportunity" observations - later
compensated by service observations on August 4, 2000.

In order to obtain a spectral resolution of about 
$R \simeq 120\,000$, we used image slicer \#3, which for an entrance 
aperture of $1.5\arcsec \times 2\arcsec$ produces five slices each 
$0.3\arcsec$ wide. Thanks to the image slicer, the efficiency of
UVES was quite good even during periods with bad seeing. In average 
conditions a $S/N$ of about
500 per spectral bin (0.00268\,nm at the \lii\ 670.8\,nm line)
could be obtained for a $V \sim 11.0$ star in 2 hours.
The total integration time was always split into at least three separate
exposures so that cosmic rays hits on the CCD could be removed by comparison
of the different spectra.

The July 2000 and February 2002 spectra were 
taken in the red arm of UVES with cross disperser \#3 and a central
wavelength of 705\,nm. With this non-standard setting, the EEV
CCD covers the spectral region 600 - 700\,nm in 15 echelle
orders, while the MIT CCD covers the region 710 - 820\,nm in 11 orders.
We note that the \lii\ 670.8\,nm line is well centered on the
EEV chip and that the orders are separated by more than
140 pixels so that despite of the large width of the orders (about 70
pixels) due to the image slicer, there is ample interorder space for
measuring the CCD background including scattered light. Furthermore,
the sky spectrum could be monitored through an extension of the entrance slit
beyond the position of the image slicer. For further details
we refer to Dekker et al. \cite{dekker02}.

The service observations in August 2004 were carried out with 
a standard UVES setting of the red arm (cross disperser \#3, central
wavelength 600\,nm). In this setting, the EEV chip covers the 500 - 595\,nm
spectral region, and the MIT from 605 to 700\,nm. The primary purpose
with this alternative setting was to check an interesting Li isotopic
abundance ratio obtained for the very metal-poor star LP\,815-43 based
on the July 2000 observations.   

Further details about the spectroscopic observations are given in 
Table \ref{t:observations}.
In addition to the 23 stars listed here, G\,271-162 is also included in the
present paper. Its spectrum was obtained during
the commissioning of UVES in October 1999 with the same setting of UVES 
as in July 2000 and February 2002 but without
the use of the image slicer, which was not available at the time. Instead
the star was continuously moved along the slit to simulate the effect
of the image slicer. This star has already been analyzed by Nissen et al.
\cite{nissen00} but is re-analyzed here for comparison purposes.

\subsection{Data Reduction}
\label{s:reduction}

The spectra have been reduced using standard tasks in the
IRAF echelle package with the following steps included:
{\em i)} bias subtraction,
{\em ii)} fitting a 2-dimensional polynomial to the flux in the interorder 
regions and subtraction of scattered light, 
{\em iii)} pixel-by-pixel division with a normalized flat field exposed without the image
slicer,
{\em iv)} tracing and extraction of the echelle orders,
{\em v)} corresponding extraction of a thorium calibration spectrum,
{\em vi)} wavelength calibration,
{\em vii)} rectification of the echelle orders by fitting cubic spline functions
with a wavelength scale of about 0.5\,nm to the continuum,
{\em viii)} correction for the Doppler shift of the spectral lines.

For some stars this procedure was carried out for the individual slicer
spectra, which were then combined to one spectrum by applying weights
according to the exposure level of the slices. The resulting spectra were
compared to spectra obtained by extraction with a long slit covering all
five slices. The difference between the two sets of spectra is marginal
in terms of $S/N$ and resolution. Hence, the simpler long slit extraction
was applied for all stars as our final reduction procedure. Furthermore,
we note that the echelle orders are not merged, which means that lines
near the edges of the orders, e.g. the oxygen triplet at 777.4\,nm, can
be measured in two consecutive orders.

The sky background spectrum turned out to be negligible for all stars
observed on July 23-25, 2000, August 4, 2000, and February 5-6, 2002. 
On these dates the Moon was close to first or last quarter. The observations 
of LP\,815-43 on August 30 - 31, 2004, however,  took place during full
Moon conditions, and the sky spectrum (corresponding to the extent of
the five slices) turned out to be about 1\% of the signal in the 
spectrum of LP\,815-43. Hence, in this case the sky spectrum was
subtracted.

\begin{figure}
\resizebox{\hsize}{!}{\includegraphics{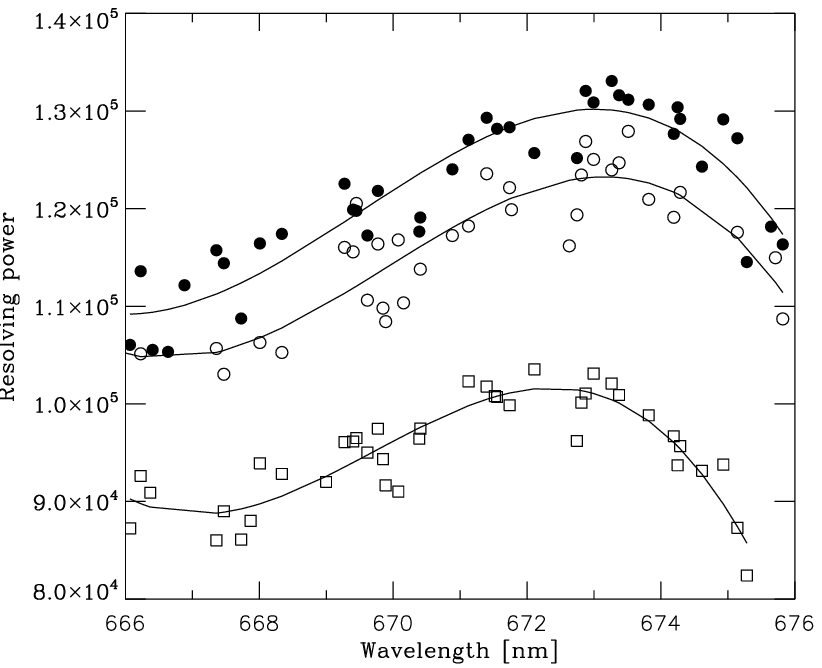}}
\caption{Measured resolving power of UVES based on the FWHM of thorium
lines in the echelle order containing
the \lii\ 670.8\,nm line. Filled circles: July 2000 (EEV CCD); Open circles:
February 2002 (EEV CCD); open squares: August 2004 (MIT CCD). The curves
are 3rd order fits to each set of data.}
\label{f:resolution}
\end{figure}

The wavelength calibration is based on 20-40 thorium lines per
echelle order. The rms of a
third order polynomial transformation between pixel and
wavelength is of the order of 0.12\,pm (1.2\,m\AA ). In addition, the Th
lines can be used to check the resolution of UVES; the intrinsic width
of a typical Th line is negligible compared to the measured width (Hobbs et al. 1999).
The profiles of the Th lines are well approximated by Gaussians, but the
resolving power as defined from the FWHM of the Th lines 
($R = \lambda / \Delta \lambda _{\rm FWHM}$) varies
along a given spectral order from about 100\,000
to 130\,000 on the EEV chip.
As can be seen in Fig. \ref{f:resolution}, we achieved slightly higher
resolving power during our first observing run in July 2000 than in
February 2002. The lower resolution
for the August 2004 spectra is caused by charge diffusion in the
MIT chip, which, as mentioned above, was used in that period.
In any case, the large number of Th lines available, enable us to map
the variation of the resolving power along an echelle order. Hence, the 
actual instrumental broadening for a given 
stellar line is known and can be applied in connection with the spectrum 
synthesis.

In addition to increasing the efficiency, the image slicer in UVES
also serves to broaden the spectrum over a larger area of the CCD.
This helps to minimize problems in the flat-fielding, i.e., to reduce
the amplitude of any residual fringing existing after the flat-fielding.
From exposures of bright B-type stars during the July 2000 observations
we estimate that the residual fringing is less than about 0.1\,\%.
For some unknown reason, the residual fringing turned out to be 
larger for the February 2002 spectra, reaching an amplitude of about 0.3\,\%
on a wavelength scale that could affect the profiles of the spectral
lines. Hence, we decided to determine a ``residual fringe function" from 
the continua of several very metal-poor stars observed in February 2002
taking advantage of the fact that these stars have only a few spectral lines, 
which are shifted to different positions in the spectra due to large 
differences in radial velocities. This residual fringe function 
is  well defined and was applied to all spectra observed in 
February 2002. A similar procedure was applied to the spectra of
LP815-43 obtained on August 30-31, 2004; in this case the spectrum of 
HD\,140283 from August 28, 2004 was used to define the residual fringe function.

\begin{figure}
\resizebox{\hsize}{!}{\includegraphics{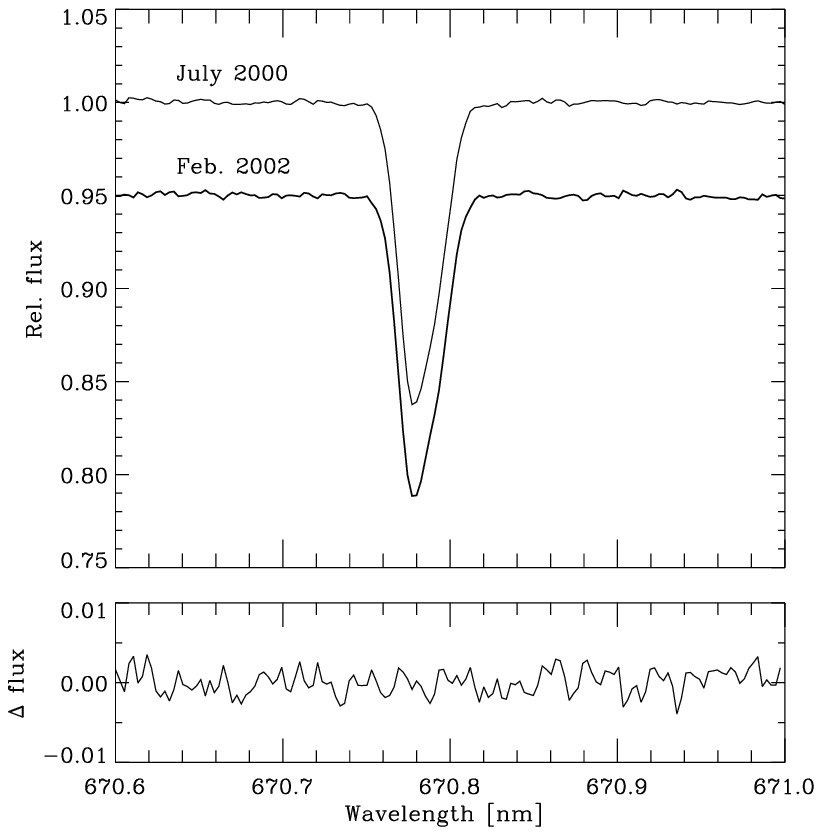}}
\caption{Comparison of the \lii\ region in the July 2000 spectrum
of HD\,140283 with the February 2002 spectrum (shifted by 0.05). The lower
panel shows the difference between the two spectra.}
\label{f:obsspec_HD140283}
\end{figure}

As a test of the quality of the spectra and of possible problems with the
correction for residual fringing in particular, we compare in 
Fig. \ref{f:obsspec_HD140283} the \lii\ spectra of HD\,140283 
([Fe/H] = $-2.4$) obtained in July 2000 and Feb. 2002. 
As seen, there is no significant trend in the residuals,
and the rms of the difference is 0.0016 corresponding to a $S/N$ of about 600.
This is close to what one would expect for the $S/N$ values of each spectrum
given in Table \ref{t:observations}. A similar comparison of the
spectra of HD\,140283 from July 2000 and August 2004 does reveal a 
significant trend in the residuals, but this is mainly due to the 
lower resolution of the August 2004 spectrum. 

\begin{figure}
\resizebox{\hsize}{!}{\includegraphics{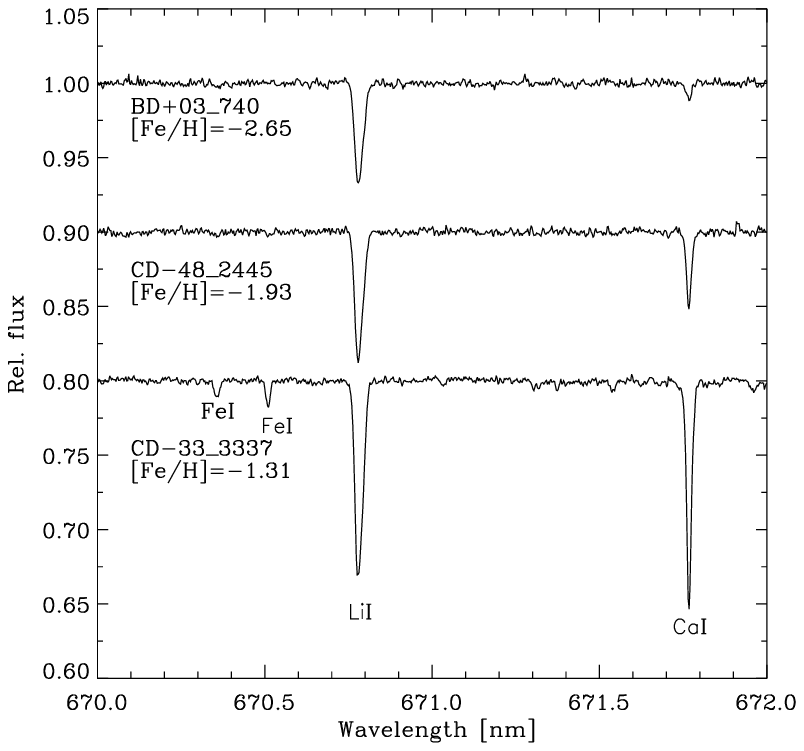}}
\caption{Sample spectra around the \lii\ 670.8\,nm line with the spectra of
CD\,$-48\arcdeg2445$ and CD\,$-33\arcdeg3337$ shifted 0.10 and 0.20,
respectively. Note, that the region
between the \lii\ and the \cai\ lines in the spectrum of CD\,$-33\arcdeg3337$
is affected by several faint lines, which are not identified on the
figure.}
\label{f:obsspec}
\end{figure}

The $S/N$ of the spectra in the region around the \lii\ 670.8\,nm
line ranges from about 400 for some of the very metal-poor stars
to more than 1000 for the relatively bright star HD\,140283.
All but two of the stars have $S/N>500$.
Representative spectra are shown in Fig. \ref{f:obsspec} for a range
of metallicities.

\subsection{Radial Velocities and Binary Stars}
\label{s:radvel}

Our conclusions regarding the lithium isotope ratio could 
be compromised if our targets are binaries.
We have searched available databases for evidence suggesting that our program
stars are not single. The extensive radial velocity survey of 
Carney et al. (1994) contains 11 of our 23 stars. One of them,
G\,126-062, is reported as a single-lined binary; the orbital solution by
Latham et al. (2002) gives a period of 219 days
and a radial velocity amplitude of $K = 4.8$\,\kms . In addition,
Carney et al. list BD\,$+03\arcdeg0740$ as a suspected binary. 
Another accurate survey, that of Nordstr\"{o}m et al. (2004) based on
radial velocities measured with the Coravel spectrometer, contains   
eight of our stars but except for G\,126-062 none of them are noted as binaries.
Finally, Ryan et al. (1999) have nine of the most metal-poor
stars in common with us, but again there is no evidence of radial velocity
variations.

To supplement these studies we have determined radial velocities
for all our stars. The values are based on the measured Doppler shifts
of four of the lines that are used in Sect. \ref{s:broadening} to determine the
macroturbulent velocity broadening. The following laboratory wavelengths 
from the NIST database are adopted: \cai\ 612.2217\,nm, \cai\ 616.2173\,nm,
\fei\ 623.0726\,nm and \cai\ 643.9075\,nm. In determining the radial velocity
one should ideally correct for gravitional redshift and convective
blueshift, but as the net effect is expected to be below 0.5\,\kms\
we have not made such a correction.

The measured radial velocities (corrected to heliocentric values using
the IRAF task {\em rvcorrect}) are given in Table \ref{t:observations}.
The accuracy of the velocities depends on the stability of UVES
because the thorium comparison spectrum was obtained only a few times per
night. Comparing the radial velocities for stars observed on different
nights we estimate that the precision is $\pm 0.3$\,\kms , i.e. similar
or slightly better than the precision obtained in the radial velocity
surveys mentioned above. A comparison to the data in these three sources
and those in the Li isotope study of Smith et al. (1998)
gives the following mean differences (us - them) and rms dispersions of the
differences
\vspace{1mm} (excluding G\,126-062): \\ 
Carney et al. (1994): $\Delta v_{\rm r} = +0.6 \pm 0.4$\,\kms\  (10), \\
Nordstr\"{o}m et al. (2004): $\Delta v_{\rm r} = -0.4 \pm 0.6$\,\kms\  (7), \\ 
Ryan et al. (1999): $\Delta v_{\rm r} = +0.5 \pm 0.3$\,\kms\ (9), \\
\vspace{1mm}Smith et al. (1998): $\Delta v_{\rm r} = -0.4 \pm 0.4$\,\kms\ (7), \\
where the number within parenthesis denotes the number of stars in each comparison. Evidently, there
are small systematic differences in the zero point of the radial
velocities, but the small rms dispersions are  as expected; hence,
no additional evidence for binarity is revealed. In particular, we note
that BD\,$+03\arcdeg0740$ that was suspected to be a binary by 
Carney et al. is included in three of the four comparisons and shows
no sign of variability.

Out of the 23 stars in Table \ref{t:observations}, there are only three
(HD\,213657, CD\,$-33\arcdeg3337$ and CD\,$-48\arcdeg2445$) for which 
our radial velocities could not be compared to other accurate sources. 
The spectra of these stars show no evidence of lines from a companion;
the same is true for the other stars including G\,126-062.
Hence, we feel confident that all stellar spectra can be used for determining
the Li isotopic ratio.

\section{Analysis: Methods, Models and Input Data}
\label{s:analysis}

\subsection{Stellar Model Atmospheres}
\label{s:model_atmospheres}

In the present study, we have made use of both 1D LTE hydrostatic
and time-dependent 3D hydrodynamical stellar model atmospheres.
One-dimensional model atmospheres with convection treated by the mixing length theory
for the particular stellar parameters of our sample
have been generated with the \marcs\ code (Gustafsson et al. 1975; Asplund et al. 1997),
which includes realistic continuous and line opacities for a wide range of
atomic and molecular species. The radiative transfer in the construction of
the model atmospheres is treated by opacity
sampling. As all our stars are metal-poor, the chemical composition of the models
have an over-abundance of the $\alpha$-elements of \alphafe\,$=+0.5$ 
and a helium abundance of \loghe\,$=10.93$ (Asplund et al. 2005).
This He abundance corresponds to a He mass fraction of $Y=0.253$, which is
close to the expected value from standard Big Bang nucleosynthesis (e.g. Steigman 2005).
For test purposes we have also employed theoretical 1D Kurucz model atmospheres 
without convective overshoot as computed by Castelli et al. (1997). Those models are based on very much the same
approximations and assumptions as our standard \marcs\ models and have
been interpolated to the specific stellar parameters of our sample from
the available grid of models.

A novel feature of our analysis compared with the pioneering works on Li isotopic
abundances in metal-poor stars of Andersen et al. (1984),
Smith et al. (1993, 1998), Hobbs \& Thorburn (1994, 1997), Nissen et al. (1999),
Hobbs et al. (1999) and Cayrel et al. (1999)
\footnote{We recognize even earlier observational studies by,
for example, Herbig (1964), Feast (1966) and Cohen (1972) who attempted
to derive Li isotopic abundances using both the center-of-gravity and profile
fitting methods. These works were  restricted to Galactic disk stars
and yielded no convincing \lisix\ detections.},
is our use also of time-dependent 3D hydrodynamical stellar
model atmospheres for the stars for which such 3D models with appropriate
stellar parameters exist. The 3D model atmospheres of metal-poor stars are taken from
Asplund et al. (1999) and Asplund \& Garc\'{\i}a P\'{e}rez (2001). In short, the
models have been constructed by solving the standard hydrodynamical equations of
conservation of mass, momentum and energy coupled with a simultaneous
solution of the 3D radiative transfer in a representative volume of the
stellar atmosphere and near sub-surface layers
(e.g. Stein \& Nordlund 1998; Asplund et al. 2000).
The 3D models are based on realistic continuous and
line opacities and equation-of-state computed for the appropriate
\feh\ under the assumption of LTE and opacity binning (Nordlund 1982).
At any given time the simulation
box is sufficiently large to cover at least ten upflowing regions (granules).
The time sequence for each simulation employed is at least one convective
turn-over time divided into thirty or more  individual snapshots to give
statistically significant results for the spatially and temporally averaged
line profiles.
In contrast to the 1D model atmospheres, no mixing length theory enters for
the calculation of the convective energy flux. Instead, the time-dependent
hydrodynamical simulation automatically and self-consistently predicts
the convection efficiency as a function of depth for the given stellar parameters.

\subsection{Spectral Line Formation}
\label{s:line_formation}

The spectral line formation is treated under the assumption of (strong) LTE making 
the source function equal to the Planck function:
$S_\nu \left(\tau\right) = B_\nu \left[T\left(\tau\right)\right]$, i.e. continuum scattering
is treated as an absorption. 
The lines used in our study are located at  
wavelengths at which
Rayleigh and electron scattering make a negligible contribution to the
source function; we have ensured that identical results are obtained with the code used
herein and the code {\sc bsyn} from the Uppsala spectrum synthesis package that
treats continuum scattering properly within the LTE framework.
The radiative transfer is solved with a Fortran implementation of the
Feautrier method to which is coupled an IDL routine for the
\chitwo -analysis to find the optimum solution in terms of
element abundance, wavelength shift, continuum normalization and extra
broadening beyond the instrumental broadening
(1D: macroturbulence and rotation, 3D: rotation only).
The rotational broadening is performed through a disk-integration
of the angle-dependent intensity profiles.
In the 1D case, microturbulence is prescribed and not part of
the \chitwo -optimization while the parameter does not enter the 3D calculations.
The same code is able to handle both 1D and 3D geometries with
otherwise identical calculations, which is a great advantage when investigating
the possible systematic errors when relying on 1D plane-parallel model atmospheres.
The background continuous opacities are taken from the \marcs\
project (Gustafsson et al. 1975; Asplund et al. 1997; Gustafsson et al.,
in preparation).
We have verified that identical \liratio\ ratios and absolute
Li abundances to within 0.01\,dex would result
with the independent spectrum synthesis code {\sc bsyn} which comes as part
of the Uppsala package.

The resulting velocity field from the predicted
convection and waves in the 3D models introduce Doppler
shifts, which translate to
shifts and asymmetries of spectral lines.
These will depend on the line formation depths and 
the atmospheric conditions (in particular temperature, pressure and velocity)
for the individual lines, making the detailed line shapes sensitive probes of
the stellar atmospheres. As will become clear in
Sect. \ref{s:li6708_3D}, this results in a significant improvement in
the ability to reproduce the detailed shapes of spectral lines
(e.g. Dravins \& Nordlund 1990; Asplund et al. 2000;
Nissen et al. 2000; Allende Prieto et al. 2002).
Since the presence of \lisix\ in the stellar atmosphere manifests
itself as an asymmetry in the \lii\ 670.8\,nm line in excess
of the unresolved fine structure of the \liseven\ line, it
is clearly important to have an understanding of other possible
mechanisms which can introduce line asymmetries, such as convection.
It should be emphasized that no microturbulence or macroturbulence
is necessary for the calculation of the line formation in 3D.

Our final \liratio\ ratios will
be taken from the 1D analysis rather than from the
3D calculations. Partly, this is due to the simple fact that
3D models do not yet exist for all combinations of stellar
parameters of our sample.
Another reason is that the very steep photospheric temperature
gradients in 3D simulations of metal-poor stars are likely conducive
to significant departures from LTE, at least for neutral minority
species (Asplund et al. 1999).
This has already been confirmed by detailed multi-level 3D non-LTE calculations
for Li (Asplund et al. 2003; Barklem et al. 2003) which showed significant
over-ionization of neutral lithium but the lithium abundance from
3D non-LTE calculations was not greatly different from 1D non-LTE
calculations.
Likewise, one may expect similarly large 3D non-LTE effects for
many of our calibration lines of for example \fei\ and \cai\ used to determine
the intrinsic atmospheric line broadening (Asplund et al. 1999; Asplund 2005).
Perhaps, more important will be the line asymmetries inherent in
predictions for 3D model atmospheres and which will depend somewhat
on the thermodynamic conditions (LTE or non-LTE).
Under these circumstances, the derived \liratio\ ratios from a 3D LTE
analysis may be prone to systematic errors.
While the above-mentioned 3D non-LTE studies based on only a couple of
snapshots from a simulation are expected to provide
a good estimate of the overall Li line {\em strength} and thus the 
Li abundance, predicting the detailed line {\em shape} requires taking
the temporal average of many more snapshots to obtain statistically significant
profiles. Unfortunately, performing
3D non-LTE calculations for long time-sequences for a stellar sample our size
even for a relatively simple atom like Li is beyond reach of current computers.
Therefore, the here presented 3D LTE calculations will mainly serve as
a test of the 1D results for the Li isotopic ratios to assess possible
systematic errors.

\subsection{Atomic Data}
\label{s:atomic_data}

The basic data for the \lii\ 670.8\,nm resonance line is fortunately
very well-known.
The wavelengths for the individual hyperfine and isotope components
are taken from the highly precise frequency measurements reported
by Sansonetti et al. (1995), which are accurate to within
0.01\,pm (0.1\,m\AA ). The $gf$-values for the different
\liseven\ and \lisix\ components are based on the calculations by
Yan et al. (1998, see also Yan \& Drake 1995) that give
$f=0.7467871\pm0.0000010$ and $f=0.7467527\pm0.0000010$ for
the \liseven\ and \lisix\ doublets.
Recent experimental values are very similar to the theoretical
calculations. For example, the radiative lifetime of the
$2p$ upper level ($27.11\pm0.06$\,ns) measured by Volz \& Schmoranzer (1996)
is very similar to the prediction by Yan et al. ($27.117301\pm0.000036$\,ns).
Our adopted wavelengths and $gf$-values are the same as in Hobbs et al. (1999)
and Smith et al. (1998).
The transition probabilities and wavelengths for the isotope
and hyperfine components for the subordinate \lii\ 610.4\,nm
were also taken from Sansonetti et al. (1995) and Yan et al. (1998).
While not important for  the \lii\ lines, we note for
completeness that we use the accurate pressure broadening
data of Barklem et al. (2000b).

The transition probabilities for the lines of other species
employed in the present study are taken from a variety of sources,
all of which are expected to be accurate and thus not introduce any
severe systematic errors. The log\,$gf$-values for the \feii\ lines
are from Bi\'{e}mont et al. (1994) while those of the \fei\ lines
are from O'Brian et al. (1991). Both of those should
be accurate to within $\pm 0.03$\,dex.
The corresponding data for the \oi\ triplet at 777\,nm comes from
Wiese et al. (1996):
log\,$gf = 0.369$, 0.223, and 0.002
The transition probabilities for the
\cai\ lines are from Smith \& Raggett (1981)
and for the \ki\ line (log$gf = -0.17$) from Reader et al. (1980).
Hyperfine and isotope splitting have been taken into account
for the \ki\ line although the effects turned out to be negligible.
We emphasize that the exact choice of transition probability for
our calibration lines used in our determination of the Li isotopic
abundances is of no consequence for the determination of the
line broadening in the \chitwo -analysis.
The excitation potential of the lower level for the lines is taken
from the NIST database while the radiative broadening comes from
the VALD database. For neutral lines we rely on the pressure
broadening from Barklem et al. (2000b) while for \feii\ 
the classical Uns\"old theory with an enhancement factor of $2.0$
has been adopted; the recent improvements in computing line
broadening for \feii\ lines by Barklem \& Aspelund-Johansson (2005)
only became available after our analysis was completed.
We emphasize that as our \feii\ lines are all 
weak ($W_\lambda \la 4$\,pm), the exact
treatment of the pressure broadening for these lines is of no importance
for our purposes.

\section{Stellar Parameters}
\label{s:parameters}

\subsection{Effective Temperature}
\label{s:teff}

In a previous paper by Nissen et al. (2002) on oxygen abundances
derived from the \oifor\ line in a subset of the same UVES spectra as 
employed here, the effective temperature
was derived from the $b-y$ and $V-K$ colour indices calibrated
on the infrared flux method (Alonso et al. 1996). The $1\sigma$ relative
error of \teff\ was estimated to be about $\pm 70$\,K with
the largest contribution coming from the uncertainty in the interstellar
reddening $\Delta (E(b-y)) \simeq \pm 0.01$. Since it is of great importance
for our discussion of the total Li abundance in the stars to obtain a higher
precision of \teff , we derive here the effective temperature
from the profile of the \ha\ line, which is very sensitive to temperature
and practically independent of interstellar reddening.

The \ha\ line is well centered in one of the echelle orders on the EEV
CCD. The main problem in getting an accurate profile is to determine
the echelle blaze function of the \ha\ order. The method of fitting
cubic spline functions to the continuum, which works well for narrow,
well isolated lines like the \lii\ line, cannnot be applied to a broad
line like \ha . Hence, we made a special reduction of
the \ha\ echelle order taking advantage of the fact that the blaze
function varies smoothly and slowly with echelle order. This allows
us to determine the blaze function of the \ha\ order by interpolating
in the blaze functions of the two adjacent orders. 
A similar technique has been successfully applied by
Barklem et al. (2002) and Korn (2002) for various
echelle spectrographs, including UVES.

The \ha\ profile is affected by a few metallic lines and several
telluric H$_2$O lines, but due to the high resolution of the spectra these
narrow absorption lines can easily be removed by interpolating the
\ha\ profile over the disturbing lines.

\begin{figure}
\resizebox{\hsize}{!}{\includegraphics{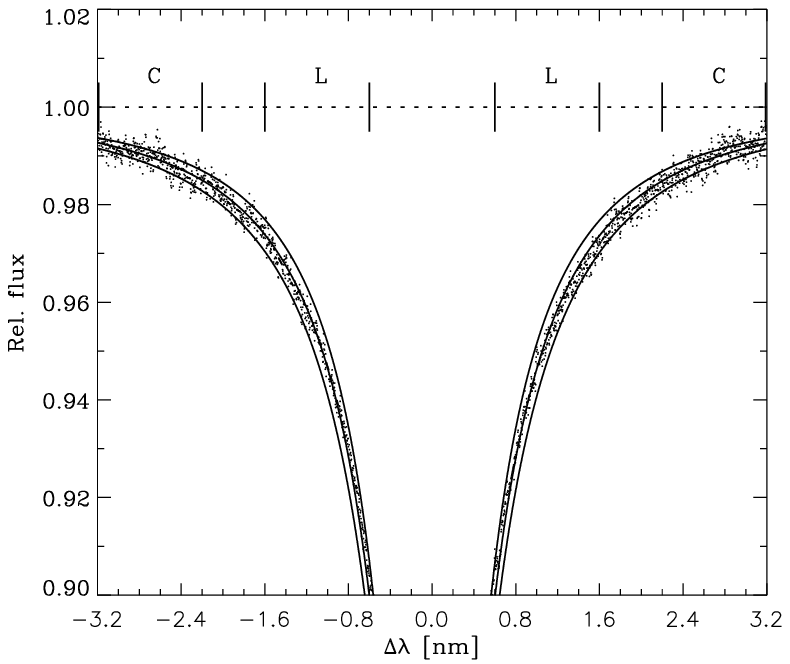}}
\caption{Observed \ha\ profile for BD\,$+03\arcdeg0740$ (dots)
compared to theoretical profiles calculated for atmospheric models
with the gravity and metallicity of the star (\logg\,$ = 4.04$,
\feh\,$ = -2.65$) and three values of the effective temperature:
\teff = 6170\,K, 6270\,K (best fit), and 6370\,K.
The spectral bands used in the definition of the \ha\ index is marked.}
\label{f:Halpha_BD03_0740}
\end{figure}

The observed \ha\ profiles have been compared to theoretical profiles
computed for \marcs\ model atmospheres using the Uppsala
LTE spectrum synthesis code {\sc bsyn}. Stark broadening is based on
calculations of Stehl\'{e} \& Hutcheon (1999) with the model
microfield method, and self-broadening of hydrogen from Barklem et al.
(2000a) is included. The profiles are computed for
models having the same gravity and metallicity as the star but with a
range of effective temperatures embracing the value estimated from the
colour of the star. An improved \teff\ is then determined from
\chitwo -fitting to the observed profile (see Fig. \ref{f:Halpha_BD03_0740}).
The region within $\pm 0.6$\,nm from the center is not included in the fit
because the core of \ha\ is affected by deviations from LTE
(Przybilla \& Butler 2004; Barklem, in preparation).
Furthermore, we note that the continuum setting is considered as a free
parameter in the fit, because the true continuum is ill-defined due to
the limited width of the \ha\ echelle order.
As an alternative to the \chitwo -fitting we have also determined \teff\
from an \ha\ index defined as the ratio between the flux in the
two regions marked by `C' in Fig. \ref{f:Halpha_BD03_0740} to the flux in the
regions marked by `L'. Comparison with model atmosphere calculations of
the index then yields \teff . As with the \chitwo -fitting, the derived
\teff\ does not depend on the continuum normalization.

The two methods to determine \teff\ give consistent results within
$\pm 20$\,K. We also note that the \teff\ derived for a given star from spectra
observed on different nights is extremely stable with variations being less
than $\pm 15$\,K. The calculated \ha\ profile is relatively
insensitive to errors in the gravity and the metallicity of the models.
Altogether, we estimate that {\em relative} temperatures
for our sample of metal-poor turnoff stars are determined with a $1\sigma$
precision of about $\pm 30$\,K. As discussed in Sect. \ref{s:feh}, the
good agreement between Fe abundances derived from \fei\ and \feii\ lines,
respectively, suggests that this small error estimate is realistic.

The {\em absolute} temperature scale is, however, more uncertain than $\pm 30$\,K,
as it depends on the theory of hydrogen line broadening among other things.
Had instead the older Ali \& Griem
(1966) theory of resonance broadening been applied instead of the
more complete self-broadening theory of Barklem et al. (2000a),
the \teff\ derived would be raised by about 100\,K. 
In addition, there are uncertainties in the absolute \teff\ values of 
at least 100\,K arising from the current generation of 1D model atmospheres,
which may not capture the effects of convection in a realistic 
manner. Detailed
statistical equilibrium calculations for \hi\  based on available
e$^-$+\hi\ and \hi +\hi\ collision cross-sections and 1D model
atmospheres reveal no
departures from LTE for the wings of \ha\ 
while the line cores are much better described
(Przybilla \& Butler 2004; Barklem, in preparation).
For completeness, we note that the adopted He abundance has no effect on
the \ha -based \teff : a 0.2\,dex lower He abundance, as perhaps induced
by elemental diffusion in the star, would only increase
\teff\ by about 8\,K.

\begin{figure}
\resizebox{\hsize}{!}{\includegraphics{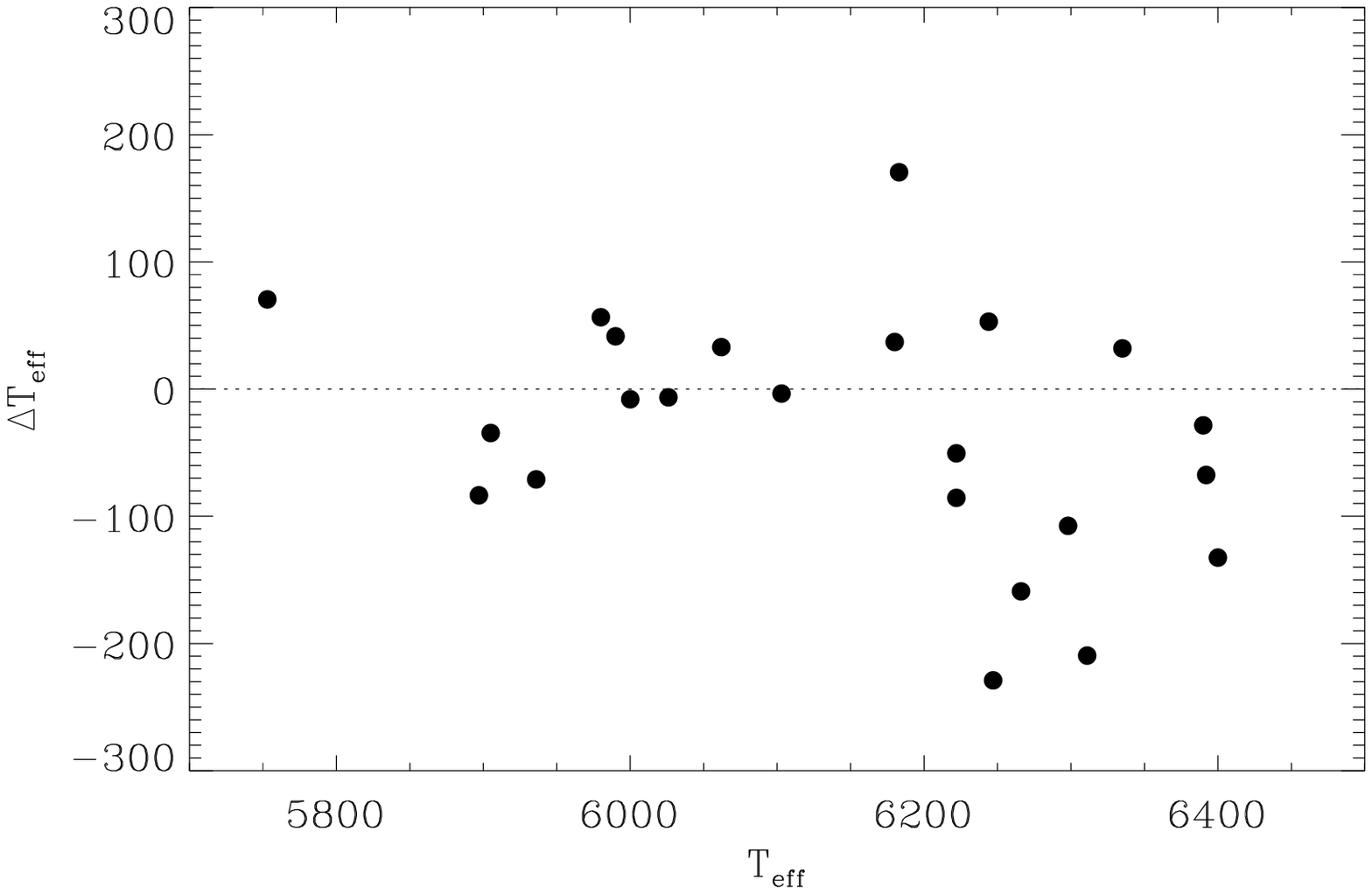}}
\resizebox{\hsize}{!}{\includegraphics{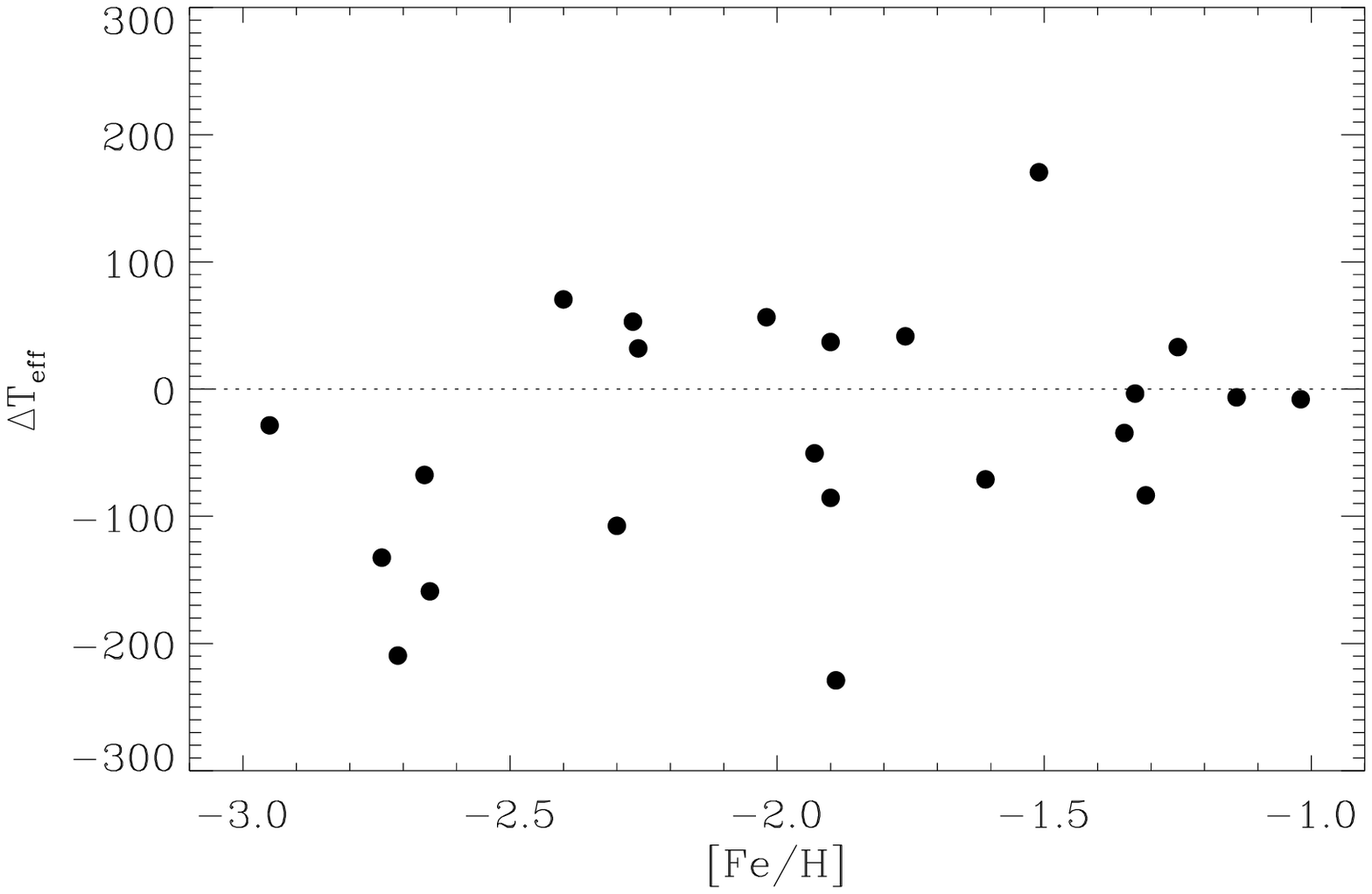}}
\caption{Difference in \teff\ as estimated from \ha\ and from \by\ and \vk\ photometry 
calibrated on the infrared flux method as a function of \teff\ ({\em upper panel}) 
and \feh\ ({\em lower panel}). }
\label{f:dteff}
\end{figure}

The temperatures derived from \ha\
are given in Table \ref{t:parameters}.
When compared to the effective temperatures in
Nissen et al. (2002, 2004) determined from $b-y$ and $V-K$ calibrated
on the infrared flux method, 
good agreement for the majority of the stars is found: 
the mean difference is  $\Delta T_{\rm eff} = -34 \pm 95$\,K.
A few stars show large differences:
the \ha -based \teff\ for G\,126-062 is 170\,K higher while
for two stars (G\,020-024 and BD\,$-13\arcdeg3442$)
the photometric temperatures are higher by about 200\,K. The reason 
for the latter is
probably an overestimate of the reddening for the two stars in
Nissen et al. (2002, 2004).
Overall the agreement is quite satisfactory, and
there are no obvious trends in the temperature differences with
\teff\ or \feh , as seen in Fig. \ref{f:dteff}. 
We are thus unable to find any evidence from \ha\ for the
very hot \teff -scale of Melend\'{e}z \& Ram\'{i}rez (2004) and
Ram\'{\i}rez \& Melend\'{e}z (2005a,b) based on their reapplication of the infrared
flux method. 

\subsection{Surface Gravity and Absolute Magnitude}
\label{s:logg}

The surface gravity is determined from the fundamental relation
\begin{eqnarray}
 \log \frac{g}{g_{\sun}} & = & \log \frac{\cal{M}}{\cal{M}_{\sun}}
 + 4 \log \frac{T_{\rm eff}}{T_{\rm eff,\sun}} + \\
    &   & 0.4 (M_{bol} - M_{bol,\sun})  \nonumber
\end{eqnarray}
where $\cal{M}$ is the mass of the star and $M_{bol}$ the absolute bolometric
magnitude.

The absolute visual magnitude \mv\ is derived from
the Hipparcos parallax (ESA 1997), if available, and also from 
Str\"{o}mgren photometry using the method and calibrations
in Nissen et al. (2002, 2004). The only exception
from the procedure described by Nissen et al. is that we used
the spectroscopic \ha\ index (Sect. \ref{s:teff}) 
instead of the photometric \hb\ index to derive the interstellar 
reddening excess. This provides a more precise estimate of $E(b-y)$,
and consequently the limit for performing a correction of the
Str\"{o}mgren photometry for interstellar absorption and reddening
was set at $E(b-y) > 0.010$ instead of $E(b-y) > 0.015$ in
Nissen et al. (2002, 2004).

If the parallax is available with an accuracy $\Delta \pi / \pi < 0.3$,
the mean value of $M_{V,phot}$ and $M_{V,par}$ is adopted for \mv ,
otherwise $M_{V,phot}$.
The bolometric correction is taken from Alonso et al. (1995), and
the mass of a star is derived by comparing its position in the 
\mv -- \logteff\ diagram with  evolutionary tracks 
calculated by VandenBerg et al. (2000).

As the determination of \logg\ depends on \teff\ directly and \feh\ indirectly
the determination of the stellar parameters has to be iterated.
The final values are given in Table \ref{t:parameters} together
with values for \feh\ and \oh\ (Sect. \ref{s:feh} and Sect. \ref{s:oh}) 
and the microturbulence \micro\ (Sect. \ref{s:broadening}).
The internal error of the derived value of \logg\ is dominated by the error
of \mv , which we have estimated to be $\Delta$\mv\,$ = \pm 0.20$.
This corresponds to $\Delta$\logg\,$= \pm 0.08$\,dex.
While the VandenBerg et al. (2000) models do not include the effects of
diffusion, this translates to the surface gravity possibly being 
overestimated by at most 0.04\,dex (e.g. Proffitt \& Michaud 1991).
We note that this effect is much too small to impact any of our
conclusions in what follows.

\begin{figure}
\resizebox{\hsize}{!}{\includegraphics{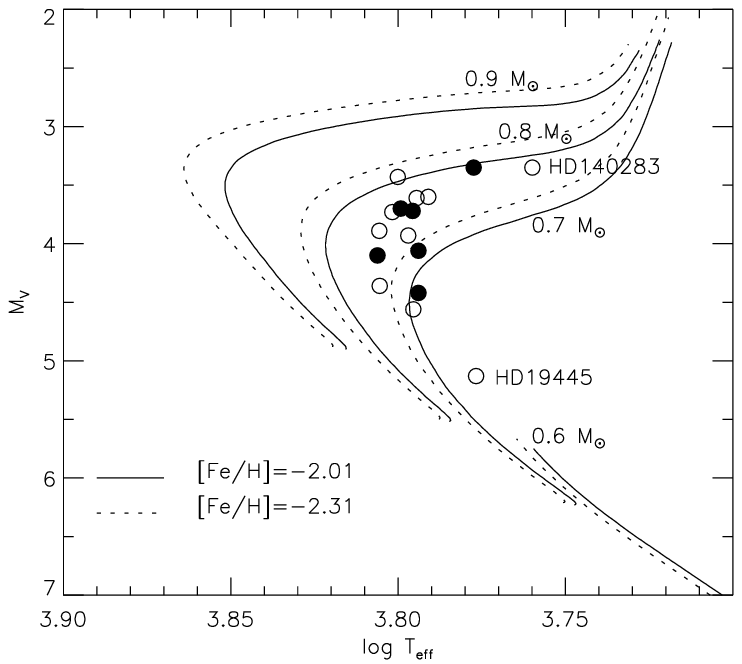}}
\resizebox{\hsize}{!}{\includegraphics{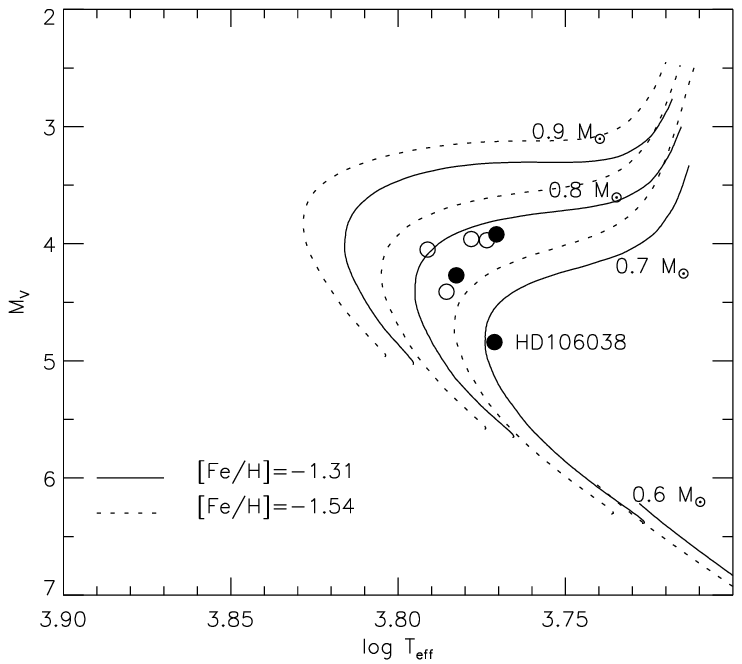}}
\caption{HR-diagram for stars with \feh $< -1.7$ ({\em Upper panel}) and 
for stars with \feh $> -1.7$ ({\em Lower panel}). Filled circles refer
to stars with a $\ge 2\sigma$ detection of \lisix , and open circles to
stars with no clear detection of \lisix . The displayed evolutionary tracks 
are from VandenBerg et al. (2000).}
\label{f:HR}
\end{figure}

The upper panel in Fig. \ref{f:HR} shows the \mv -- \logteff\ diagram for stars with 
\feh $< -1.7$ compared to two sets of evolutionary tracks from 
VandenBerg et al. (2000) corresponding to 
\feh = $-2.31$ (the mean metallicity of the stars) and $-2.01$, respectively,
and an $\alpha$-element enhancement of \alphafe\ = +0.3. 
Except HD\,19445 (a main sequence star) and HD\,140283 (a subgiant),
the stars are close to the turnoff. Some of the scatter in \teff\
around the turnoff can be attributed to differences in \feh\
with the most metal-poor stars being warmest. The lower panel of
Fig. \ref{f:HR} shows the location of the eight stars with \feh $> -1.7$
in the HR-diagram.
They have a mean metallicity of \feh = $-1.33$. Again all stars 
except perhaps HD\,106038 are classified as turnoff stars. As
discussed in Sect. \ref{s:trends}, HD\,106038 has a peculiar high
Li abundance.

\subsection{Metallicity}
\label{s:feh}

Iron abundances have been determined as described in detail by
Nissen et al. (2002) from 13 \feii\ lines, six of which
are in our spectra. The remaining seven lines lie in the spectral
region 490 - 550\,nm and were measured from UVES spectra with
$R \, \simeq 55\,000$ and $S/N \, \simeq 300$ obtained with the primary
aim of determining beryllium abundances (Primas et al., in preparation).
For the very metal-poor stars BD\,$+09\arcdeg2190$,
BD\,$+03\arcdeg0740$ and BD\,$-13\arcdeg3442$ only two
of the \feii\ lines (519.76 and 523.46\,nm) could
be measured reliably. For these stars, we added the somewhat stronger
\feii\ 492.32\,nm line to improve the precision of \feh .
For the most metal-poor star in our sample, CD\,$-33\arcdeg1173$,
we do not have a spectrum in the 490 - 550\,nm region. Hence,
\feh\ could not be determined from \feii\ lines, but was
estimated from \fei\ lines taking into account the offset of 0.08 dex
between \feh\ from \feii\ and \fei\ lines discussed below.
The \feii\ lines were also measured in the solar flux spectrum
(Kurucz et. al. 1984), allowing differential iron abundances \feh\ to be
determined without knowledge of the exact oscillator strengths of the lines.
We note  that for an adopted solar iron abundance of
\logfe\,$ = 7.50$, the $gf$-values derived from the solar spectrum
agree well with those of Bi\'emont et al. (1991) using 
the {\sc marcs} solar model atmosphere.

The Fe abundances for the Sun and our sample of stars were determined from
the observed \feii\ line strengths using the Uppsala {\sc eqwidth} program
together with 1D \marcs\ model atmospheres having the appropriate stellar
parameters. As \teff , \logg\ and \feh\ are interlinked,
the determination of \feh\ is an iterative procedure, with the final values listed
in Table \ref{t:parameters}.
The microturbulence \micro\ used in these calculations was adjusted such that
the derived Fe abundances show no trend with line strength. For \feh\,$\la -2.0$,
the employed \feii\ lines are weak and, hence, the derived Fe abundances are
practically independent of \micro . For those stars, a microturbulence of 1.5\,\kms\
has been adopted.

The error in \feh\ induced by the uncertainties in \teff\ is small:
$\Delta$\feh\,$\la \pm 0.02$\,dex.
Since \feii\ is the dominant ionization stage in the line formation region of these stars,
the derived \feii\ based abundance is dependent on \logg : an error in
\logg\ of $\pm 0.1$ translates to an error in \feh\ of $\pm 0.035$. Adding in quadrature
the error indicated by the line-to-line scatter with those arising from
the uncertainties in  \teff\ and \logg\ implies a statistical error of $\pm 0.05$\,dex in \feh .
The systematic error in \feh\ may be somewhat larger.

\begin{figure}
\resizebox{\hsize}{!}{\includegraphics{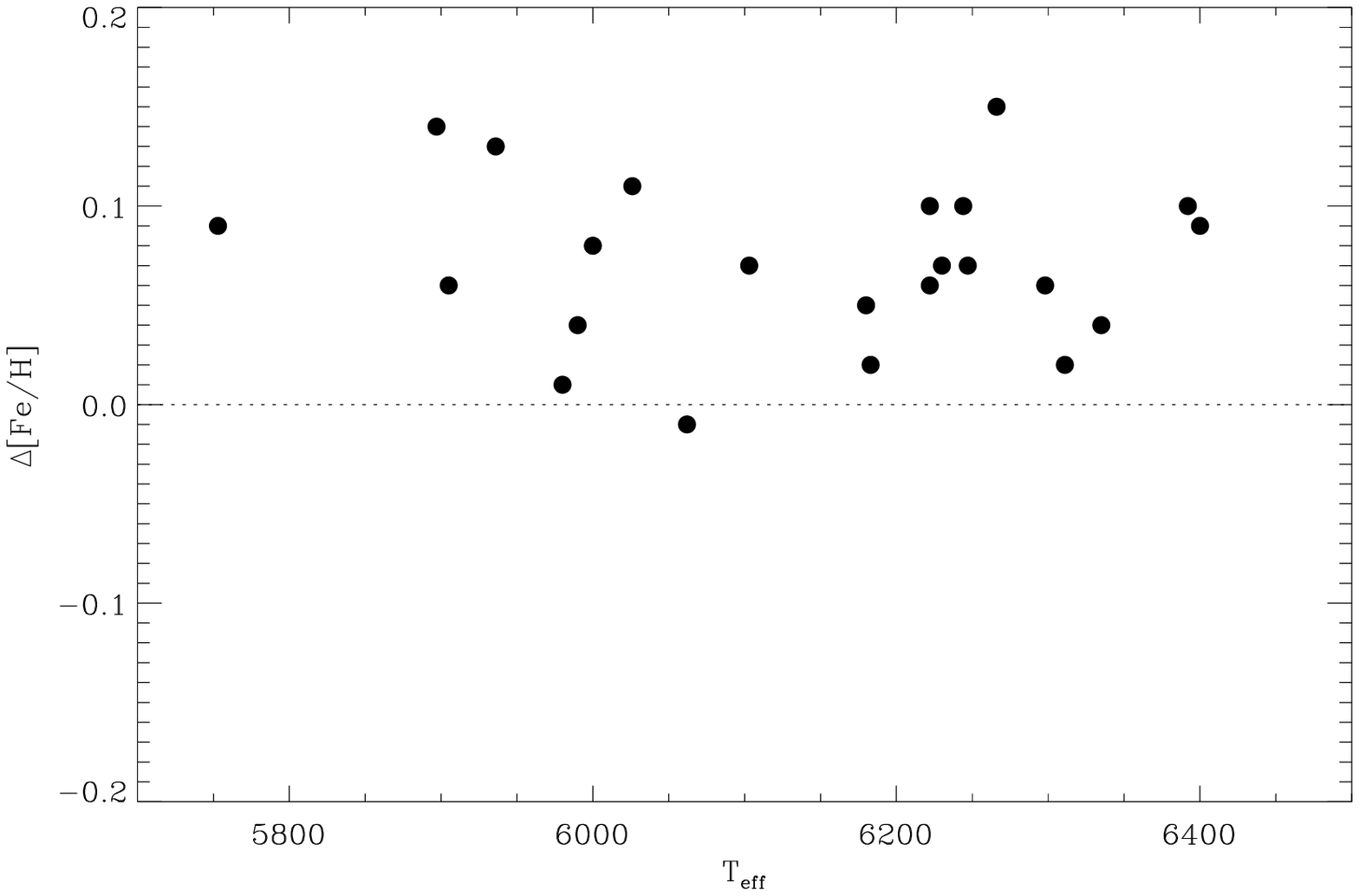}}
\resizebox{\hsize}{!}{\includegraphics{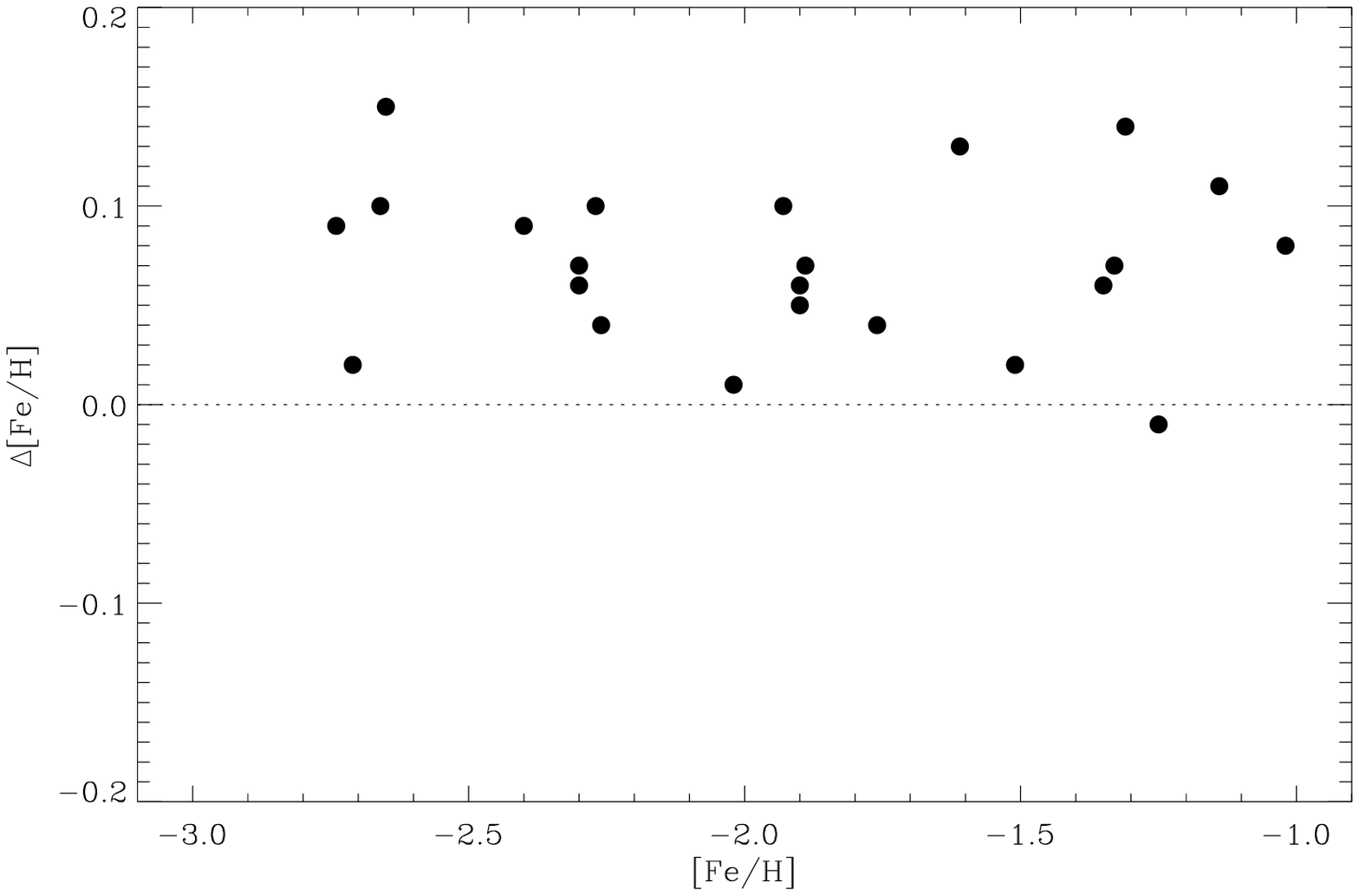}}
\caption{Differences in \feh\ estimated from \feii\ and \fei\ lines as a function of
\teff\ ({\em upper panel}) and \feh\ ({\em lower panel}). 
While there is an offset of about 0.1\,dex, there are no 
obvious trends with stellar parameters. This difference may signal 
differential non-LTE effects for \fei\ relative to the Sun or slightly too
low adopted \teff\ by about 100\,K.}
\label{f:fei_feii}
\end{figure}

In addition, we have measured Fe abundances based on up to 11
weak \fei\ lines using $gf$-values from O'Brian et al. (1991).
As shown in Fig. \ref{f:fei_feii}, the \fei\ lines systematically give
\feh\ values slightly lower than those from \feii\ lines. The average difference,
\feiih - \feih, is 0.08\,dex with an rms scatter of only 0.041\,dex.
The expected scatter arising from the adopted errors,
$\Delta$\teff\,$=\pm 30$\,K and $\Delta$\logg\,$=\pm 0.08$\,dex,
is 0.046 dex, i.e., not significantly different from the observed scatter.
This suggests that the estimated precision of \teff\ and \logg\
is realistic.

The difference in the derived Fe abundances
from \fei\ and \feii\ lines could reflect departures from
LTE, most likely for \fei\ (see discussion in Asplund 2005)
or erroneous stellar parameters (e.g., \teff\ too low by about
100\,K).
The small abundance differences and the
 lack of a trend in \feiih -\feih\ with either \teff\ or \feh\ is
a strong argument against pronounced departures from LTE for 1D
atmospheres, as claimed
for example by Th\'evenin \& Idiart (1999).

Our \feh\ values are systematically
 higher by about 0.2\,dex than
those of Ryan et al. (1999) for the 11 stars in common. Ryan et al. adopted
literature values of \feh\,
 mainly based on \fei\ lines, with a supporting role played by
a calibration of medium resolution spectra. It is likely that this 0.2\,dex
difference is largely attributable to the discrepancy
between \fei\ and \feii\ lines mentioned above, the lower \teff -scale of
Ryan et al. and/or differences in the adopted $gf$-values.
Given the remaining uncertainties surrounding \fei\ lines in terms of
possible departures from LTE, we believe it is wise to rely on
\feii\ lines for the estimate of \feh .

\subsection{Oxygen abundance}
\label{s:oh}

Oxygen may be a preferred alternative to iron in tracking the
evolution of the lithium isotopes in the early Galaxy. Oxygen is synthesised
primarily in the most massive stars and distributed when 
those stars explode as Type II
supernovae; \liseven\ may also be made in these
supernovae by the $\nu$-process (Woosley et al. 1990). Also,
when spallation is an effective mode of lithium
synthesis, spallation of oxygen nuclei is the dominant mechanism.
Iron is also a product of Type II supernovae but in contrast to
oxygen its yield
is dependent on the mass cut between the stellar remnant
(neutron star or black hole) and the ejecta and iron is also a
product of Type Ia supernovae. The latter factor is not a
major concern here as our stars have a metallicity below that
at which Type Ia supernovae are generally considered to
begin their contributions. 

The O abundances for our program stars have been derived
from the \oi\ triplet at 777\,nm.
Unfortunately, the alternative abundance indicator, the \oifor\ line at 630\,nm,
is undetectable for \feh\,$\la -2.5$, even with the very high $S/N$ of our spectra.
We have, therefore, opted for using the same diagnostic for the entire sample.
The \oi\ triplet typically yields somewhat higher O abundances than
the forbidden \oifor\ line at low metallicity (see discussion in Asplund 2005
and references therein).

The  \oi\ 777.1, 777.4 and 777.5\,nm lines are located towards the
edges of two consecutive echelle orders from which the equivalent widths
were measured and averaged (weighted by their respective $S/N$).
The results for the July 2000 spectra have already been presented in 
Nissen et al. (2002) to which we now add nine stars from February 2002.
From the equivalent widths the 1D LTE O abundances were derived
using the stellar parameters
listed in Table \ref{t:parameters}; it should be noted that the O abundances
for the July 2000 targets differ slightly from those given in Nissen et al. (2002)
due to our use of \ha -based \teff\ rather than the photometric values employed in
our previous study.
We have also computed 1D non-LTE abundance corrections
for the \oi\ triplet lines in an identical fashion to the calculations presented
in Nissen et al. (2002). These non-LTE corrections are always negative and
typically $-0.1...-0.2$\,dex but reach $-0.3$\,dex at the lowest \feh . 
Our final adopted O abundances given in Table \ref{t:parameters} include
these non-LTE corrections. These abundances are referenced relative to
the same solar O abundance from a 1D solar model atmosphere 
as adopted in Nissen et al. (2002), i.e. 
\logo\,$=8.74$, which is 0.08\,dex higher than the current best estimate
based on 3D hydrodynamical model atmospheres (Asplund et al. 2004, 2005).

Nissen et al. (2005a) have discussed in some detail the trend and scatter
of \ofe\ as a function of \feh\ for our sample. Here we just note that \ofe\
increases from about 0.4~dex at \feh\ = $-1.0$ to 0.6~dex at \feh\ = $-3.0$. 
The  scatter in \ofe\ around a linear fit to the data
(\ofe = $-0.11$\,\feh\ + 0.31) is 0.08~dex
with a large contribution coming from a relatively low \ofe\ in HD\,160617, 
a star that is known to belong to the rare class of nitrogen-rich subdwarfs 
(Bessell \& Norris 1982).

\subsection{Spectral Line Broadening}
\label{s:broadening}

In order to extract the lithium isotopic ratio from the
profile of  \lii\ 670.8\,nm line, 
it is paramount to have an accurate
understanding of the intrinsic spectral line broadening.
This broadening comes in various forms all of which must be taken into
account: thermal, pressure, radiative, rotational and Doppler broadening.
Thermal broadening can be directly computed from the known temperature structure
of the stellar atmosphere while the pressure and radiative broadening are,
in principle at least, known
atomic properties specific for each transitions. 

It is well-known that the observed spectral lines are always broader than
 predicted solely from the thermal and atomic broadening.
All stars are expected to rotate, which broadens
the line profiles but does not modify the line strength.
 As our observations are unable
to disentangle readily
 rotational broadening from other atmospheric broadening agents,
we adopt in all cases \vsini\,$=0.5$\,\kms\ given the great age
 of these metal-poor
halo stars (for comparison, the Sun has \vrot\,$=1.8$\,\kms\ at the equator).
We have confirmed that our derived results are immune to this particular choice of \vsini .
In addition, atmospheric velocity fields from convection, wave motion, turbulence and
the like introduce Doppler shifts which
affect the line profiles. In analyses based on 1D hydrostatic model atmospheres,
one normally introduces two
free parameters: microturbulence \micro\ and macroturbulence \macro , acting
on small and large spatial scales, respectively.
 Following standard practice we determine
\micro\ by ensuring that the derived abundances from \fei\ and \feii\ lines
are independent of line strength while \macro\ is estimated from the
observed line shapes.
The form of \macro\ is assumed to be Gaussian. This choice was made mainly
for convenience as the effects of \macro\ can then be easily combined with
the line broadening due to the finite instrumental resolution which is also
described by a Gaussian distribution.
We have performed various tests ensuring that our results in terms of
Li isotopic abundances are not affected by the use of a Gaussian
\macro\ instead of, for example, a radial-tangential version (e.g. Gray 1992).

The lines used for determining \macro\ were \ki\ 769.9\,nm, \cai\ 612.2\,nm,
\cai\ 616.2\,nm, \cai\ 643.9\,nm, \fei\ 623.0\,nm,
\feii\ 624.7\,nm and \feii\ 645.6\,nm. In some cases we also added the
\fei\ 610.2\,nm and \fei\ 613.6\,nm lines when there was a lack of calibration
lines of suitable strength compared with the \lii\ line.
For the August 2004 observations of LP\,815-43 and HD\,140283, additional lines
could be used since the spectra extended further to the blue, enabling the
\feii\ 501.8\,nm, \fei\ 523.3\,nm, \fei\ 532.4\,nm, \fei\ 539.7\,nm, \fei\ 538.3\,nm, 
\fei\ 543.4\,nm, \fei\ 542.4\,nm, \mgi\ 552.8\,nm, \fei\ 561.2\,nm, \cai\ 558.9\,nm
and \fei\ 558.7\,nm lines to be covered in exchange for the  \ki\ 769.9\,nm line.  
The \ki\ line was  broader than other lines,
such as \fei\ 623.0\,nm,
of comparable strength in most stars. This may signal significant
departures from LTE which have indeed been predicted for this resonance
line for metal-poor stars (Ivanova \& Shimanskii 2000; Takeda et al. 2002).
Good agreement was found, however, when the \ki\ line had an
equivalent width of about 2\,pm, i.e., similar to that of the \lii\ line itself.
This is consistent with the
finding of Smith et al. (2001) in the case of HD\,84937.
We, therefore, decided not to include this line in the final estimates
of \macro\ and  used it only as a consistency check on the mean value
and scatter implied by the other lines.

\begin{figure}
\resizebox{\hsize}{!}{\includegraphics{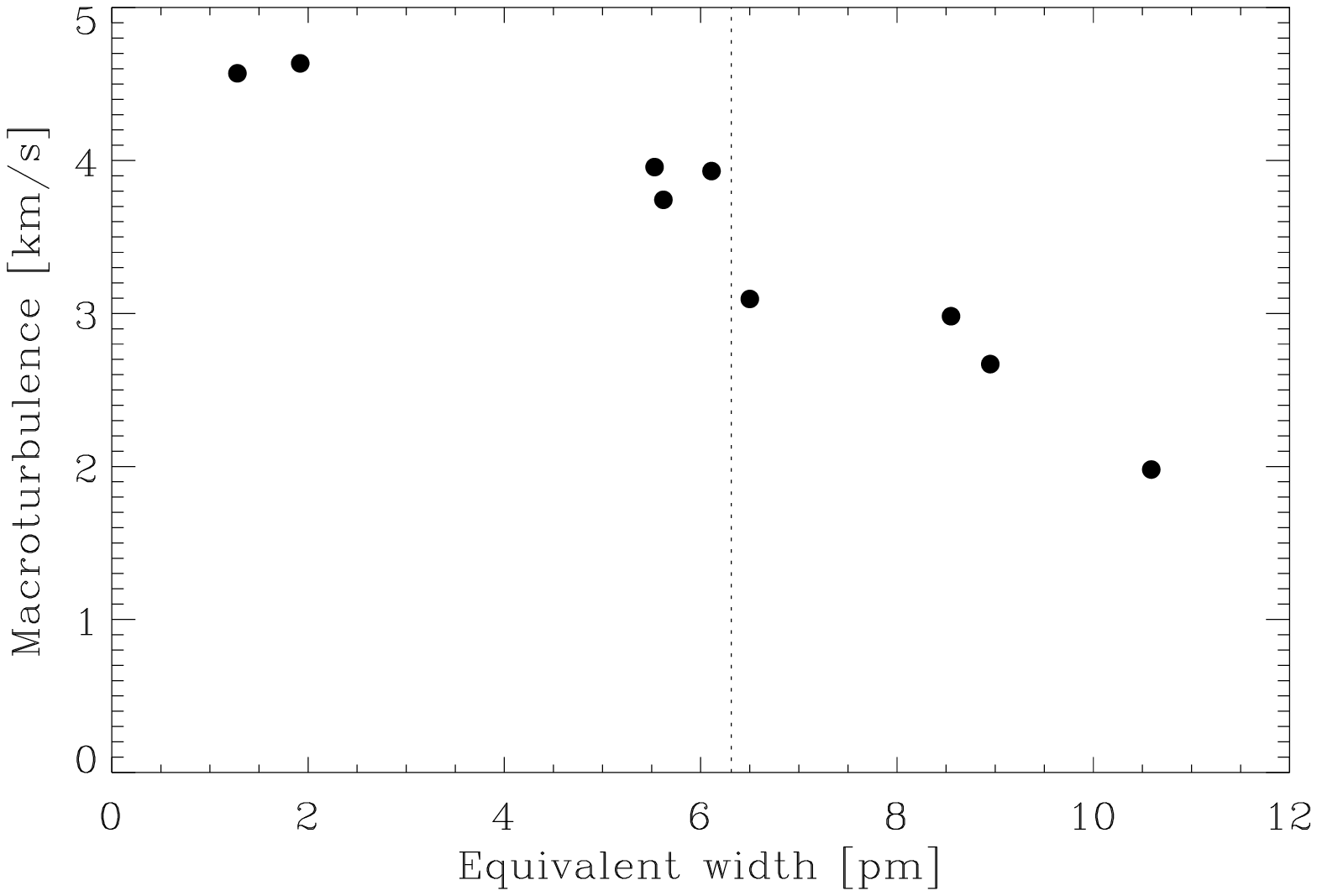}}
\caption{Estimated macroturbulence \macro\ (assumed to be described by a Gaussian)
from different lines of varying strengths in HD\,106038; in addition
the lines were broadened by rotation (assumed to be \vsini\,$=0.5$\,\kms ) and instrumental
broadening (measured from Th lines). Note the strong correlation between \macro\ and
equivalent width, which emphasizes the importance of having calibration lines of similar
strengths as the \lii\ 670.8\,nm line when estimating the intrinsic line broadening. 
The equivalent width for the Li line in HD\,106038 is denoted by the vertical dotted line.}
\label{f:FWHM}
\end{figure}

When computing the average broadening parameters from the calibration lines,
we have  considered only
 lines with  strengths $0.9 \le W_\lambda \le 7.5$\,pm.
The calibration lines included in the computation
of the mean broadening are of similar line strength to
the \lii\ 670.8\,nm line, which should minimize possible systematic differences.
This point is important since we have noticed that the derived \macro\
is dependent on the strength of the line, as illustrated in Fig. \ref{f:FWHM}
for the case of HD\,106038 (a Li-rich star with a significantly
stronger \lii\ 670.8\,nm line than in the typical star of our sample:
$W_\lambda = 6.32$\,pm rather than $\approx 2$\,pm).
In this particular instance, \macro\ was computed as a mean of the four lines
with $W_\lambda = 6\pm1$\,pm, which is expected to give a good
estimate of the intrinsic broadening for the \lii\ line.

\section{Li Isotopic Abundances from the \lii\ 670.8\,nm Resonance Line}
\label{s:li6708}

\subsection{Li Isotopic Abundances}
\label{s:li6708_chi2}

\begin{figure}
\resizebox{\hsize}{!}{\includegraphics{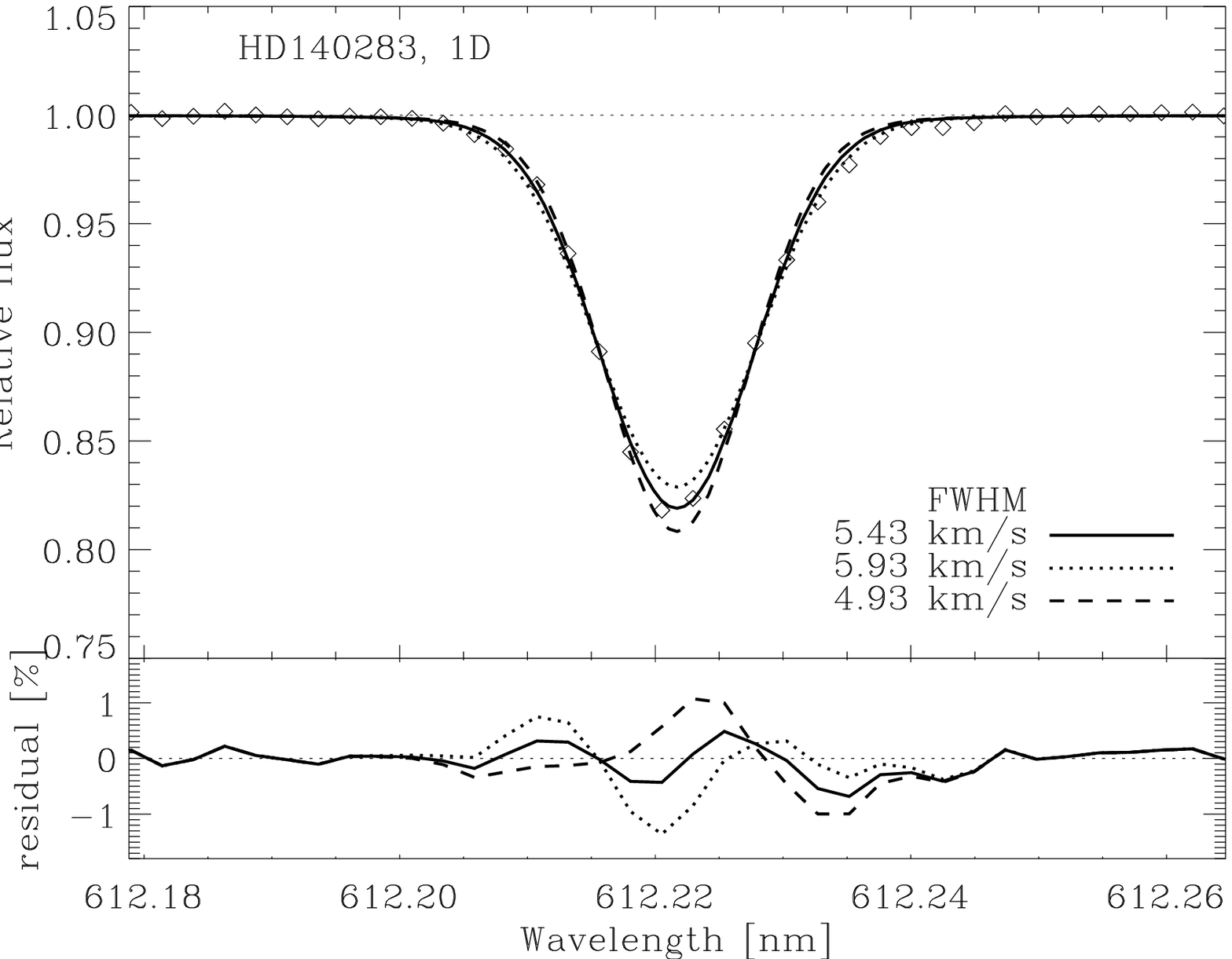}}
\resizebox{\hsize}{!}{\includegraphics{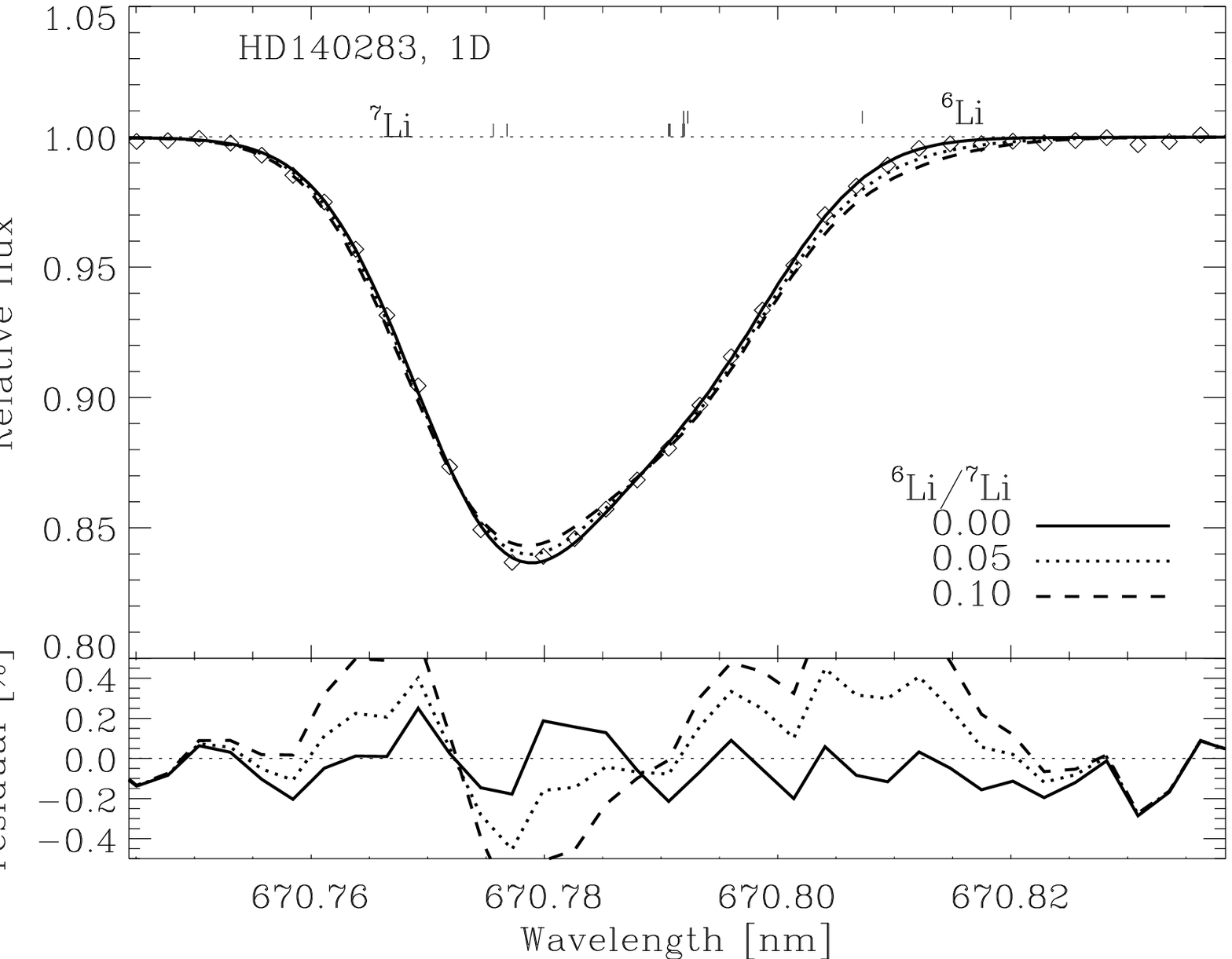}}
\resizebox{\hsize}{!}{\includegraphics{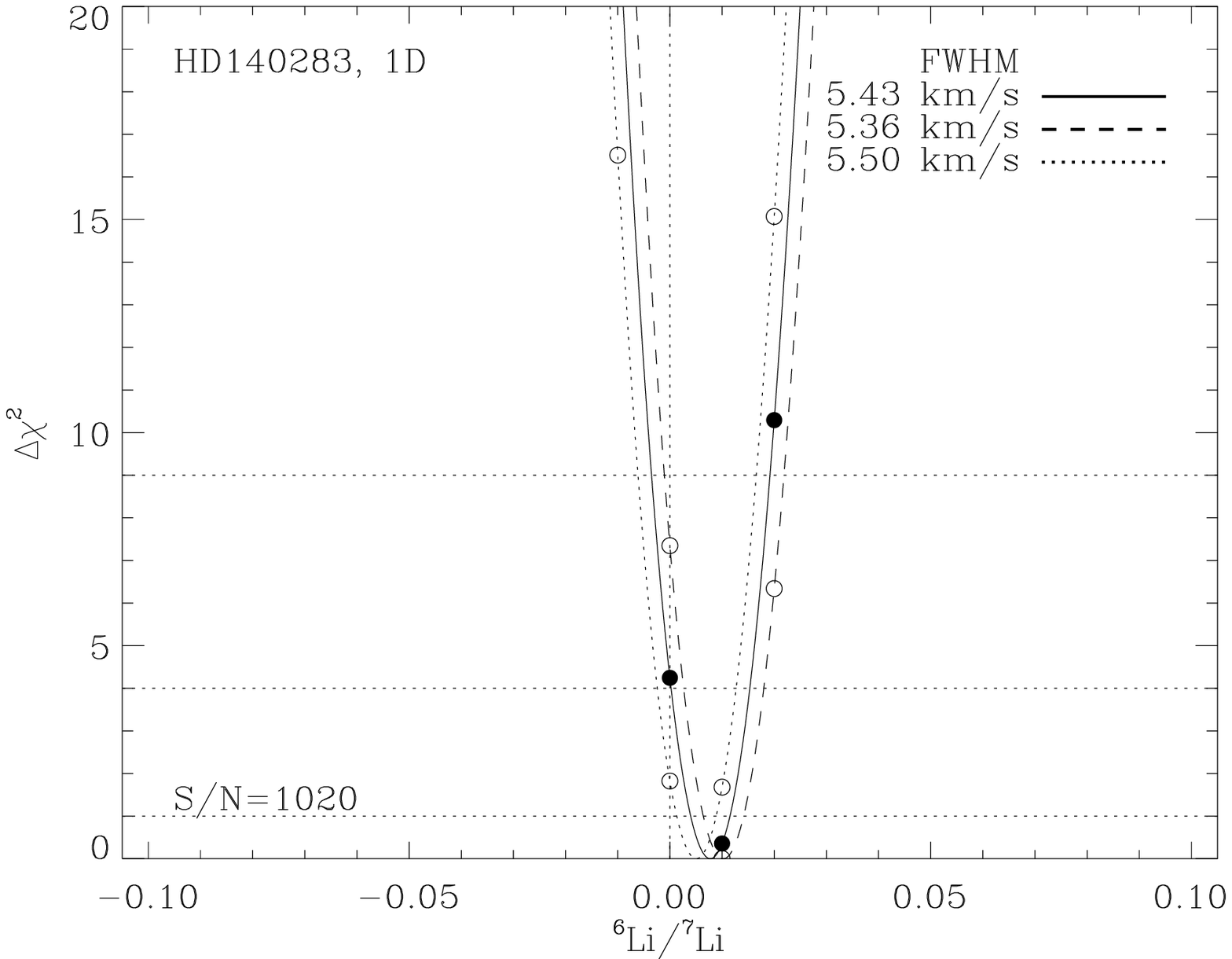}}
\caption{{\em Upper panel:} Observed \cai\ 612.2\,nm profile (rhombs) in HD\,140283 
together with the best fit theoretical 1D LTE profile (solid line). Also shown are the predictions with
the macroturbulence \macro\,$\pm0.5$\,\kms .
The Gaussian FWHM is here the combination of \macro\ and the instrumental broadening; 
the profiles have also been convolved with
a stellar rotation velocity of \vrot\,$=0.5$\,\kms . 
{\em Middle panel:} Computed \lii\ 670.8\,nm profile for three different values for 
\liratio\ (0.00, 0.05 and 0.10) shown together with the observed line (rhombs). 
In each case, the Li abundance, wavelength shift and continuum level have been allowed to vary. 
The location of the various \liseven\ and \lisix\ 
components are indicated.
{\em Lower panel:} Resulting $\Delta$\chitwo\ for the combined Gaussian
of the mean \macro\ and instrumental broadening; also shown are the cases 
for \macro\,$\pm \Delta$\macro . In this star, no significant detection of \lisix\ can be claimed.}
\label{f:li6708_HD140283} 
\end{figure}

\begin{figure}
\resizebox{\hsize}{!}{\includegraphics{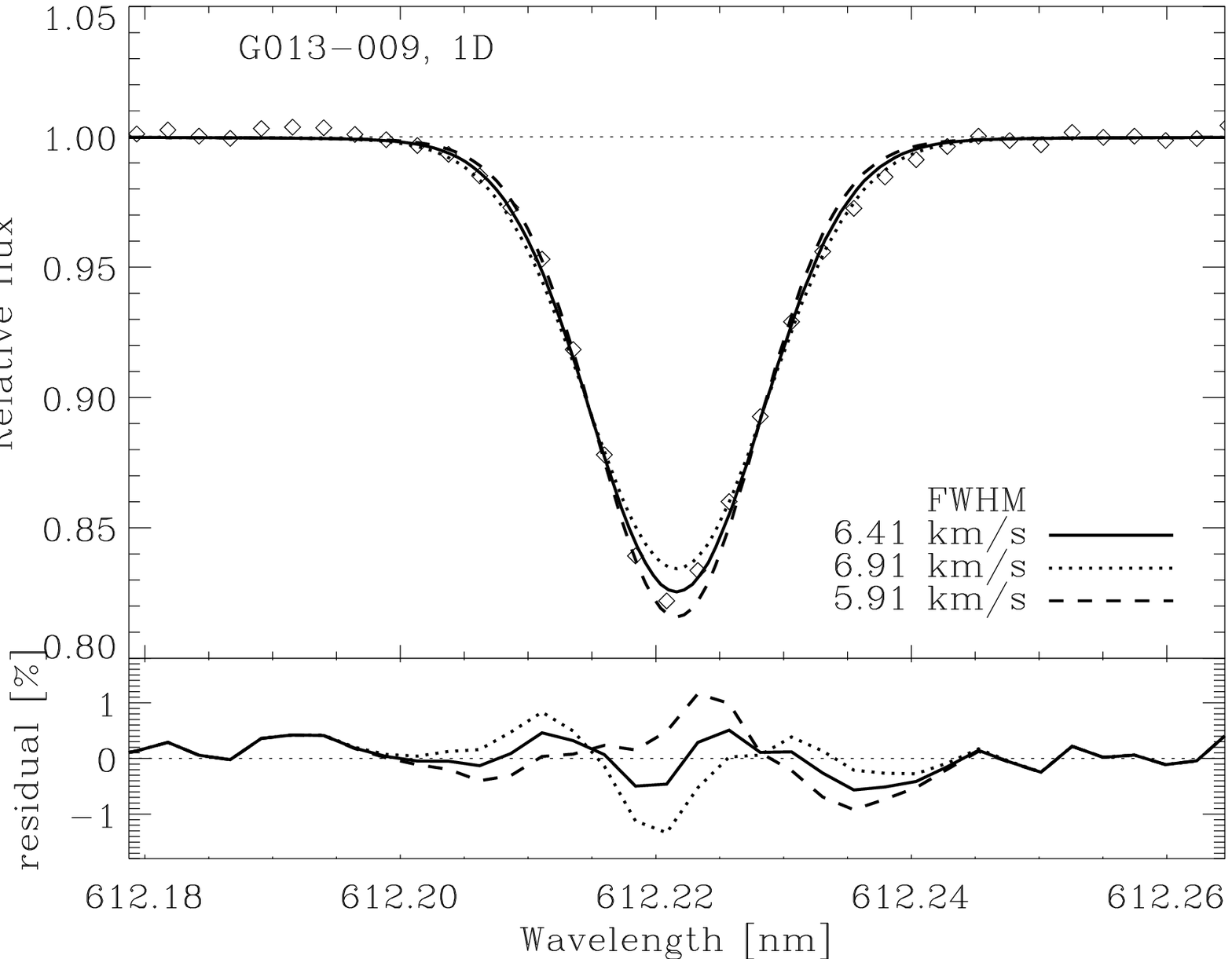}}
\resizebox{\hsize}{!}{\includegraphics{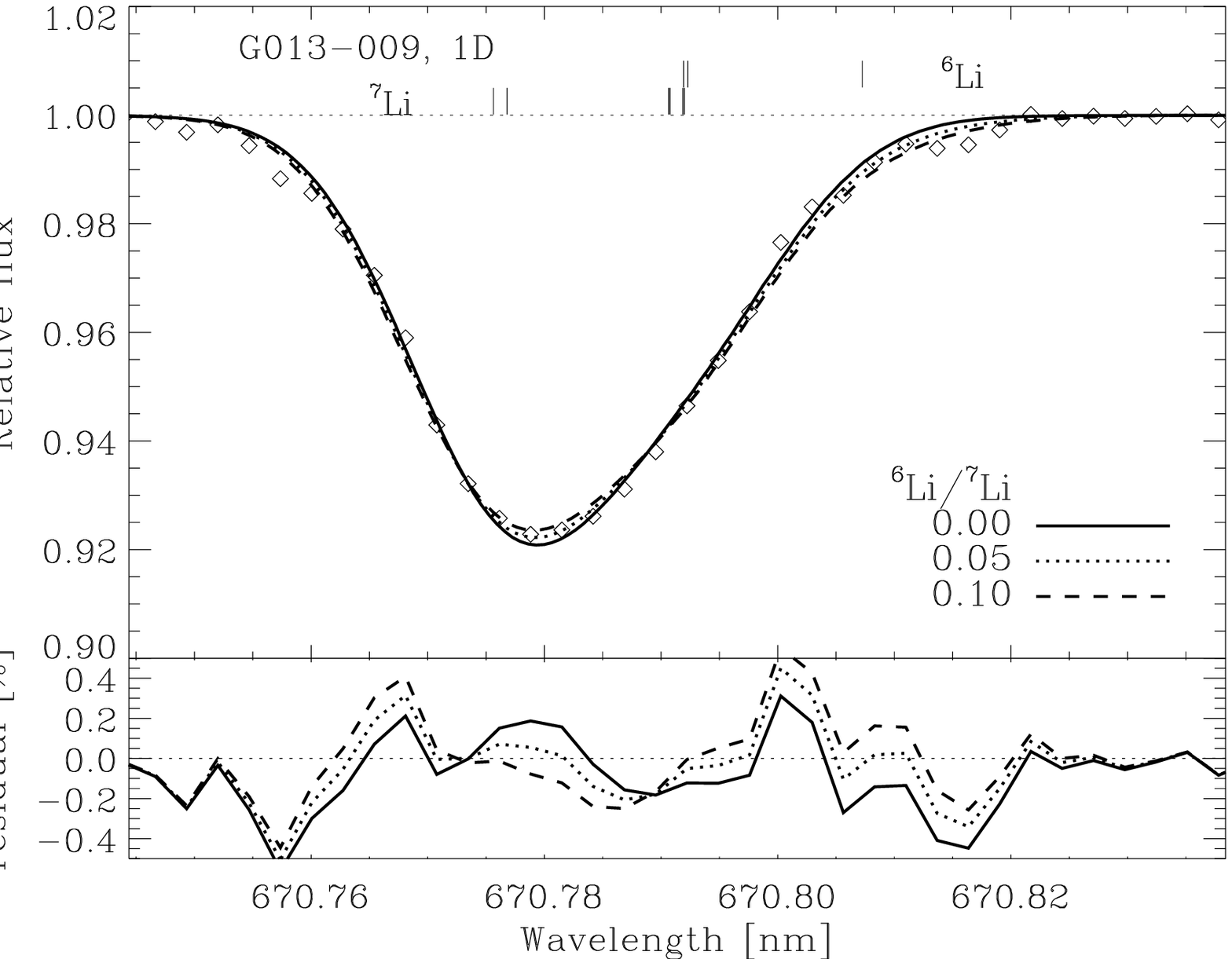}}
\resizebox{\hsize}{!}{\includegraphics{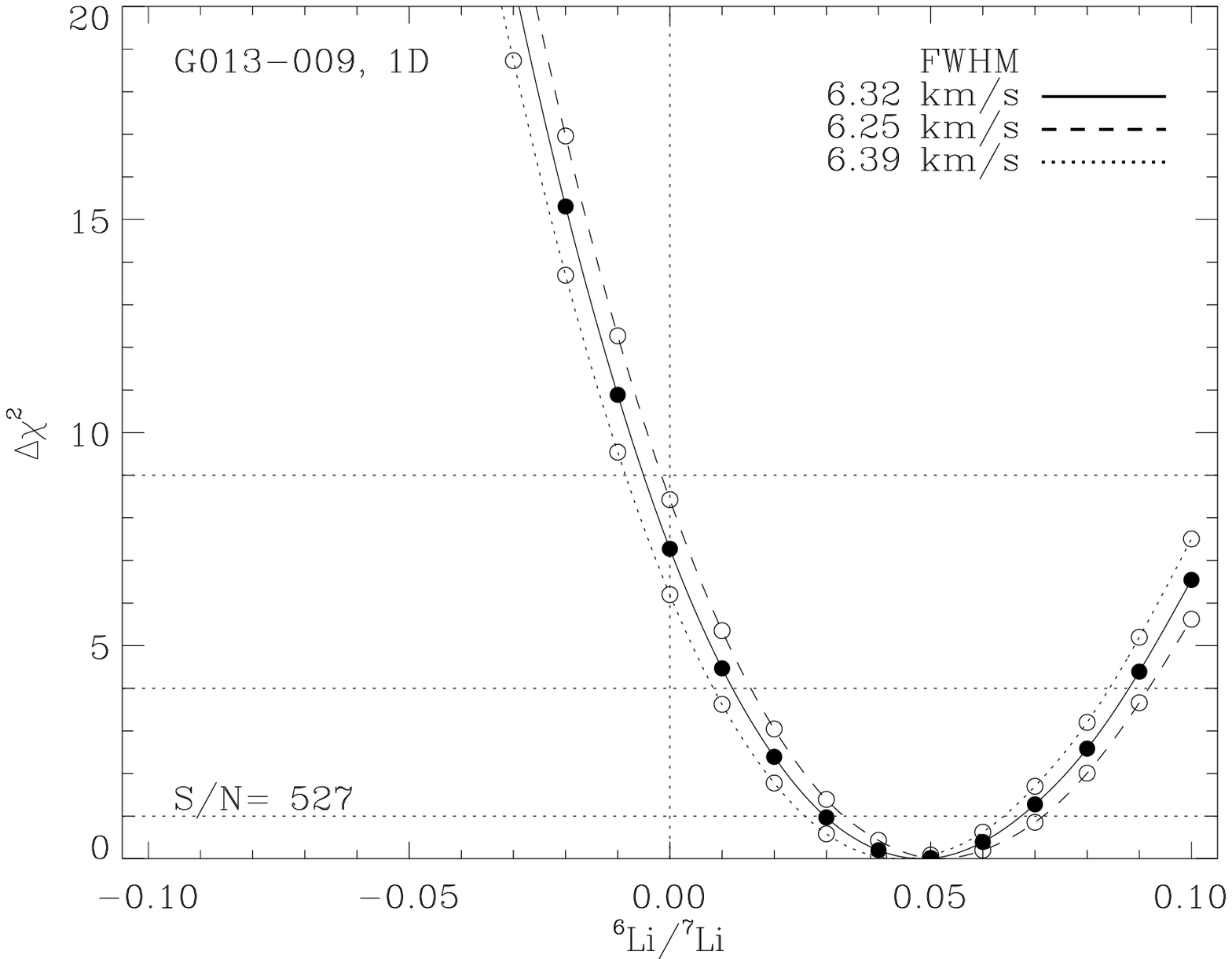}}
\caption{Same as Fig. \ref{f:li6708_HD140283} but for G\,013-009.
In this star, the \chitwo -analysis
indicate a $>2\sigma$ detection of \lisix\ at the level of \liratio\,$=0.05$. }
\label{f:li6708_G013-009}
\end{figure}

\begin{figure}
\resizebox{\hsize}{!}{\includegraphics{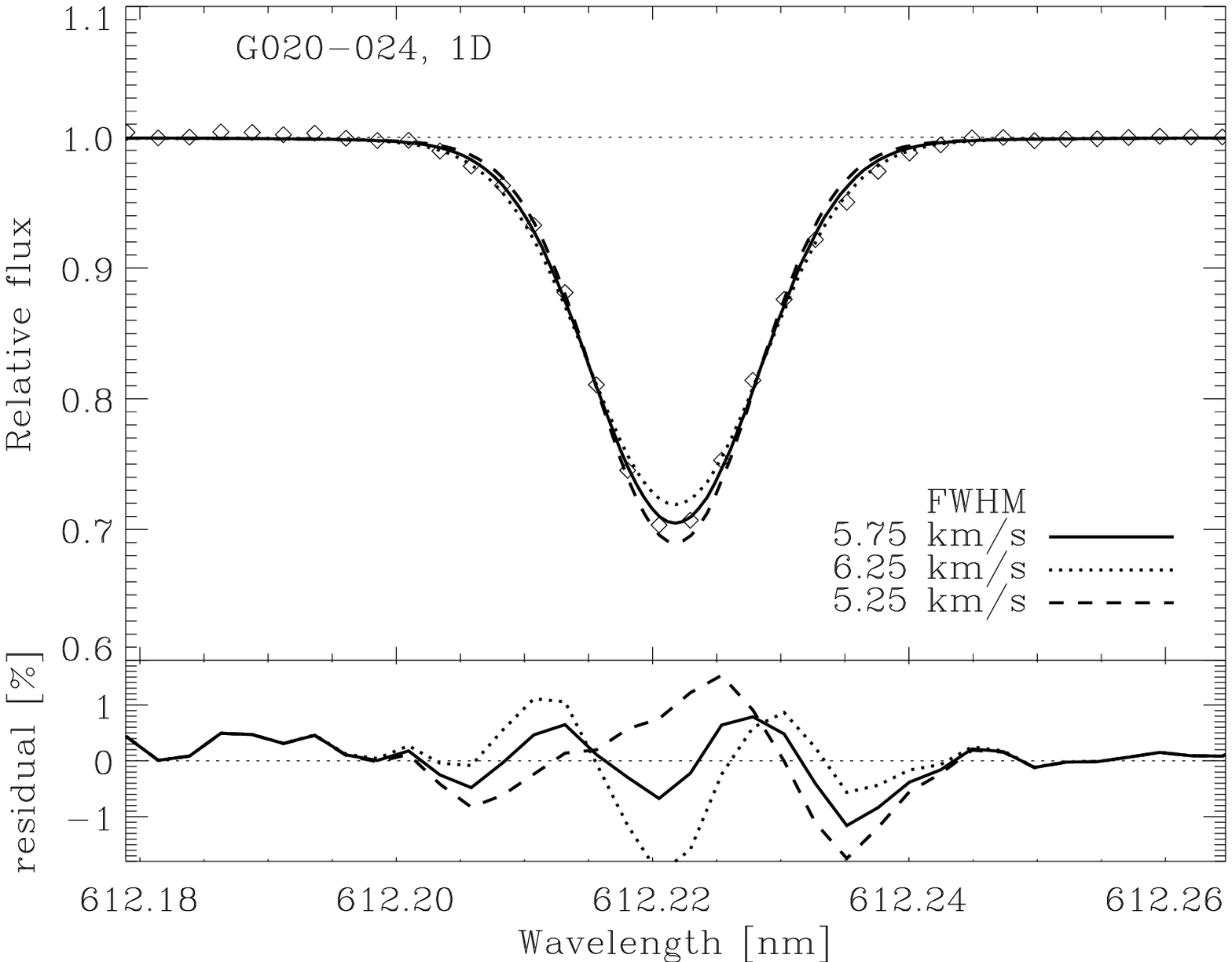}}
\resizebox{\hsize}{!}{\includegraphics{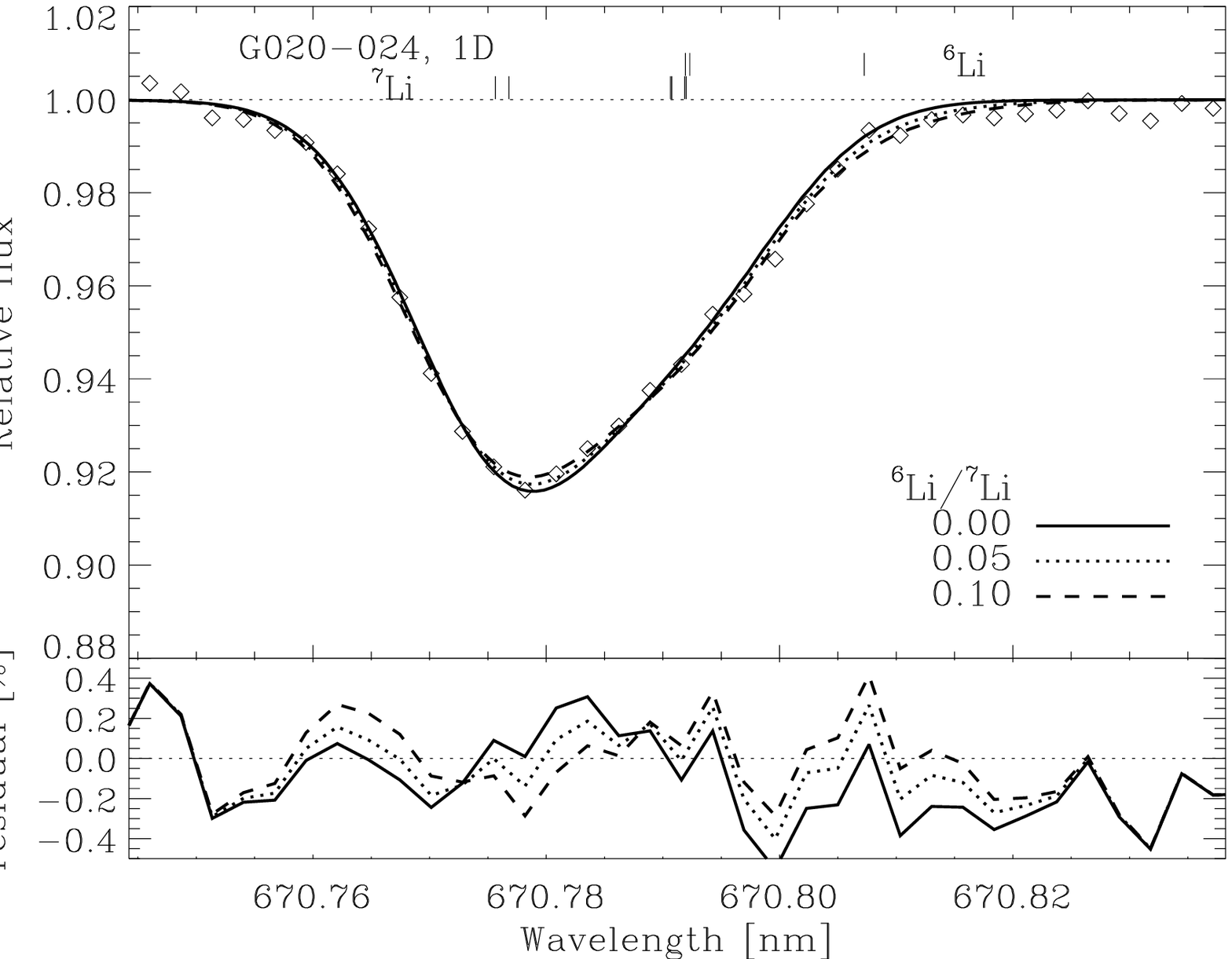}}
\resizebox{\hsize}{!}{\includegraphics{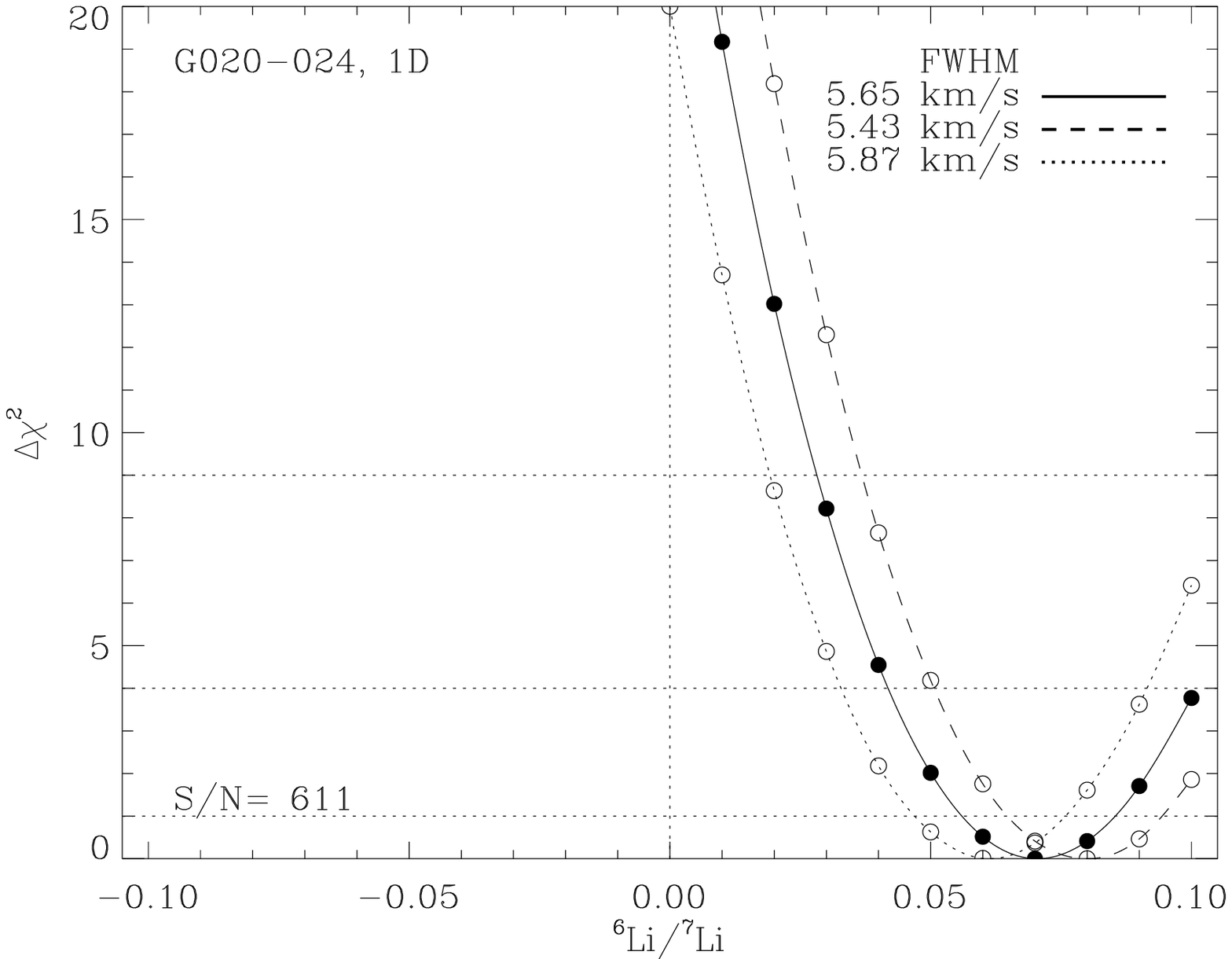}}
\caption{Same as Fig. \ref{f:li6708_HD140283} but for G\,020-024.
In this star, the \chitwo -analysis
indicate a $4\sigma$ detection of \lisix\ at the level of \liratio\,$=0.07$. }
\label{f:li6708_G020-024}
\end{figure}

\begin{figure}
\resizebox{\hsize}{!}{\includegraphics{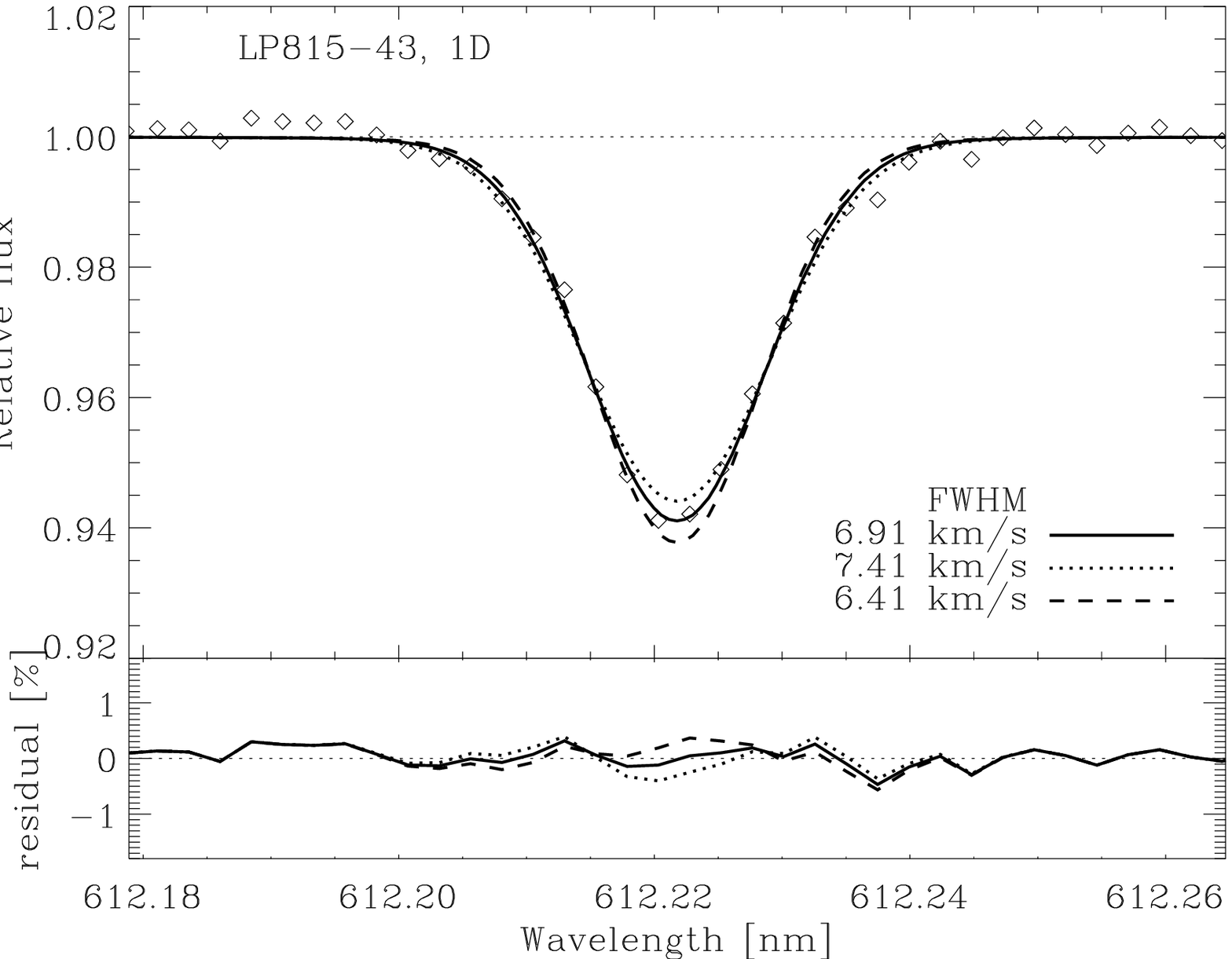}}
\resizebox{\hsize}{!}{\includegraphics{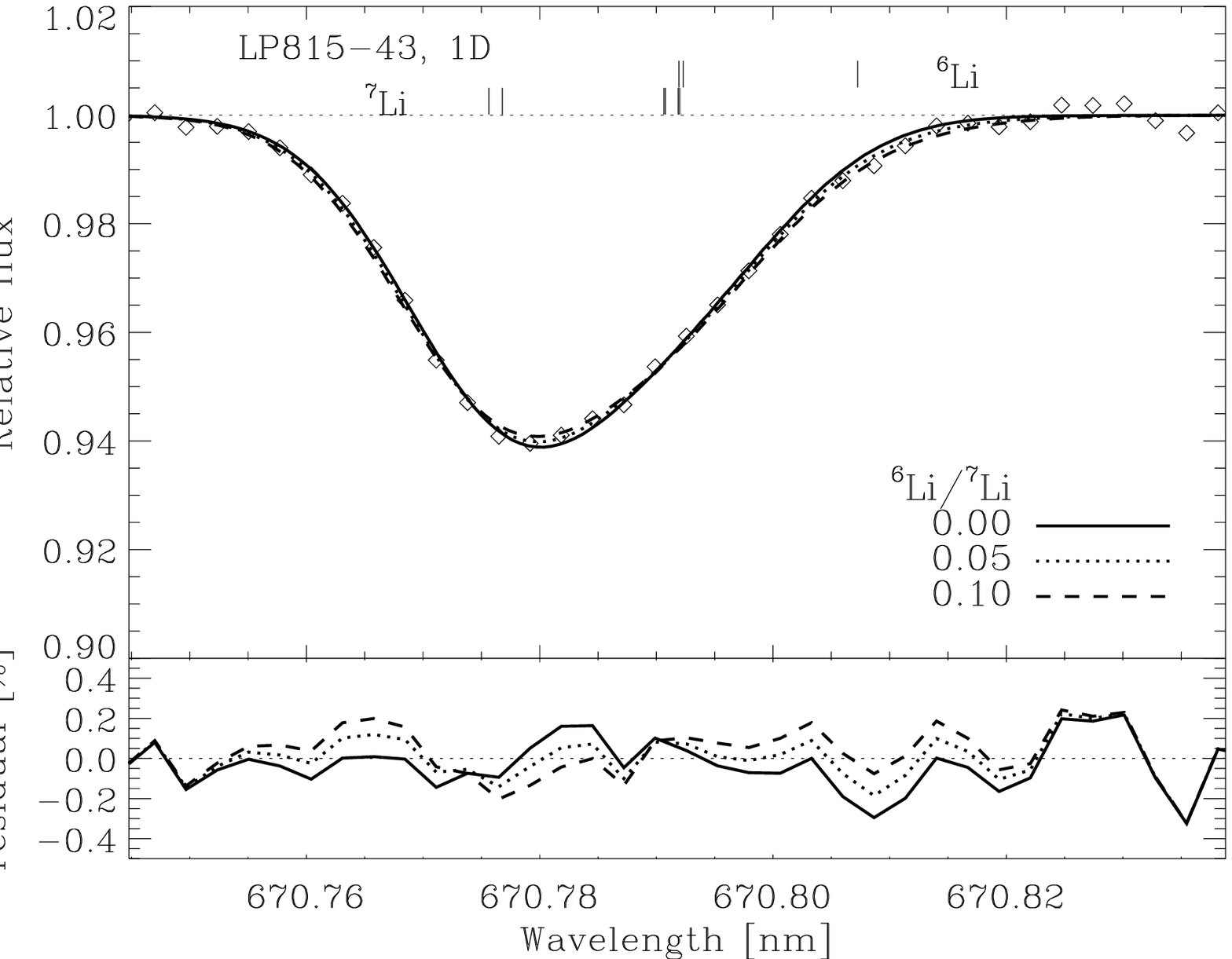}}
\resizebox{\hsize}{!}{\includegraphics{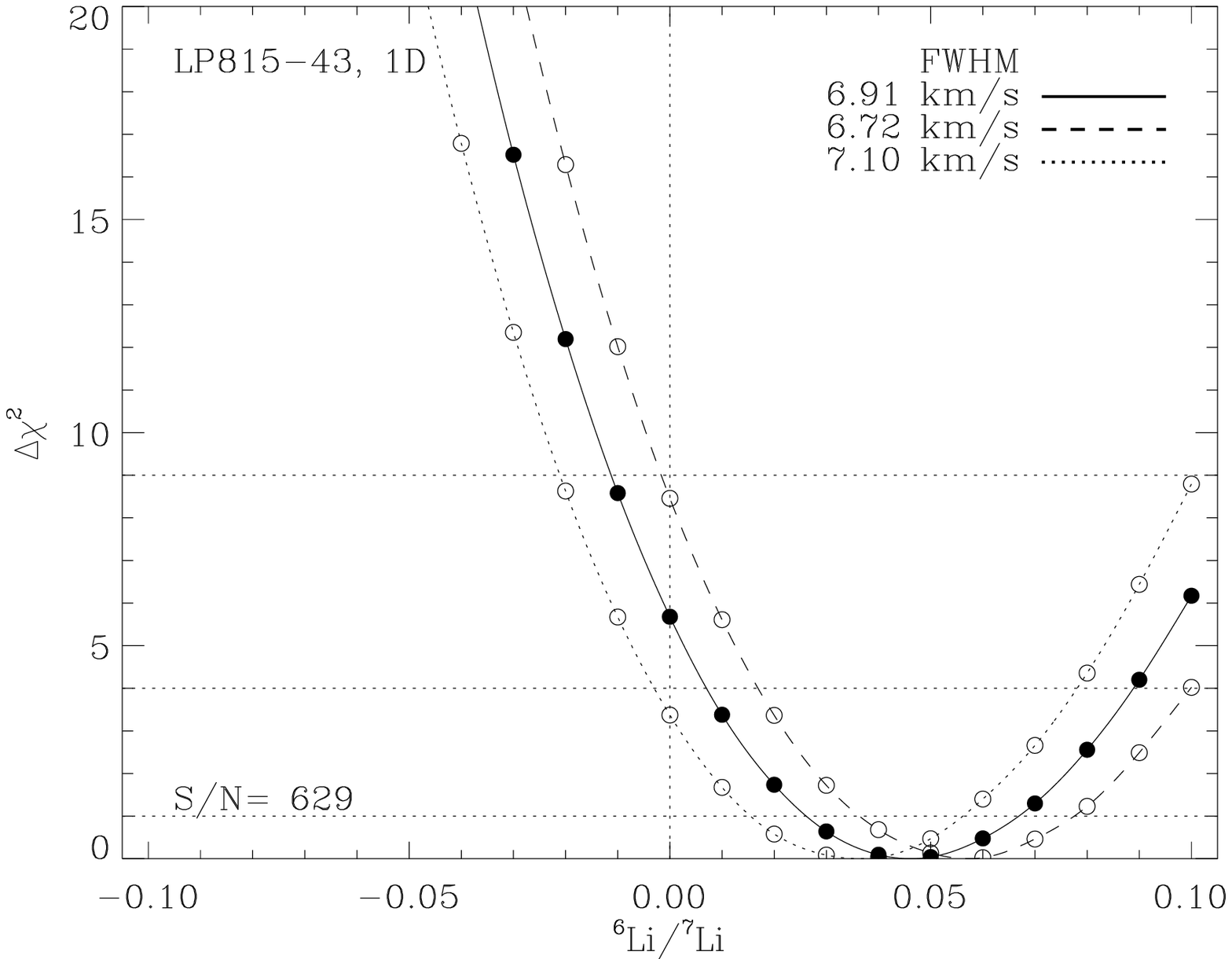}}
\caption{Same as Fig. \ref{f:li6708_HD140283} but for LP\,815-43.
In this star, the \chitwo -analysis
indicate a $2\sigma$ detection of \lisix\ at the level of \liratio\,$=0.05$. The observed spectrum shown here
is the one from August 2004.}
\label{f:li6708_LP815-43}
\end{figure}

The presence of \lisix\ alongwith \liseven\ in the stellar atmosphere
introduces additional width and asymmetry of the
line profile\footnote{We will not consider here the center-of-gravity method
which is based on the wavelength shift introduced by the addition
of \lisix . The
nearby \cai\ 671.7\,nm line has served as a reference line.
As discussed by Smith et al. (1998) and
Cayrel et al. (1999), the presence of differential line shifts arising from
convection (see, for example, Reddy et al. 2002),
and the remaining uncertainties in the laboratory
wavelength of the \cai\ line render this method very prone to
systematic errors.}.
As outlined  in Sect. \ref{s:broadening},
we first determine \macro\ from
a number of \ki , \cai , \fei\ and \feii\ lines.
This is achieved by a \chitwo -analysis of the predicted and observed
line profiles allowing the element abundance, wavelength zero-point
and continuum normalization to vary besides \macro .
In most cases, \macro\ could be well determined with a standard
deviation of $\Delta$\macro$=\pm0.1-0.2$\,\kms . 
Larger uncertainties of $\pm 0.3-0.6$\,\kms\
resulted for four stars: G\,126-062, CD\,$-33\arcdeg1173$,
HD\,59392 and BD\,$+09\arcdeg2190$. 
The estimated broadening parameters are listed in Table \ref{t:li_abundances}, while
Figs. \ref{f:li6708_HD140283}--\ref{f:li6708_LP815-43} show 
the resulting profiles for one of these calibration lines
(\cai\ 612.2\,nm) for a few stars.

\begin{figure}
\resizebox{\hsize}{!}{\includegraphics{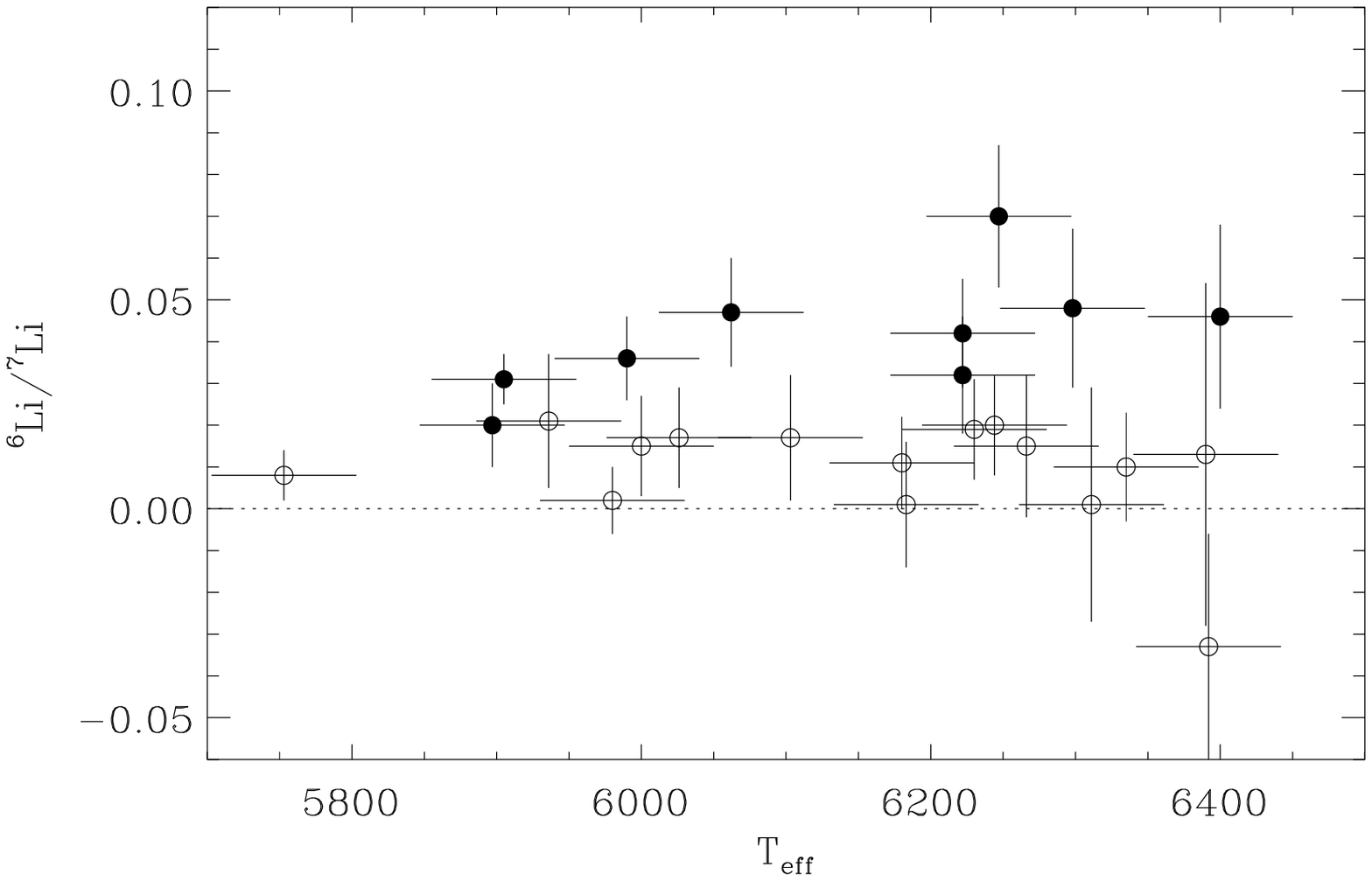}}
\resizebox{\hsize}{!}{\includegraphics{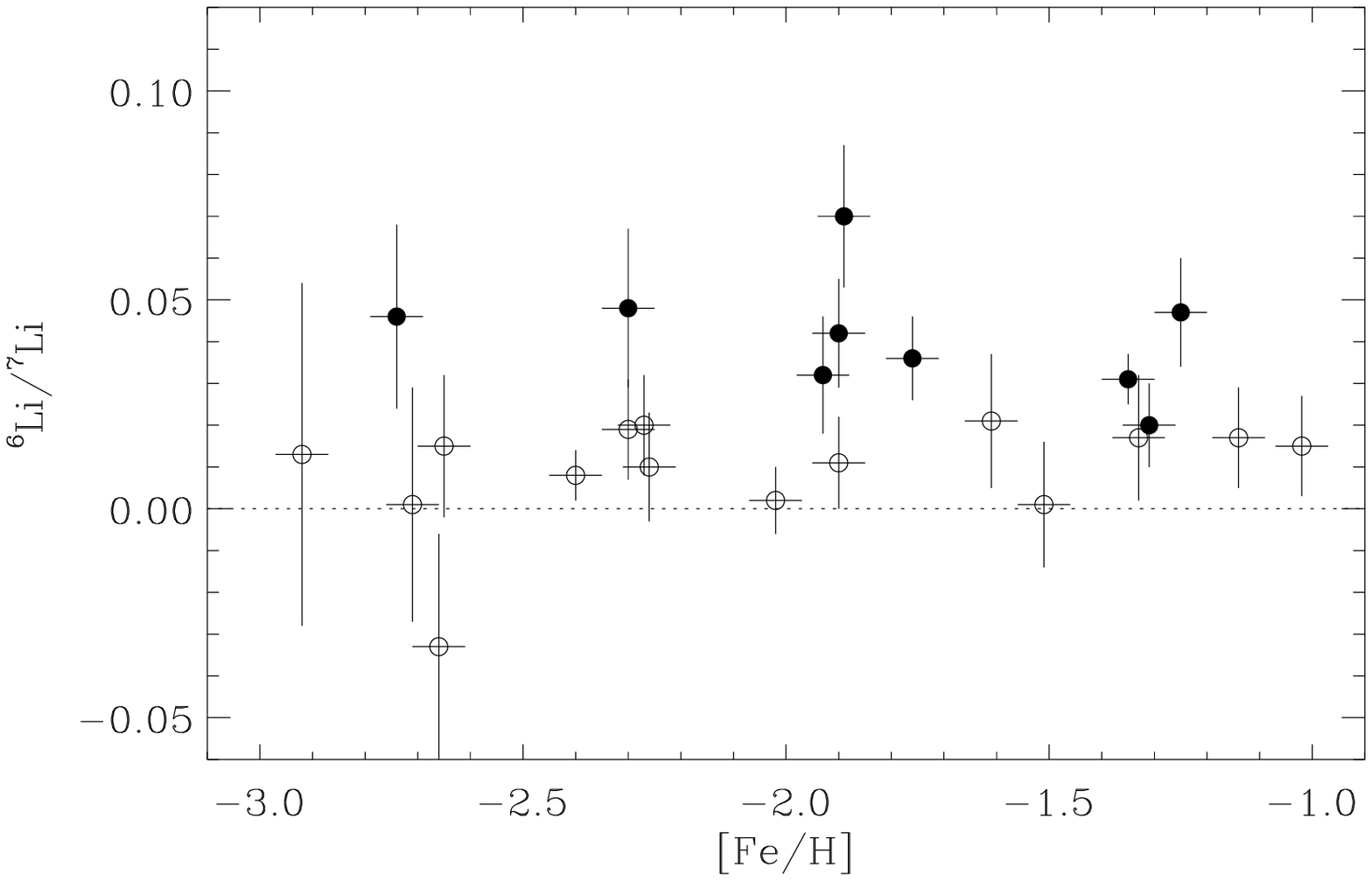}}
\resizebox{\hsize}{!}{\includegraphics{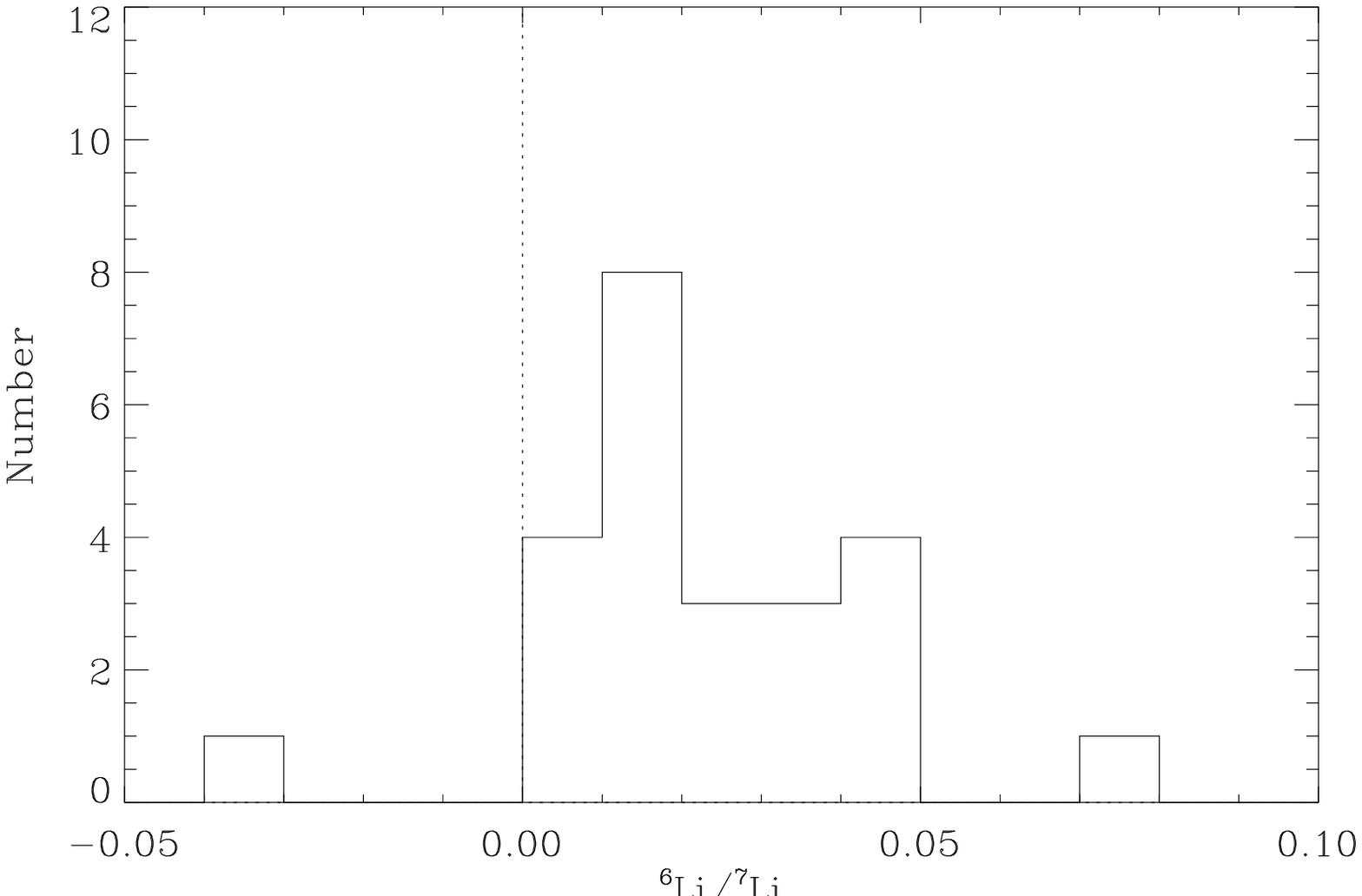}}
\caption{Derived \liratio\ as a function of \teff\ ({\em Upper panel}) and \feh\ ({\em Middle panel}).
The stars considered to have a significant detection ($\ge 2\sigma$) of \lisix\ are shown as 
solid circles while non-detections are plotted as open circles. Note that the non-detections do
not cluster around \liratio\,$=0.00$ but rather at $\approx 0.01$. This is also illustrated as a
histogram ({\em Lower panel}). The observed 
distribution is clearly not representative of a Gaussian centered
at \liratio\,$=0.00$.}
\label{f:li6708_summary}
\end{figure}

Equipped with the estimate of the macroturbulence \macro\ for each star, the adopted
\vsini\ (0.5\,\kms ) and the measured instrumental broadening at 670.8\,nm from Th lines, 
synthetic profiles of the 
\lii\ 670.8\,nm line for different Li isotopic ratios were generated for the program stars
using the stellar parameters listed in Table \ref{t:parameters}.
Calculations were done for \liratio\ between $-0.10$ and $0.10$ by step of $0.01$.
Negative \liratio\ ratios, of course,
 do not exist in reality but were included in order 
to obtain  estimates as accurate as
 possible for stars with very small or vanishing \lisix\ content. 
These were produced by allowing the \lisix\ components of the \lii\ 670.8\,nm line
to be in emission rather than in absorption as is the case for positive \liratio\ ratios.
At the low metallicity of the program stars, one does not need to worry about potential
blends from various atomic and molecular lines that can perturb the Li feature at
solar metallicity (Nissen et al. 1999; Israelian et al. 2001, 2003; Reddy et al. 2002)
and we thus computed the theoretical profile  taking into account only
 the various 
components of the Li line. Test calculations revealed that this was fully justifiable for
the target stars. 

The comparison between the theoretical and observed profiles was quantified 
using a \chitwo -analysis similar to that used in 
most recent investigations of \liratio\ in late-type stars (e.g. Smith et al. 1998;
Nissen et al. 1999, 2000; Israelian et al. 2001, 2003; Reddy et al. 2002; Aoki et al. 2004).
The \chitwo\ was computed according to 
\chitwo\,$=\sum \left(O_{\rm i}-S_{\rm i}\right)^2/\sigma^2$, where $O_{\rm i}$ and
$S_{\rm i}$ denote the observed and synthetic flux at wavelength point $i$, respectively,
and $\sigma = \left(S/N\right)^{-1}$ as estimated in three nearby continuum windows.
Typically 26 wavelength points were considered across the Li line to estimate \chitwo .
For each \liratio , the total Li abundance, the wavelength zeropoint of the observed spectrum
and the continuum level were allowed to vary in order to optimize the fit and thus 
minimize \chitwo . The most probable value for \liratio\ 
was obtained by cubic spline interpolation of the resulting \chitwo\ values, from which 
the minimum \chitwomin\ could be determined. The $1\sigma$, $2\sigma$ and $3\sigma$
confidence limits of the determinations correspond to
$\Delta \chi^2 = \chi^2 - \chi^2_{\rm min} = 1, 4$ and 9, respectively.
These uncertainties are determined by the noise of the spectra, which as mentioned
above are better than $S/N >  500$ for all but two stars that have 400 and 470, respectively.
The reduced $\chi^2_{\rm r}$ is in all cases close to one with the variations between
stars as expected from statistical fluctuations.
In addition, the \chitwo -analysis was repeated for each star with different adopted macroturbulence
parameters, $\zeta_{\rm macro} \pm \Delta \zeta_{\rm macro}$, where $\Delta \zeta_{\rm macro}$
was given by the standard deviation from the various Ca and Fe lines. 
Examples of the observed and predicted \lii\ line profiles and the corresponding
\chitwo\ values can be found in Figs. \ref{f:li6708_HD140283}--\ref{f:li6708_LP815-43} for a few stars. 
The resulting \liratio\ ratios and \liseven\ abundances are listed in Table \ref{t:li_abundances}.

The errors in the derived \liratio\ ratios are quantified by taking the uncertainties from 
the finite $S/N$ as well as the adopted \macro\ and stellar parameters  in quadrature.
In essentially all cases, the first source dominates the error budget in spite of
the extremely high $S/N$ achieved in these observations, but
for a couple of stars with poorly determined \macro\ such as CD\,$-33\arcdeg$1173 
the second uncertainty drives the final error.
The errors arising from the uncertainties in the stellar parameters are in all cases
very minor, as discussed in detail in Sect. \ref{s:li6708_parameters}. 
We note that our error estimates are slightly more conservative than previous studies of
the Li isotopic ratios in metal-poor stars, which, for example, often have relied on only
one or two calibration lines to determine the line broadening.

As seen in Table \ref{t:li_abundances}, nine of the 24 stars fulfill our criteria of a $\ge 2\sigma$ result
to be considered a significant \lisix\ detection: HD\,102200, HD\,106038, HD\,160617, G\,013-009, G\,020-024,
CD\,$-30\arcdeg$18140, CD\,$-33\arcdeg$3337, CD\,$-48\arcdeg$2445 and LP\,815-43.
This should be contrasted with the previous single confirmed detection of
\lisix\ in HD\,84937 (Smith et al. 1993, 1998;
Hobbs \& Thorburn 1997; Cayrel et al. 1999). 
The stars with detected \lisix\ are all very near the turn-off for their respective metallicities
(Fig. \ref{f:HR}), in accordance
with the suggestion of Smith et al. (1993, 1998). 
In addition, we find a large number of borderline cases
at the $1-2\sigma$ level that we do not consider significant enough to claim a positive detection. 
In fact, there is a tendency for the non-detections to cluster at \liratio\,$\approx 0.01$, as illustrated
in Fig. \ref{f:li6708_summary}. 
This could be taken as a warning sign that our analysis contains an unidentified systematic error
that bias the derived isotopic ratio towards too high values. 
We have therefore carried out  tests in an attempt to identify such a problem,
which are described in detail in the following subsections. 
In summary, no serious problem that could affect our analysis has been uncovered:
no differences are found when taking into account possible ISM absorption in the Li line
(Sect. \ref{s:li6708_lp815}), adopting an alternative \chitwo -analysis
(Sect. \ref{s:li6708_chi2_cayrel}), changing the stellar parameters within
reasonable bounds (Sect. \ref{s:li6708_parameters}), employing a different suite of
1D model atmospheres (Sect. \ref{s:li6708_kurucz}) or using 3D hydrodynamical model 
atmospheres (Sect. \ref{s:li6708_3D}). Furthermore, we independently obtain 
the same results as some previous investigations when relying on their observations
(Sect. \ref{s:li6708_literature}). 
Finally, our confidence in the derived results is boosted by the analysis of HD\,19445, which
was included in our program to provide a consistency check (see, also,
Smith et al. 1993). Since this star 
is a relatively cool (\teff\,$=5980$) metal-poor dwarf (\logg\,$=4.42$),
standard models of stellar evolution predict that  \lisix\ should
be thoroughly destroyed on the star attaining the main sequence (see discussion
in Sect. \ref{s:depletion}).
Indeed, an analysis identical to that for all of our other stars indicate no presence of \lisix :
\liratio\,$=0.002\pm0.008$.

We conclude that the Li isotopic ratios presented in Table \ref{t:li_abundances} 
are reliable. 
We claim significant \lisix\ detections only in the above-mentioned nine stars,
but  we can not rule out the possibility that the majority of turn-off halo stars 
in fact have \liratio\,$\approx 0.01$ (Fig. \ref{f:li6708_summary}). To confirm an isotopic ratio
as low as 0.01 for an individual star is extremely challenging (c.f. our analysis of HD\,140283 
with $S/N \approx 1000$),
but confirmation should be attemptable  in at least a statistical sense
 for a sample like ours but with
even higher $S/N$ spectra and 
additional calibration lines to nail down the line broadening accurately. 

The Li abundances derived from the profile fitting of the \lii\ 670.8\,nm line
have been obtained under the assumption of LTE.
We have applied the non-LTE abundance corrections computed by Carlsson et al. (1994)
interpolated to the stellar parameters of our sample to the LTE Li abundances.
These non-LTE calculations are particularly appropriate as they have
been performed using similar \marcs\ model atmospheres as employed here.
The resulting non-LTE \liseven\ abundances 
are given in Table \ref{t:li_abundances}.
The non-LTE abundance corrections are in all cases very small
($\Delta$\logli$\la 0.02$\,dex) and generally negative.
The scatter in Li abundances is not significantly modified by the application
of these non-LTE corrections.

\subsection{Is the High \liratio\ in LP\,815-43 Real?}
\label{s:li6708_lp815}

Due its low \feh , LP\,815-43 will feature prominently in the discussions in
Sect. \ref{s:evolution}. It is  natural to inquire whether its
high \lisix\ abundance determined from our \chitwo -analysis is real.
We note that both the spectrum of LP\,815-43 obtained in July 2000
and that acquired in August 2004 imply a detection of \lisix\ at the $2\sigma$ level
or more. The former indicated \liratio\,$=0.078\pm0.033$ based on an
observed spectrum with $S/N=540$ but with only three \cai\ lines to
estimate the intrinsic line broadening. This result prompted us to
request additional observing time for this star to improve the $S/N$ and
include more calibration lines. The combined spectrum from two nights 
in August 2004
has $S/N=630$ as well as covering as many as 10 calibration lines of the right strength
which importantly
beat down the uncertainty in the macroturbulence to $0.19$\,\kms .
The resulting isotopic ratio is slightly less but still implies a significant detection
by our criteria:
\liratio\,$=0.046\pm0.022$. The result from the July 2000 spectrum is fully consistent with this
improved value taking the uncertainties into account.
The individual spectra obtained on August 30 and
31, 2004, have $S/N = 420$ and 490, respectively. Both of these imply 
\liratio\,$\approx 0.03-0.04$ but each allows only
a $1\sigma$-detection due to the reduced $S/N$.
Finally, we note that had we adopted the better determined macroturbulence value from
the August 2004 spectrum for the analysis of the July 2000 spectrum, the derived
\liratio\ would have increased to $0.089\pm0.027$, i.e. a $3\sigma$-detection. 

\begin{figure}
\resizebox{\hsize}{!}{\includegraphics{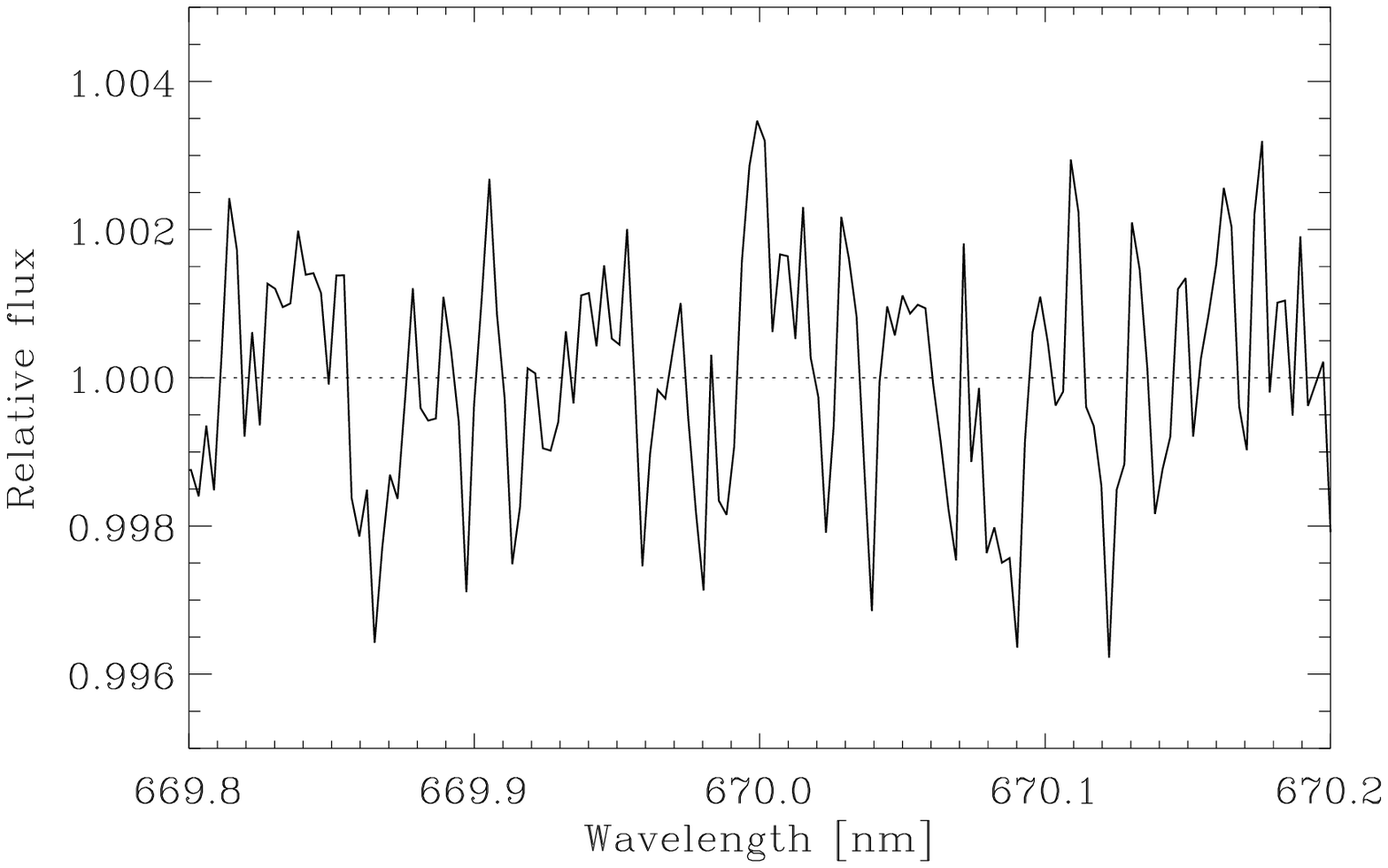}}
\resizebox{\hsize}{!}{\includegraphics{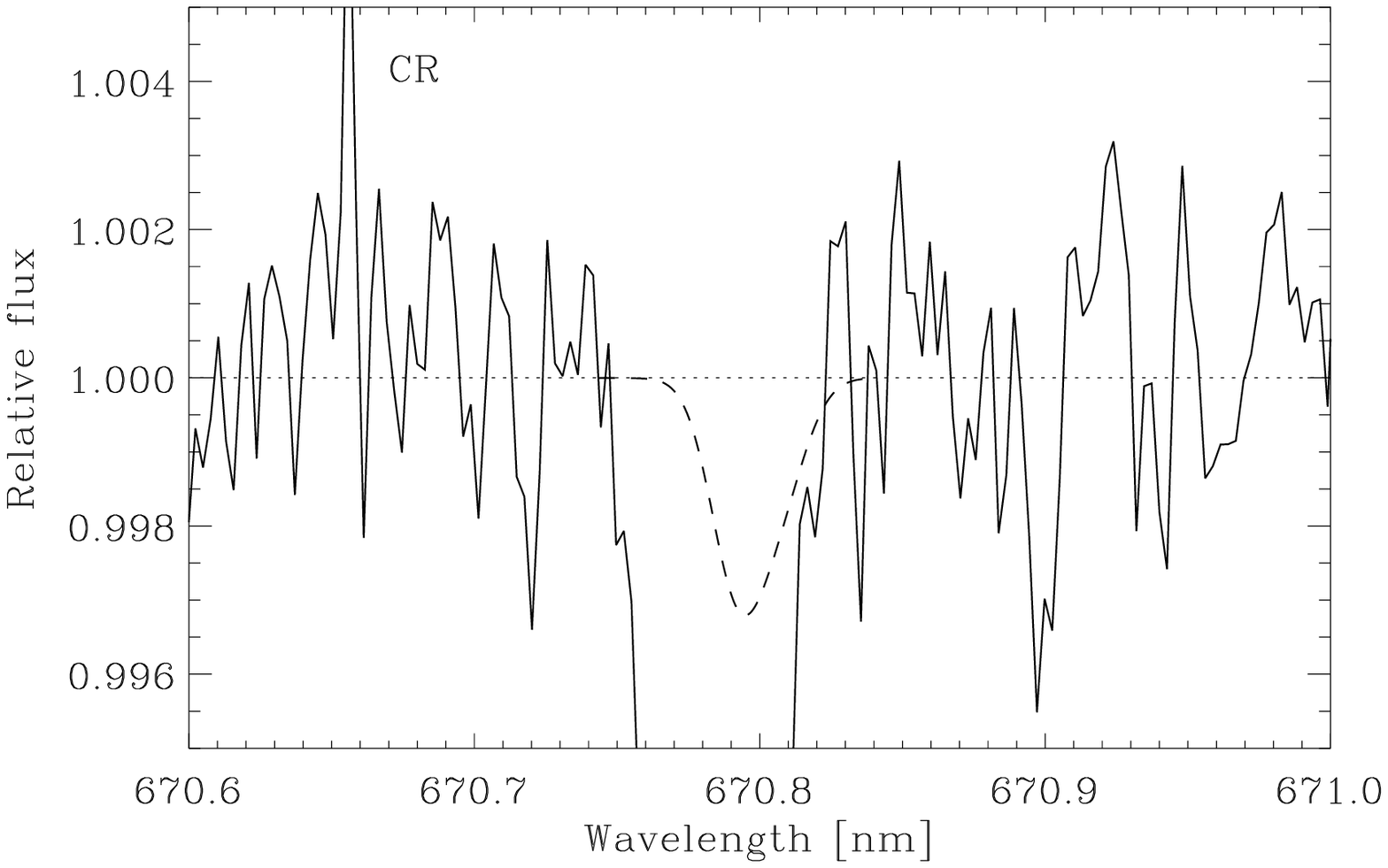}}
\resizebox{\hsize}{!}{\includegraphics{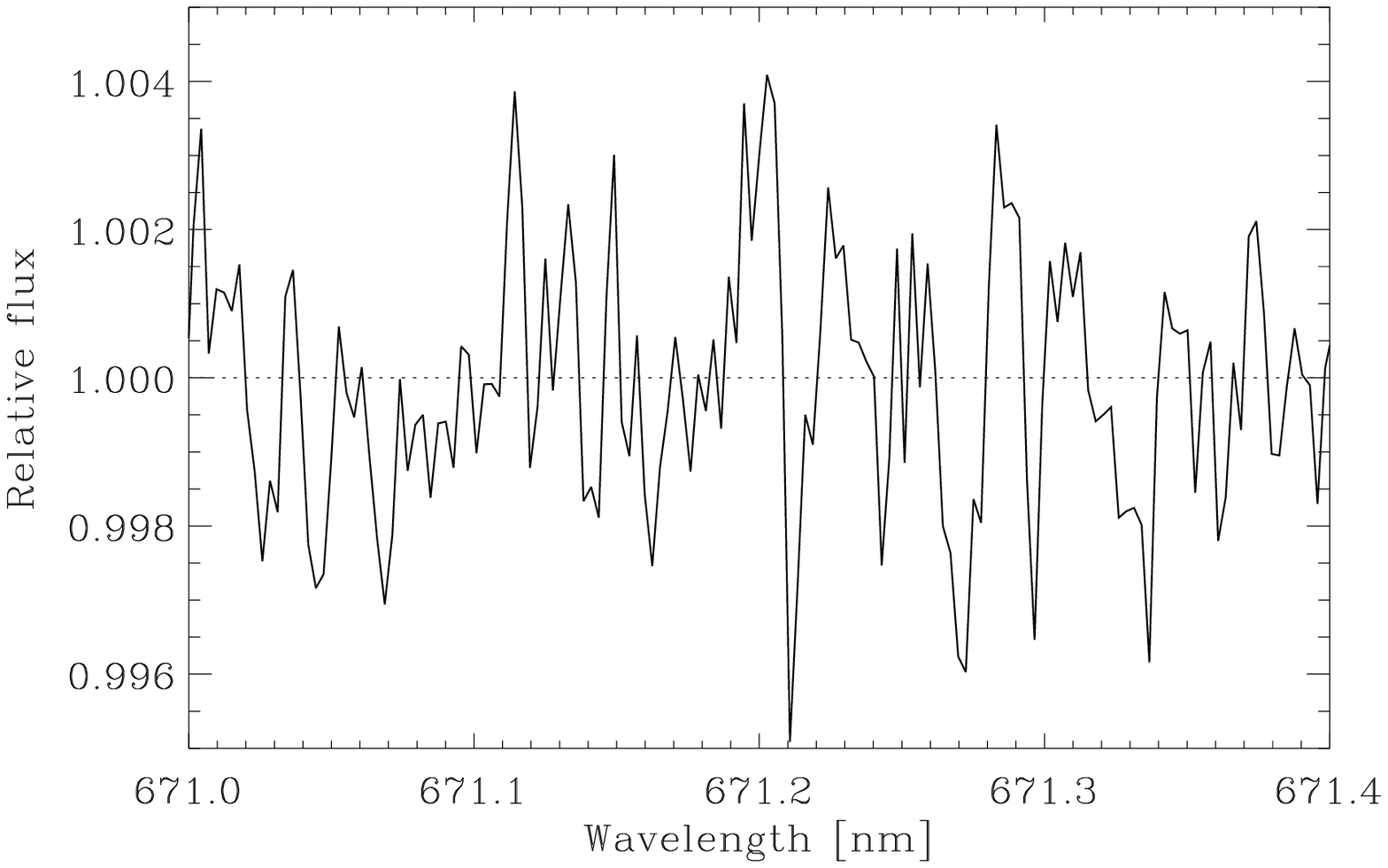}}
\caption{Enlargement of the August 2004 spectrum of LP\,815-43 around the
\lii\ 670.8\,nm line. Note that the spike at 670.655 is a cosmic ray hit. Also
shown with a dashed line is the predicted contribution of \lisix\  to the \lii\ 670.78\,nm
line for \liratio\,$=+0.05$.}
\label{f:LiI6708_obs_LP815-43}
\end{figure}

\begin{figure}
\resizebox{\hsize}{!}{\includegraphics{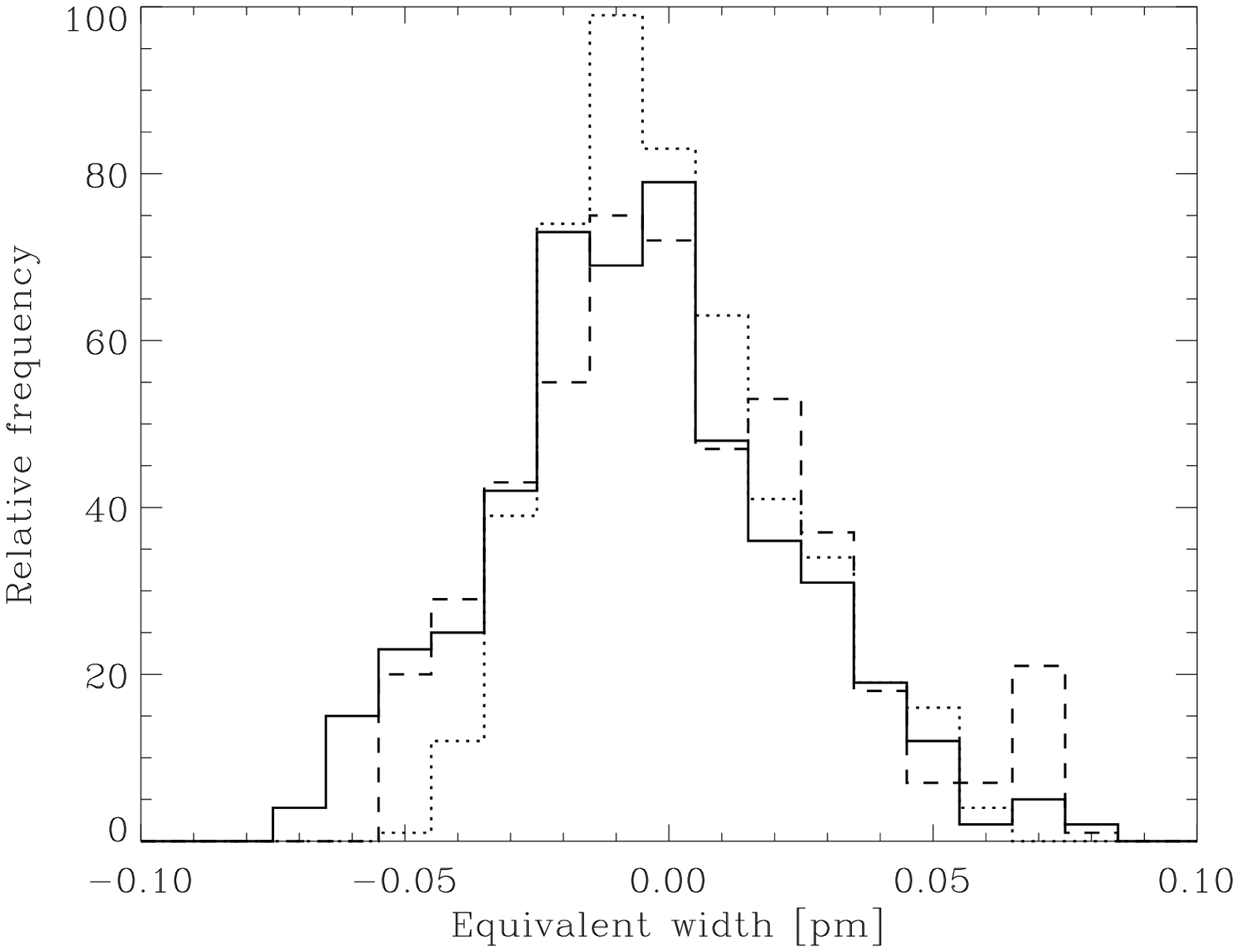}}
\caption{Histogram of the measured equivalent widths over a  0.08\,nm window in
the continuum near the \lii\ 670.8\,nm line for the August 2004 spectrum of LP\,815-43 (solid line). 
Also shown is the same distribution for pure random noise with the same $S/N$ (630) as the
observed spectrum (dotted line). The dashed line shows the distribution of equivalent 
widths for the random noise spectrum with an added sinusoidal component of amplitude 0.05\%
(see text for details).
The similarity with the observed distribution suggests the presence of residual fringing
at the 0.05\% level for this star while fringing at the 0.1\% level can be ruled out (not shown).}
\label{f:LiI6708_obs_LP815-43_histogram}
\end{figure}

As described in Sect. \ref{s:reduction}, there may be residual fringing in the reduced
spectrum at the level of $<0.1$\% of the continuum level with a typical wavelength scale of 0.1-0.2\,nm. 
If it is as high as $0.1$\%  and one of those depressions falls on top
of the Li line, it could mean that the \liratio\ has been overestimated. 
An ad-hoc adjustment of the local continuum level by this amount would change
\liratio\ for LP815-43  to only a $1\sigma$ detection and thus not
significant by our standards
(of course if instead of a depression there is a peak in the residual fringing
it would increase \liratio\ by the corresponding amount). 
This illustrates clearly the challenges posed by this method to determine
the Li isotopic ratio in metal-poor turn-off stars. 
An enlargement of the observed August 2004 spectrum is shown in Fig. \ref{f:LiI6708_obs_LP815-43}.
While there may be hints of a residual fringing at some
wavelengths, there are no traces thereof in other parts of the spectrum near the Li line.

We have tried to quantify the amount of fringing by computing the equivalent width directly
integrated over a 0.08\,nm window (the width of the Li line) along the echelle order with the Li line.
The wavelength variation of these equivalent width measurements is slightly more pronounced than
expected from pure random noise with the same $S/N$, suggesting the presence
of fringing with a quasi-regular wavelength dependence. 
Adding a sinusoidal component with an amplitude of 0.05\% and wavelength scale
of about 0.17\,nm to such pure random noise produces a similar equivalent width
distribution to that observed in August 2004, as
shown in Fig. \ref{f:LiI6708_obs_LP815-43_histogram}; 
fringing as large as 0.1\% can clearly be ruled out. A similar exercise for
the July 2000 spectrum suggests residual fringing with an amplitude of 0.08\%.
It can also be noted that the expected \lisix\ contribution to the equivalent width 
(0.1\,pm for \liratio\,$=+0.046$) is very unlikely to be produced purely by chance as the
observed equivalent width distribution of the nearby continuum have a typical 
uncertainty of 0.03\,pm, consistent with our claimed $2\sigma$ detection for this star. 
We stress that the presence of fringing at the 0.05\% level does not
automatically imply that our derived \liratio\ for this star is overestimated as it
also requires that a fringing depression rather than a peak falls on top of the Li line, something
we are not able to evaluate the likelihood of.
The fact that independent observations acquired in 2000 and 2004 both yield
a significant presence of \lisix\ in the star increases our confidence that
our \lisix\ detection in this star is real. 
We nevertheless strongly encourage further observations of this  star
to confirm our conclusions.  

A  further concern in the analysis of LP\,815-43
is that lithium interstellar (ISM)  
absorption line may mimic the
presence of \lisix\ considering that the \ki\ 769.8\,nm is clearly distorted by
a {\em blue-shifted} ($-2.07$\,\kms ) ISM line.
Welty \& Hobbs (2001) found in their survey of ISM absorption that
$N$(\lii )/$N$(\ki )\,$= 0.0054$. Together with the difference
in $gf$-values of the two resonance lines, the ISM \lii\ profile
should be about 100 times weaker than the corresponding ISM \ki\ line.
The ISM \ki\ line has a line depth of about 8\% and an equivalent width of
0.97\,pm. The ISM \lii\ line should, therefore, have
an equivalent width of less than 0.01\,pm (0.1\,m\AA ).
It is, therefore,
unlikely that an ISM lithium line  will have a significant influence
on the derived \liratio . 

To quantify the above estimates, 
we first subtracted from the \ki\ 769.8\,nm profile
the expected stellar \ki\ line as predicted using an K abundance similar to
the other four stars with similar \feh\ and \macro\ and \vrad\ from
the \cai\ and \fei\ lines. We then fitted a Gaussian to this ISM \ki\ profile,
which was subsequently converted to a corresponding ISM \lii\ profile while
taking into account the hyperfine structure of the Li line and the
difference in overall line strength. The difference in line width
of the Gaussians due to different thermal broadening of Li and K
was also included.
After subtraction of this blue-shifted ISM profile,
we again performed a \chitwo -analysis of the remaining stellar \lii\ line.
As expected from the above estimates, the resulting \liratio\ was
indistinguishable from the previous value.
We  conclude that our derived high \liratio\ is not an artifact
of ISM Li absorption.

\subsection{Estimating Uncertainties Using Monte Carlo Calculations}
\label{s:monte_carlo}

As an independent means of estimating the uncertainties attached to 
our Li isotopic abundance analysis, we have carried out 
extensive Monte Carlo simulations for HD\,140283.
A theoretical profile of the \lii\ 670.8\,nm line with
an adopted \liratio\,$=0.02$ was treated as an observed line,
to which noise corresponding to various $S/N$ values (400, 500, 600,
700, 800, 900 and 1000) were added.
All line profiles were convolved with a Gaussian of width 5\,\kms\
and projected stellar rotational velocity of 0.5\,\kms\ prior to
introducing the effects of finite $S/N$. 
These artificial observed lines had the same wavelength sampling
as our UVES spectra. 
We then performed a \chitwo -analysis of these profiles in an
identical fashion as for our real observations with the exception
that the line broadening parameters were assumed to be known beforehand.
For each $S/N$, 100 realisations of the noisy synthetic profile were
generated and exposed to the \chitwo -analysis.
The scatter in the  estimated \liratio\ gives an estimate of
the accuracy with which the isotopic ratio can be determined 
for a given $S/N$. To these uncertainties should still be added the
possible errors arising from uncertainties in the line broadening,
continuum placement and stellar parameters. 

The estimated \liratio\ deviated from
the input value with a standard deviation of  0.010, 0.007,
0.005 and 0.004 for $S/N = 400$, 600, 800 and 1000, respectively. 
These estimates are consistent with the \chitwo\ results using the true
observed profile for this star
when ignoring the uncertainty stemming from the estimated 
errors in the adopted macroturbulence and stellar parameters.
We  conclude that the error estimates in Sect. \ref{s:li6708_chi2}
are indeed realistic.

\subsection{Effects of Choice of a \chitwo\ Analysis}
\label{s:li6708_chi2_cayrel}

In their analysis of the Li isotopic ratio in HD\,84937,
Cayrel et al. (1999) adopted a slightly different technique than the one employed by us.
Rather than determining the intrinsic line broadening
due to macroturbulence and rotation from lines of other species, they introduced 
this as an additional
 free parameter in the \chitwo -analysis of the \lii\ 670.8\,nm line.
Only after the analysis  did they confirm that the  derived
broadening parameter was consistent with values  estimated from
the \cai\ 616.2 and 671.7\,nm lines.
The \liratio\ ratio Cayrel et al. obtained for this star ($0.052\pm0.019$ at $1\sigma$)
is in agreement with the value ($0.06\pm0.03$ at $1\sigma$) of Smith et al. (1998)
from an independent observational data set.
Smith et al. used the same approach as us but with fewer calibration lines.

We have investigated how our conclusions would be modified by employing
Cayrel et al.'s  approach for our full sample.
Our results are reasonably unchanged by adopting this alternative approach.
 Only in
six of our 24 stars is the derived \liratio\ ratio changed by more than $\pm 0.02$.
Of our nine detections at the $\ga 2\sigma$ level, seven 
(HD\,102200, HD\,106038, HD\,160617, G\,020-024,
CD\,$-30\arcdeg18140$, CD\,$-33\arcdeg3373$, LP\,815-43) 
are still detections under the same criterion. 
While G\,013-009 and CD\,$-48\arcdeg2445$ would no longer be significant detections, 
three other stars would in this framework be classified as
$\ga 2\sigma$ detections: HD\,213657, HD\,298986 and BD\,$+03\arcdeg0740$, which
otherwise  were only $1\sigma$ results.  BD\,$+03\arcdeg0740$
(\feh\,$=-2.65$)
is noteworthy given its low \feh . 

We believe that  Cayrel et al.'s method  is inferior to ours
as it does not make full use of the
information content in the observed spectra. Due to the limited
wavelength coverage of their spectra, Cayrel et al.  had access to only
two calibration lines (\cai\ 616.2 and 671.7\,nm) which in HD\,84937 are
of quite different strength compared with the Li line. Under those circumstances
it may be justified to allow also the broadening parameters to vary in
the \chitwo -analysis of the \lii\ 670.8\,nm profile when the $S/N$ is very
high as in their spectrum. One of the great advantages of the
VLT/UVES is the wide wavelength coverage that provides
several lines  for estimating the
intrinsic line broadening.
We also note that the uncertainties in the estimated \liratio\ ratios
are in general larger when Cayrel et al.'s method is used in place of
our standard analyis presented in Sect. \ref{s:li6708_chi2}, in
some cases significantly so (e.g. BD\,$-13\arcdeg3442$: \liratio$=0.001\pm0.028$
and $0.042\pm0.050$, respectively), while in no case was   the
uncertainty reduced.
We  conclude that the method of Cayrel et al. is less appropriate than
the one employed in the present study. Nevertheless, our results would
have been quite similar had we adopted this approach.

\subsection{Effects of Choice of Stellar Parameters}
\label{s:li6708_parameters}

As described in Sect. \ref{s:parameters}, the estimated uncertainties in the
adopted stellar parameters in absolute terms are $\Delta$\teff$=\pm100$\,K,
$\Delta$\logg$=\pm0.2$, $\Delta$\feh$=\pm0.2$ and $\Delta$\micro$=\pm0.2$\,\kms .
We have repeated our Li isotopic abundance determinations for all stars
using \marcs\ model atmospheres
with modified stellar parameters by deriving new line broadening parameters
and then repeating the  \chitwo -analysis of the \lii\ 670.8\,nm profile.
The resulting \liratio\ ratios when changing the stellar parameters by
$+100$\,K, $+0.2$\,dex and $+0.5$\,\kms , respectively, are given in Table \ref{t:li_parameters}.
Clearly, the \liratio\ ratios are basically immune to the uncertainties at this
level in the stellar parameters: the slightly different line broadening
parameters derived from the calibration lines also apply to the \lii\ line, leaving
\liratio\ essentially intact. Very rarely is
$\Delta$\liratio\ as large as 0.003 with the difference being $\le 0.001$ in the
majority of cases. This finding is in line with those reported previously
by Smith et al. (1993, 1998).

While the effects of the uncertainties in \logg , \feh\ and \micro\ are
negligible for the derived \liseven\ and \lisix\ absolute abundances
($\Delta$\logli$<\pm 0.01$\,dex for our estimated stellar parameter uncertainties),
this is not true when changing \teff . Our calculations reveal that
\logli\ increases with about 0.073\,dex for $\Delta$\teff$=+100$\,K,
necessarily quite
similar to the estimate by Ryan et al. (1999).

\subsection{Effects of Choice of Suite of 1D Model Atmospheres}
\label{s:li6708_kurucz}

The tendency for our derived \liratio\ ratios to be slightly but positively offset from zero,
may raise concern that our analysis suffer from an unidentified systematic error.
One possibility could be our use of \marcs\ model atmospheres, since
Nissen et al. (1999) found a {\em smaller} isotopic ratio (0.033 instead
of 0.041) for the disk star HD\,68284 when switching from a \marcs\ to
a Kurucz model atmosphere with identical stellar parameters.
To investigate whether the choice of suite
of 1D model atmospheres has any effect on the derived Li isotopic
abundances, we have repeated the analysis for all stars using Kurucz
model atmospheres without convective overshoot and for the same stellar parameters,
as computed by Castelli et al. (1997).
The resulting \liratio\ ratios are presented in Table \ref{t:li_parameters}
together with the corresponding values with \marcs\ model atmospheres.
The changes to the isotopic ratios
 resulting from substitution of the Kurucz for the \marcs\
models are, as anticipated,   negligible:
$\Delta$\liratio$\le \pm 0.001$ and in no case larger than 0.003.
Furthermore, the small differences can be both positive and negative.

Similarly, the effect of the Kurucz models on the absolute Li abundances
is marginal\footnote{Differences in absolute Li abundances approaching
0.1\,dex would have been found had we resorted to the convective overshoot
models of Kurucz (1993) instead of using the  models with this
option switched off (Ryan et al. 1996). These overshoot models are, however, significantly
less successful in reproducing observed stellar flux distributions and their
shallower temperature gradients find no support in our detailed 3D hydrodynamical
simulations of stellar atmospheres with surface convection. 
We conclude that the overshoot models are less appropriate for abundance
analyses of halo stars.}.
The average difference compared with the \marcs\ based results
is $\Delta$\logli$ = 0.015\pm0.018$\,dex with the maximum difference being
0.048 for HD\,102200. 

For completeness, we note that the above test  is based on the
assumption that the stellar parameters of the model atmospheres
need no adjustments when
switching to a different suite of model atmospheres. Given the almost identical
behaviour of our \lii , \ki, \cai , \fei\ and \feii\ lines in the two cases,
we expect insignificant differences for both the photometric calibrations
or the \ha\ line profile fitting.
In summary, our conclusions about the lithium isotopic ratio are immune and the
lithium abundance insensitive to the particular brand of
theoretical 1D, hydrostatic, LTE model atmospheres employed in the analysis.

\subsection{Effects of 3D Hydrodynamical Model Atmospheres}
\label{s:li6708_3D}

In addition to the tests described above, we have also performed a similar
\chitwo -analysis using the time-dependent 3D hydrodynamical model atmospheres described in 
Sect. \ref{s:model_atmospheres} under the assumption of LTE. 
Since such 3D models still only exist for relatively few stellar parameter combinations,
only some of our target stars could be exposed to a 3D analysis. 
The 3D models assigned to the various stars were based on similarity in stellar parameters;
we note therefore that the here presented 3D results are obtained with somewhat different
stellar parameters than adopted for the 1D calculations described in Sect. \ref{s:li6708_chi2}.
As explained in Sect. \ref{s:line_formation}, the main advantages with a 3D approach is
that neither micro- nor macroturbulence enter the analysis and that the intrinsic line asymmetries
arising from convective motions follow directly from the 3D line formation calculations. 
The projected stellar rotation velocity \vsini\ must however be determined
from other lines. Thus, in the \chitwo -analysis, \vsini\ replaces \macro\ as a free variable;
the Gaussian instrumental broadening measured for each line is known from Th lines.

\begin{figure}
\resizebox{\hsize}{!}{\includegraphics{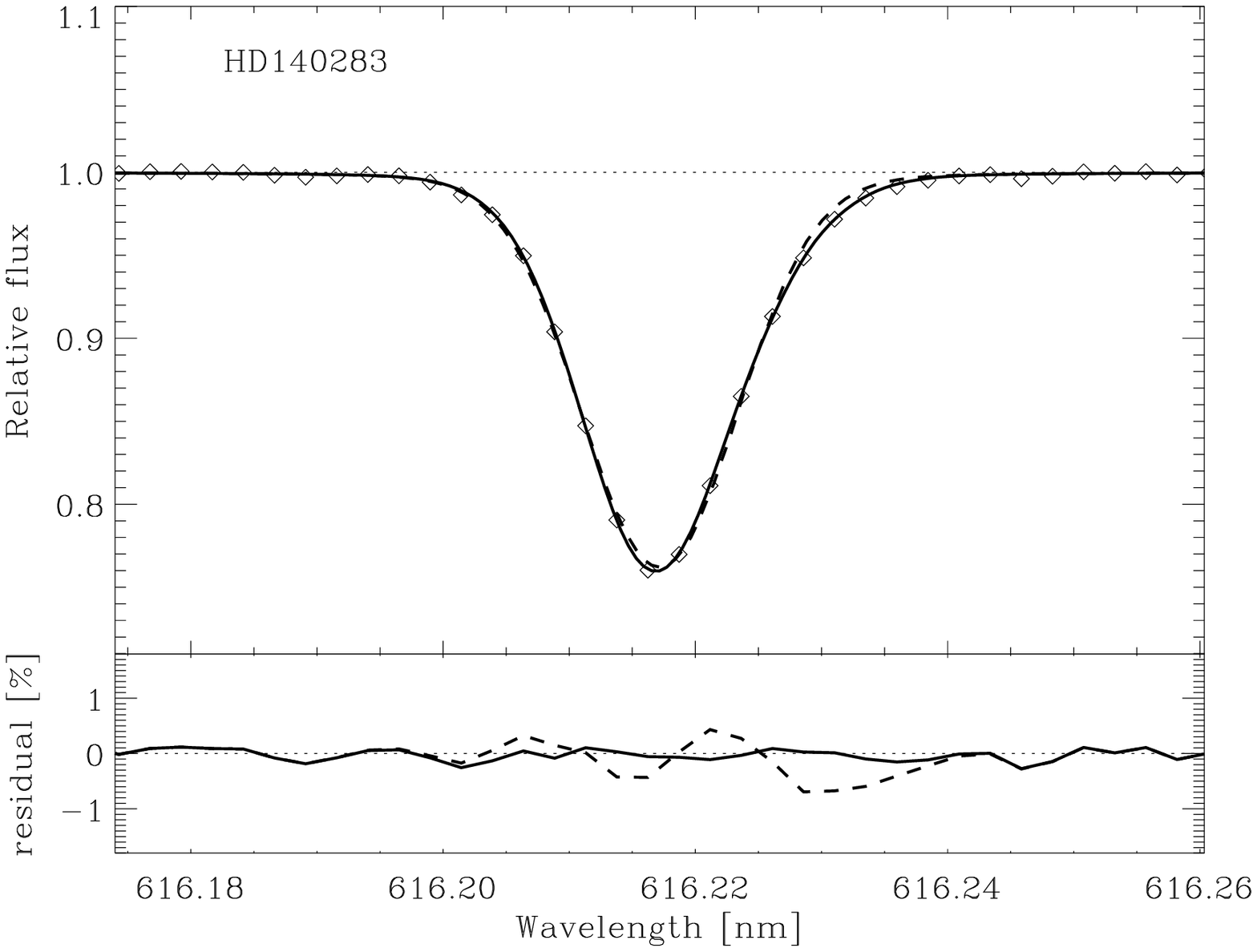}}
\resizebox{\hsize}{!}{\includegraphics{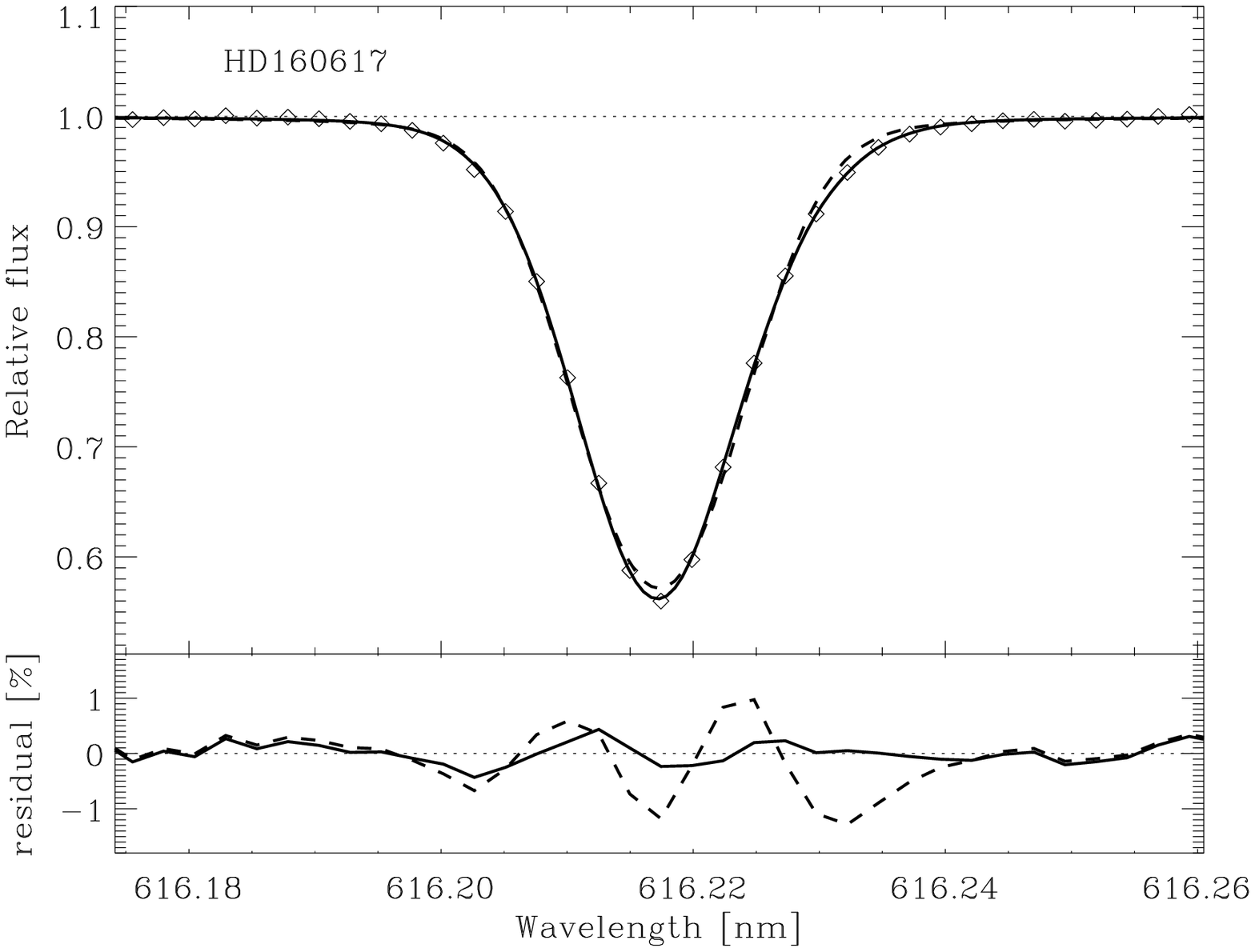}}
\caption{\cai\ 616.21\,nm line in HD\,140283 ({\em upper panel}) and HD\,160617 ({\em lower panel}).
The solid lines correspond to the best fitting 3D profiles from a \chitwo\-analysis while
the dashed lines are for the 1D profiles; the rhombs are the observed profiles. Also shown 
are the residuals between observed and predicted line profiles.}
\label{f:ca6162_3D_1}
\end{figure}

\begin{figure}
\resizebox{\hsize}{!}{\includegraphics{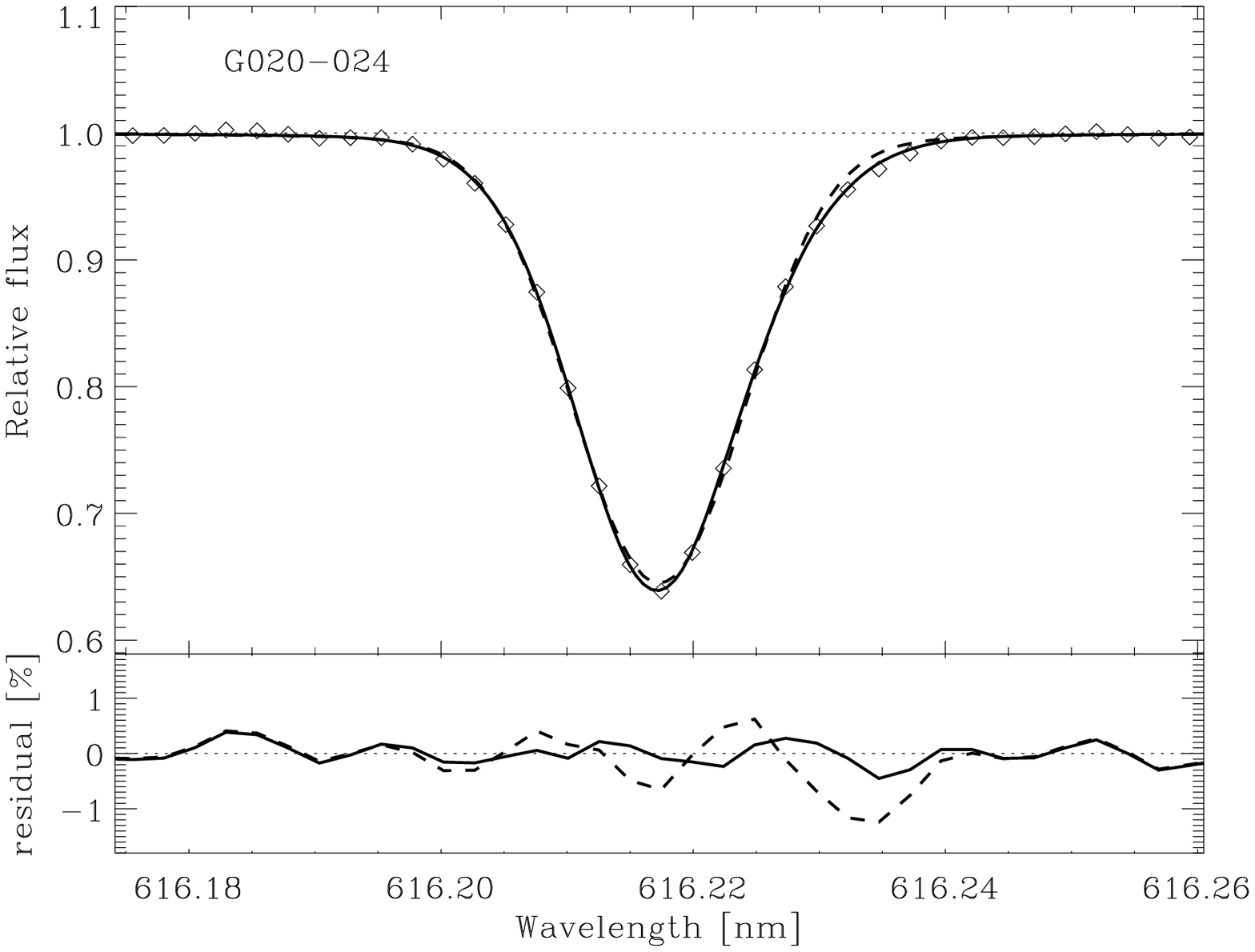}}
\resizebox{\hsize}{!}{\includegraphics{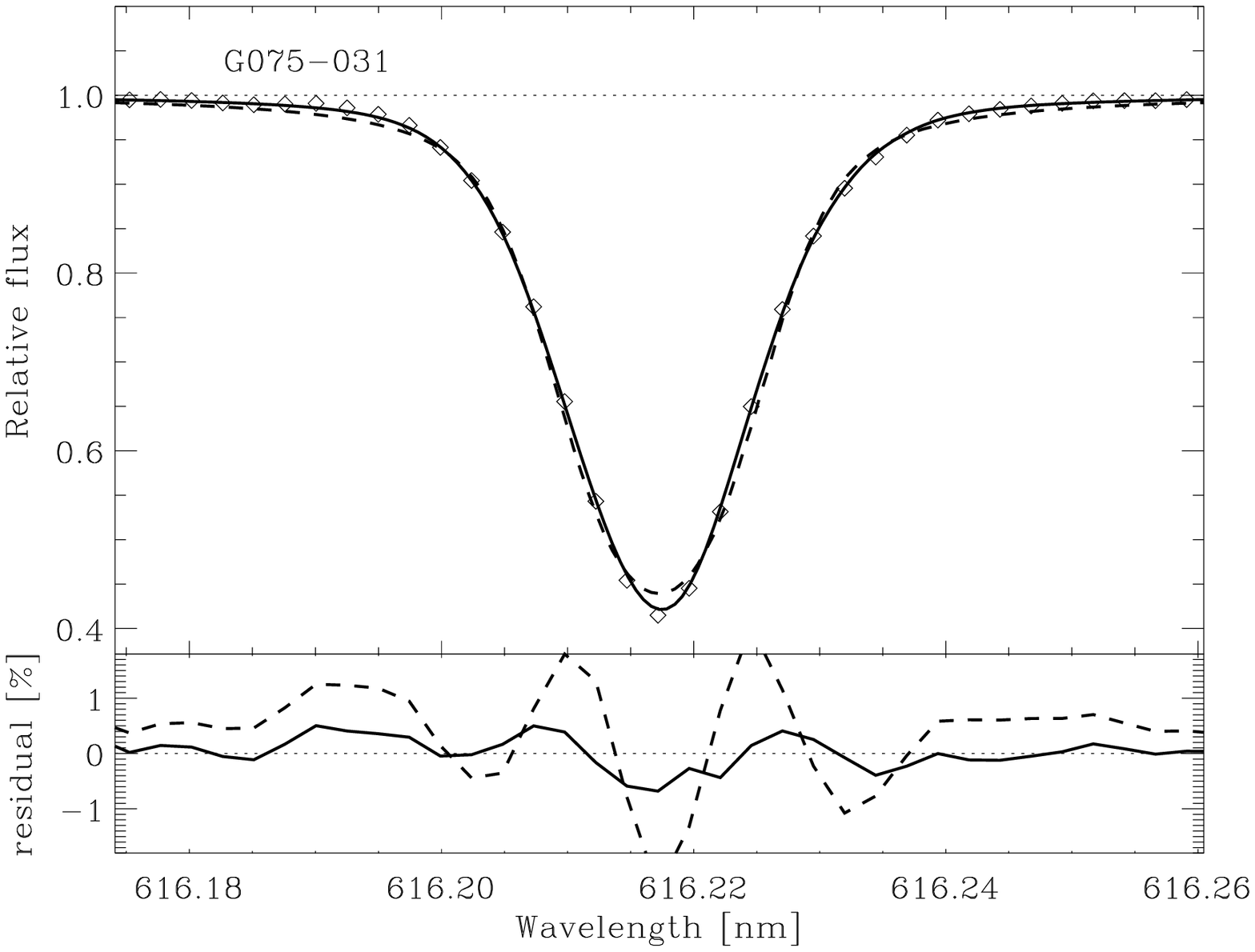}}
\caption{Same as Fig. \ref{f:ca6162_3D_1} but for G\,020-024 ({\em upper panel}) 
and G\,075-031 ({\em lower panel}).}
\label{f:ca6162_3D_2}
\end{figure}

The results for the stars for which a 3D analysis has been carried out are given 
in Table \ref{t:li6_3D}, while Fig. \ref{f:ca6162_3D_1} and Fig. \ref{f:ca6162_3D_2}
show examples of the achieved profile fits 
for the \cai\,616.21 lines for a few of the stars. 
It is noteworthy that the 3D analysis yields a substantial improvement in the agreement
with the observed profiles of Ca and Fe lines, as also previously reported by Nissen et al. (2000). 
In almost all cases, the resulting \chitwo\ for the calibration lines are
smaller than for the corresponding 1D cases;
the only exceptions are for the \ki\,769.89, \feii\,624.76 and/or \feii\,645.64\,nm lines for
a few of the stars. 
The derived \vsini\ values are between 0.1 and 3.8\,\kms . The typical uncertainties in \vsini\
are 0.3\,\kms\ but in some cases can be as large as $\approx 1$\,\kms . 
Very little is known about the observed rotation properties of metal-poor halo stars but due to their
high age one would expect them to rotate slowly (the Sun has $v_{\rm rot}=1.8$\,\kms ). 
It is therefore quite possible that the estimated rotational velocities towards the higher end are 
overestimated. For example, Smith et al. (1998) suggest \vsini\,$<2-3$\,\kms\ 
for a typical metal-poor turn-off star. It can, however,
not be ruled out at present that some of these stars actually
have \vsini\,$\approx 3$\,\kms .  
Part of the explanation for some of the high estimated \vsini\ values is that
the adopted stellar parameters for the 3D models are inappropriate for some of the stars:
if a too low \teff\ or too high \logg\ are used then the convective motions and thus
the Doppler line broadening 
will be too small, requiring large \vsini\ to compensate when fitting the observed
spectral lines. 
Furthermore, we remind the reader that LTE has been assumed for these 3D line 
formation calculations while the steep temperature gradients in low metallicity
3D models can be expected to result in significant non-LTE effects
(Asplund et al. 1999; Asplund 2005). 
We suspect that this may be the explanation for the slight correlation between
\vsini\ and \feh\ in our sample, although we note that \vsini\ is particularly
uncertain for the two stars with \feh\,$\approx -1$ (HD\,3567 and G\,075-031)
because of the quite strong calibration lines employed.
We therefore urge these \vsini\ values not to
be taken too literally at this stage. In fact, it is encouraging that the resulting rotational
velocities are at least reasonable given that there are no broadening mechanisms
like micro- and macroturbulence as in 1D that can be tuned. 

\begin{figure}
\resizebox{\hsize}{!}{\includegraphics{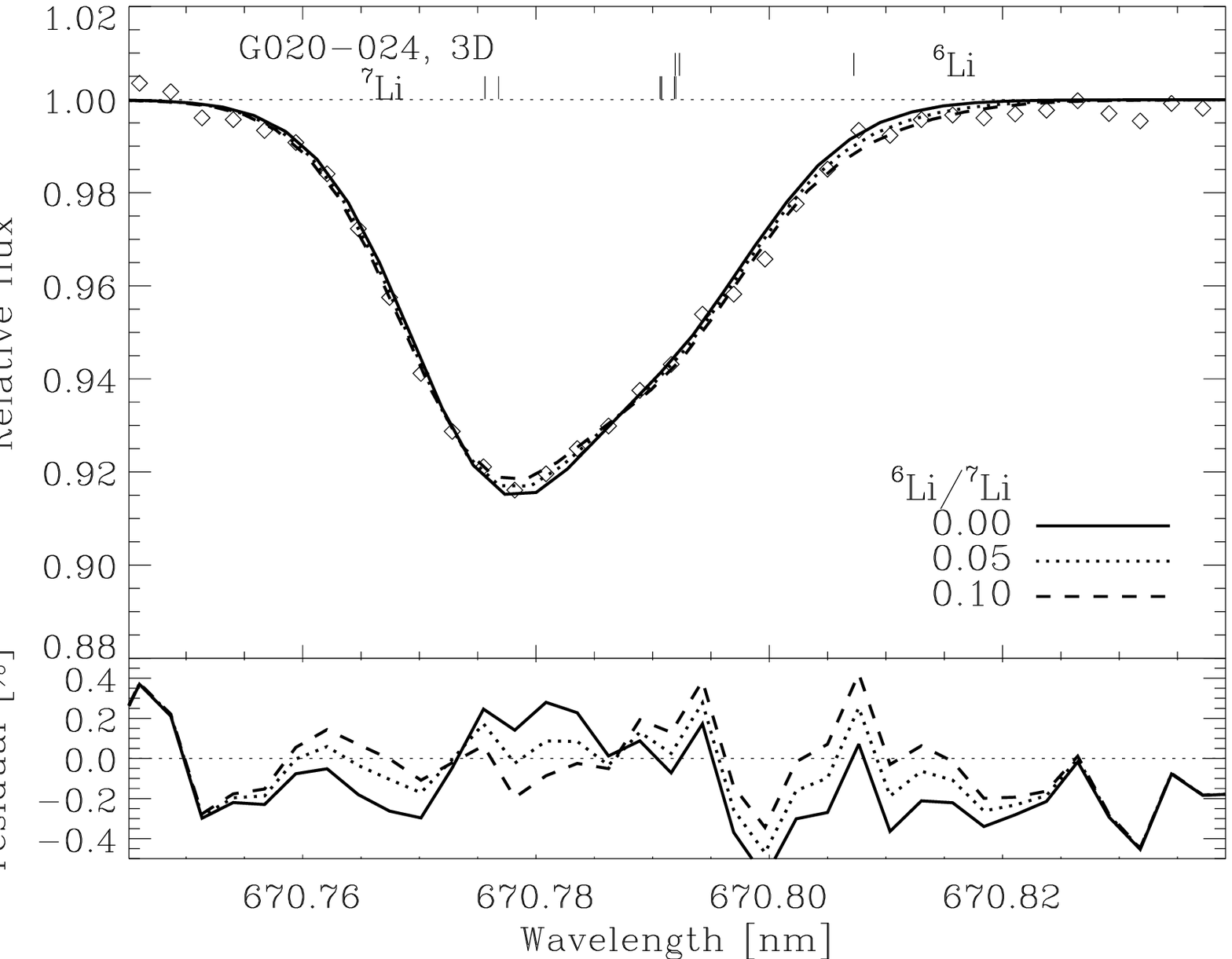}}
\resizebox{\hsize}{!}{\includegraphics{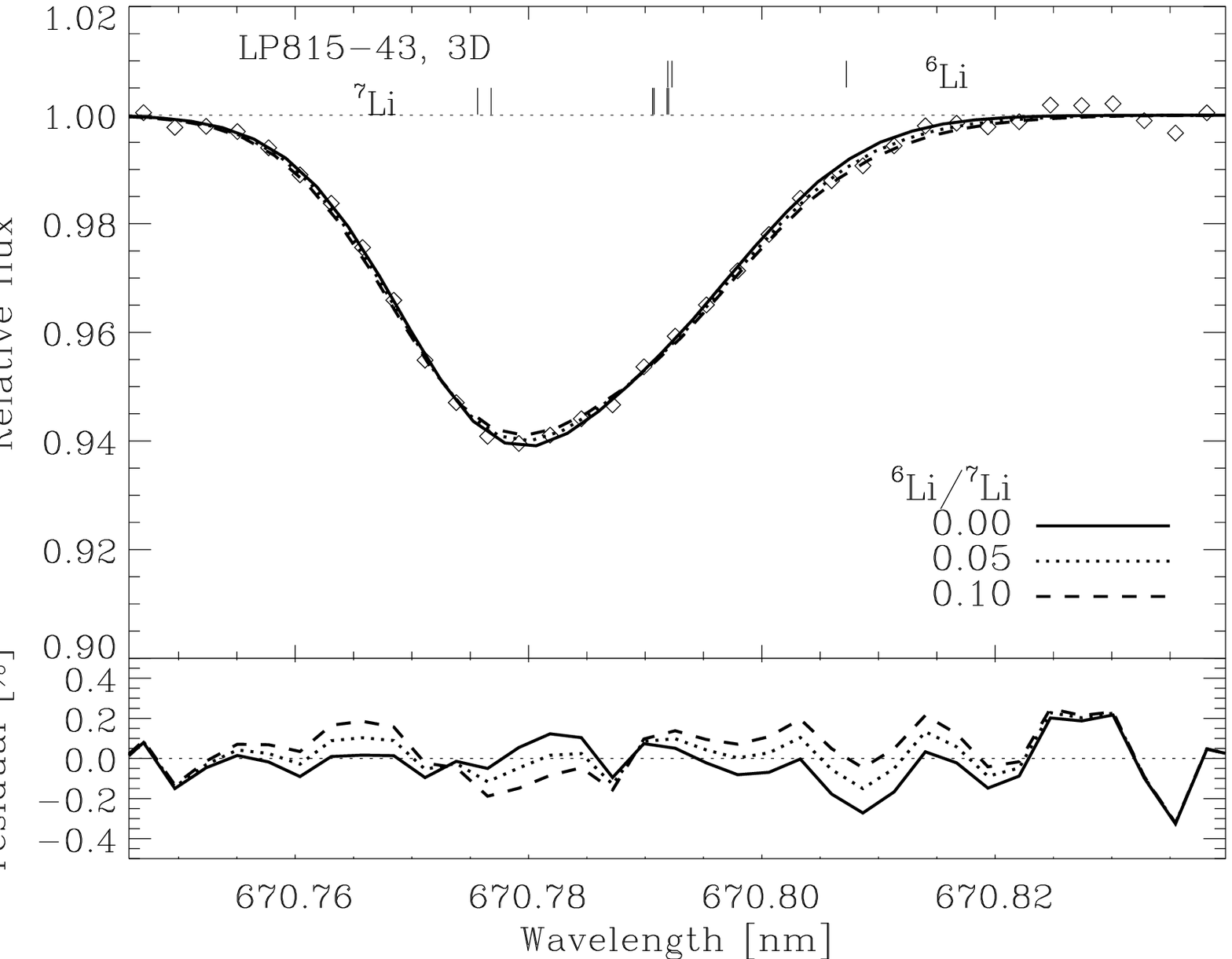}}
\caption{\lii\ 670.78\,nm line in G\,020-024 ({\em upper panel}) and LP815-43 ({\em lower panel})
based on a 3D LTE analysis. The best fitting line profiles as judged by a \chitwo\-analysis 
for \liratio\,$=+0.00, +0.05$ and $+0.10$ are shown.}
\label{f:li6708_3D}
\end{figure}

Interestingly, in general the 3D LTE calculations do not improve the agreement with 
the \lii\,670.78\,nm line as judged by the \chitwo -statistics. For more than half of
the stars, the 3D analysis yield slightly higher minimum \chitwo-values than in 1D. 
The most likely explanation in our opinion for this contrasting  behaviour compared with 
the Ca and Fe lines is that LTE is a particularly poor assumption for
the \lii\ line in metal-poor stars; pronounced over-ionization of \lii\ is predicted from 
3D non-LTE calculations
due the steep temperature gradients of the 3D model atmospheres 
(Asplund et al. 2003; Barklem et al. 2003). 

An inspection of Table \ref{t:li6_3D} reveal that the 1D \liratio\ ratios appear to be 
similar or lower than the ratios derived from 3D LTE calculations. Only in three stars
are the 3D results noticeably smaller than for the corresponding 1D case.
The smaller \liratio\ ratios in HD\,3567 and G\,075-031 are likely a reflection of the relatively
high \feh\ of these two stars, which renders the calibration lines used to determine \vsini\ 
quite strong (equivalent widths up to 12\,pm); in the case of the Sun, 
the impressive agreement between predicted and observed line profiles found for
weak lines break down for such strong lines (Asplund et al. 2000).
The difference between 3D and 1D for LP\,815-43 is of particular interest given the prominent
status this star will have when later discussing the implications of our \lisix\ survey.
In 3D, the derived isotopic ratio is \liratio\,$=0.036\pm0.021$ to be compared with
the 1D value of \liratio\,$=0.046\pm0.022$ (both results are for the August 2004 observations). 
Taking these 3D LTE calculations at face value, LP\,815-43 would no longer be 
considered a significant \lisix\ detection,
although it would not be far from making the $2\sigma$ criteria. 

Considering the expected severe 3D non-LTE effects on the \lii\ line in metal-poor stars 
that are not accounted for here, the 1D results are to be preferred. Indeed,
one would expect also the calibration lines used to estimate \vsini\ to be vulnerable
to departures from LTE, in particular minority species like \cai\ and \fei\ (Asplund 2005).
Carrying out detailed
3D non-LTE calculations for both the calibration lines and the \lii\ line 
for the necessary many simulation snapshots is unfortunately extremely computationally demanding.
We will  therefore have to postpone investigating this remaining issue to a future date. 
The preliminary 3D LTE analysis carried out here, however, 
has not uncovered any significant systematic errors that would
suggest that our 1D-based Li isotopic ratios are overestimated in general.

\subsection{Comparison with Previous Studies}
\label{s:li6708_literature}

A number of our targets have been investigated for their
Li isotopic abundances in previous studies.
Our non-detection of \lisix\ is consistent with the upper limits
reported for HD\,19445, HD\,140283 and BD\,$+03\arcdeg0740$ by
Smith et al. (1993, 1998) and Hobbs \& Thorburn (1994, 1997).
Our derived value for G\,271-162 (\liratio\,$=0.019\pm0.012$) does not
constitute a
detection of \lisix\ but is very close to 
Nissen et al.'s  (2000) result of \liratio\,$=0.02\pm0.01$
based on the same VLT/UVES commissioning spectra,
in particular as we have here adopted a slightly more conservative
estimate of the uncertainties.

More noteworthy  is the result that we do not confirm the \lisix\ detection
for HD\,338529 (also known as BD\,$+26\arcdeg3578$) 
by Smith et al. (1998) who found \liratio\,$=0.05\pm0.03$.
Our value of $0.010\pm0.013$ is based on a higher quality spectrum
($S/N \simeq 520$ vs 420) but the main difference can be
traced to the adopted line broadening parameters.
Smith et al. combined the effects of instrumental 
broadening and macroturbulence by convolving the synthetic spectra with 
a Gaussian profile, but used only the \cai\ 671.78\,nm line to determine
the width of this Gaussian; however, this line is quite weak 
in HD\,338529 having a line depth of only 3\% at $R = 110,000$.  
As a result, Smith et al. could not constrain the macroturbulent velocity parameter 
for this star very well
(this can be seen in their Figure 6 for HD\,338529, where 
the \chitwo\ as a function of macroturbulent velocity is very shallow). 
Our higher broadening parameter is significantly better determined as it
is based on four \cai\ and \fei\ lines, all in very good agreement.
 We would also have found a significant
\liratio\ ratio had we adopted the same macroturbulence as Smith et al. (1998).
In addition, for two of our stars with significant \lisix\ detections,
Smith et al. could only obtain upper limits: $\le 0.06$ and $\le 0.05$ ($2\sigma$)
for HD\,102200 and HD\,160617, respectively. Again, the main difference
appears to be the adopted macroturbulence parameter, which in our case
is smaller than their value. Also, our spectra for these stars  have 
significantly higher $S/N$.
The differences in \liratio\ between us and Smith et al. for HD\,338529 
certainly serve as a warning when trying to interpret the results
in terms of Li isotopic abundances for this extremely challenging method.
It is clearly paramount to have exceptionally high $S/N$ as well as
well-determined intrinsic broadening parameters based on a multitude
of spectral lines with similar strength as the \lii\ 670.8\,nm line. 
While we have tried to accomplish both of these factors
and believe that we have set a new standard in this respect,
we urge the reader to exercise some 
caution when using our quoted Li isotopic abundances.

Deliyannis \& Ryan (2000) reported a firm detection of \lisix\ in the
subgiant HD\,140283: \liratio\,$=0.040 \pm 0.015$. 
No details of this analysis have as yet appeared in the literature.
We are, therefore, not able to identify  reasons for
the discrepancy with our non-detection of \liratio\,$=0.008 \pm 0.006$.
Our result for HD\,140283 is, however, consistent with the  upper limit
of \liratio\,$<0.018$ estimated recently by Aoki et al. (2004) based
on a Subaru/HDS spectrum of the quality of  our VLT/UVES spectrum.
Wako Aoki has kindly made available their Subaru spectrum of HD\,140283
for an independent analysis by us. 
As we did not have access to their measured instrumental profile,
we performed the analysis adopting a combined Gaussian for the
instrumental and macroturbulence broadening, a case also considered
by Aoki et al. Our result of \liratio\,$=0.003$ is in perfect agreement
with their corresponding result of 0.002 when using their choice of
calibration lines. 
We  note that when we perform an analysis in the same
way as for our own sample (i.e. using the same  
calibration lines and taking into account uncertainties
in line broadening, $S/N$ and stellar parameters), we find 
\liratio\,$=0.006\pm0.004$, which is in excellent agreement with 
the UVES result. When expanding the set of calibration lines
to include in total 14 lines between 500 and 650\,nm from the 
Subaru spectrum, the 
$1\sigma$ error doubles: \liratio\,$=0.007\pm0.009$. 
The exact results and uncertainties thus depend (slightly) on the 
particular choice of lines to estimate the line broadening,
which must be borne in mind when comparing different
\liratio -analyses.

\section{Li Abundances from the \lii\ 610.4\,nm Subordinate Line}
\label{s:li6104}

\subsection{Analysis}
\label{s:li6104.anal}

\begin{figure}
\resizebox{\hsize}{!}{\includegraphics{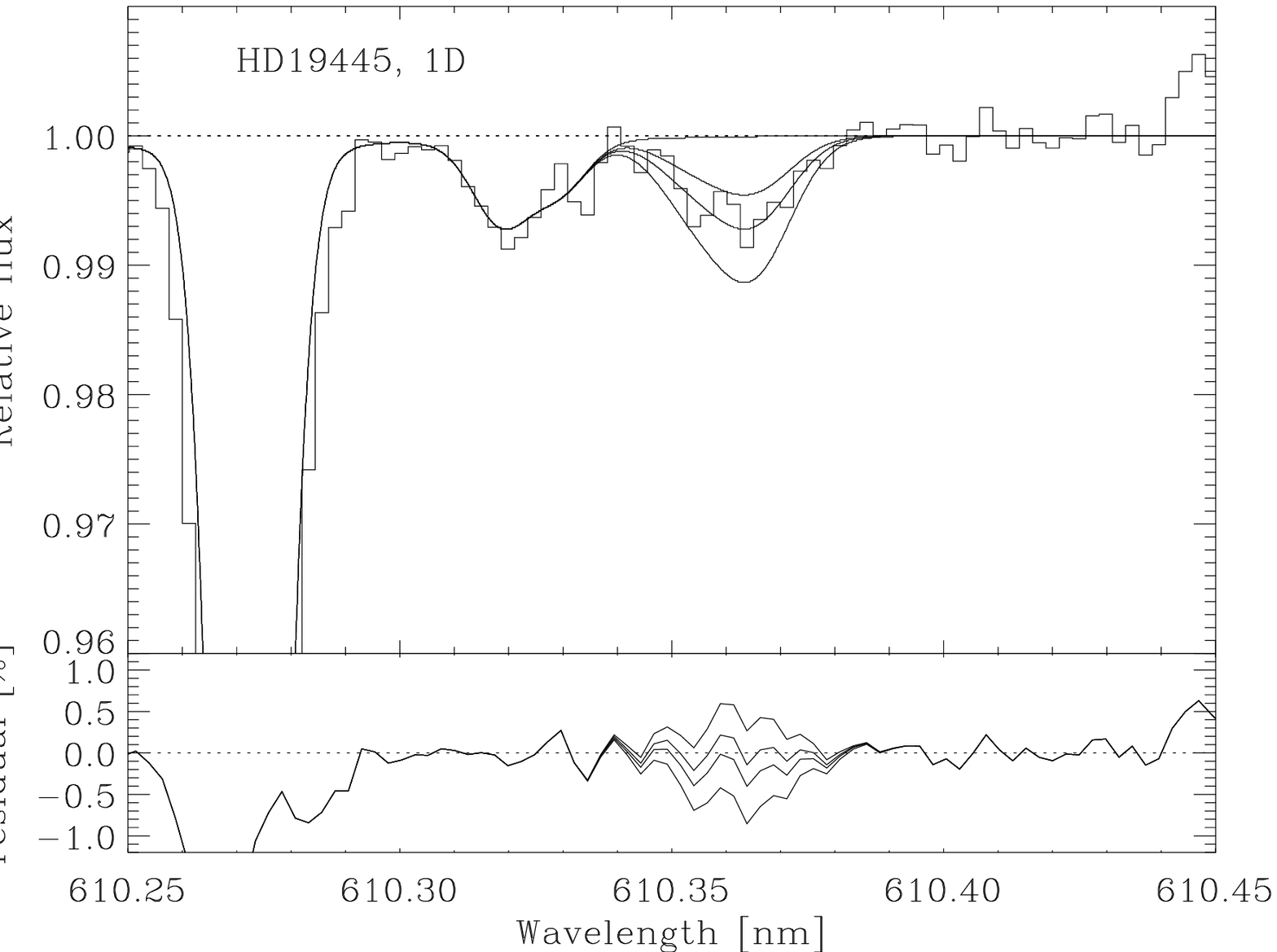}}
\resizebox{\hsize}{!}{\includegraphics{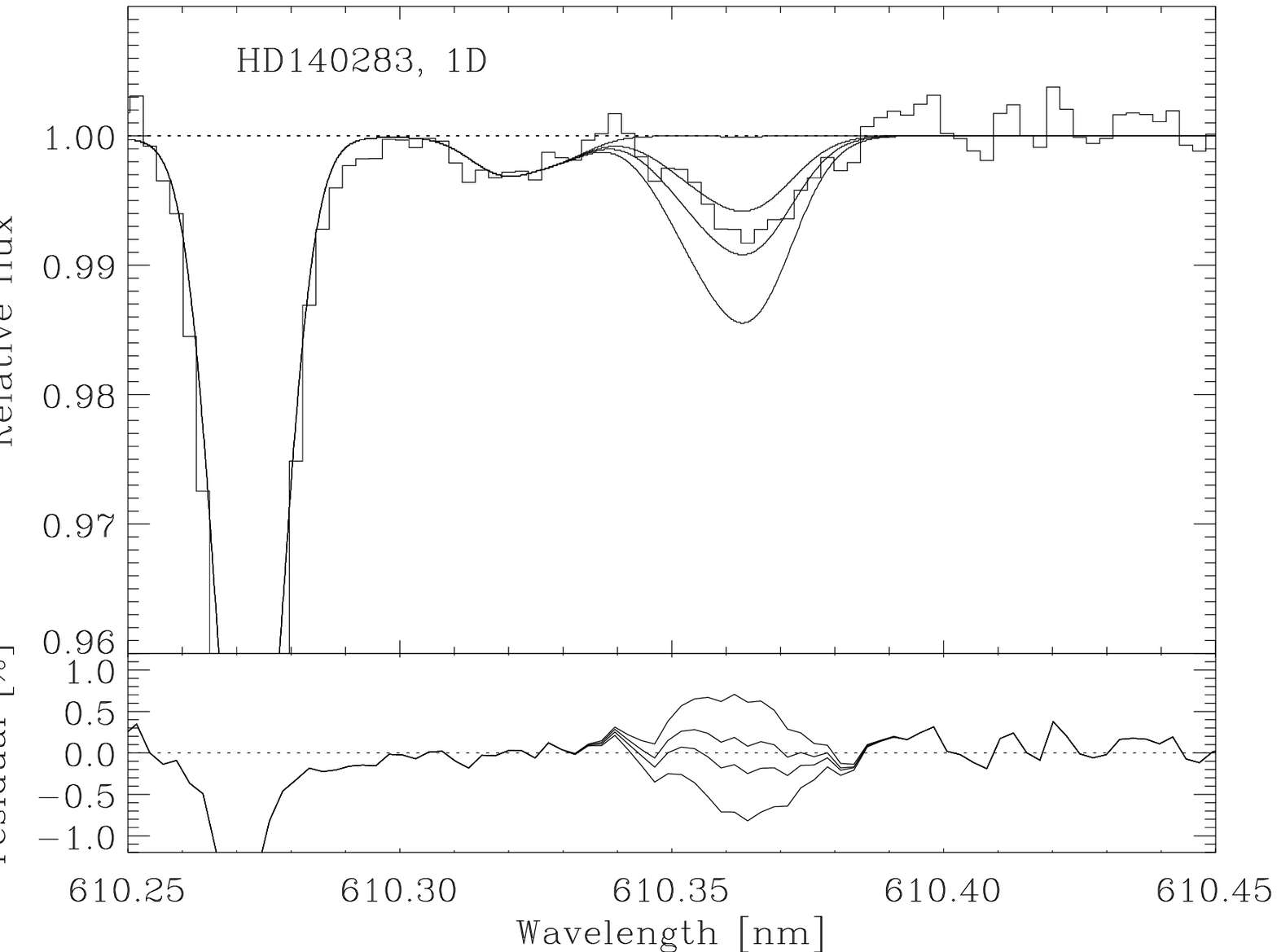}}
\caption{Observed and theoretical \lii\ 610.4\,nm profiles for HD\,19445 (upper panel)
and HD\,140283 (lower panel). The four theoretical 1D LTE profiles are computed with
no Li, \logli\,$=2.0$, 2.2 and 2.4, respectively, with the stellar parameters listed
in Table \ref{t:parameters}. For each star, the residuals between the synthesis 
and observations are also shown.}
\label{f:li6104_spec1}
\end{figure}

\begin{figure}
\resizebox{\hsize}{!}{\includegraphics{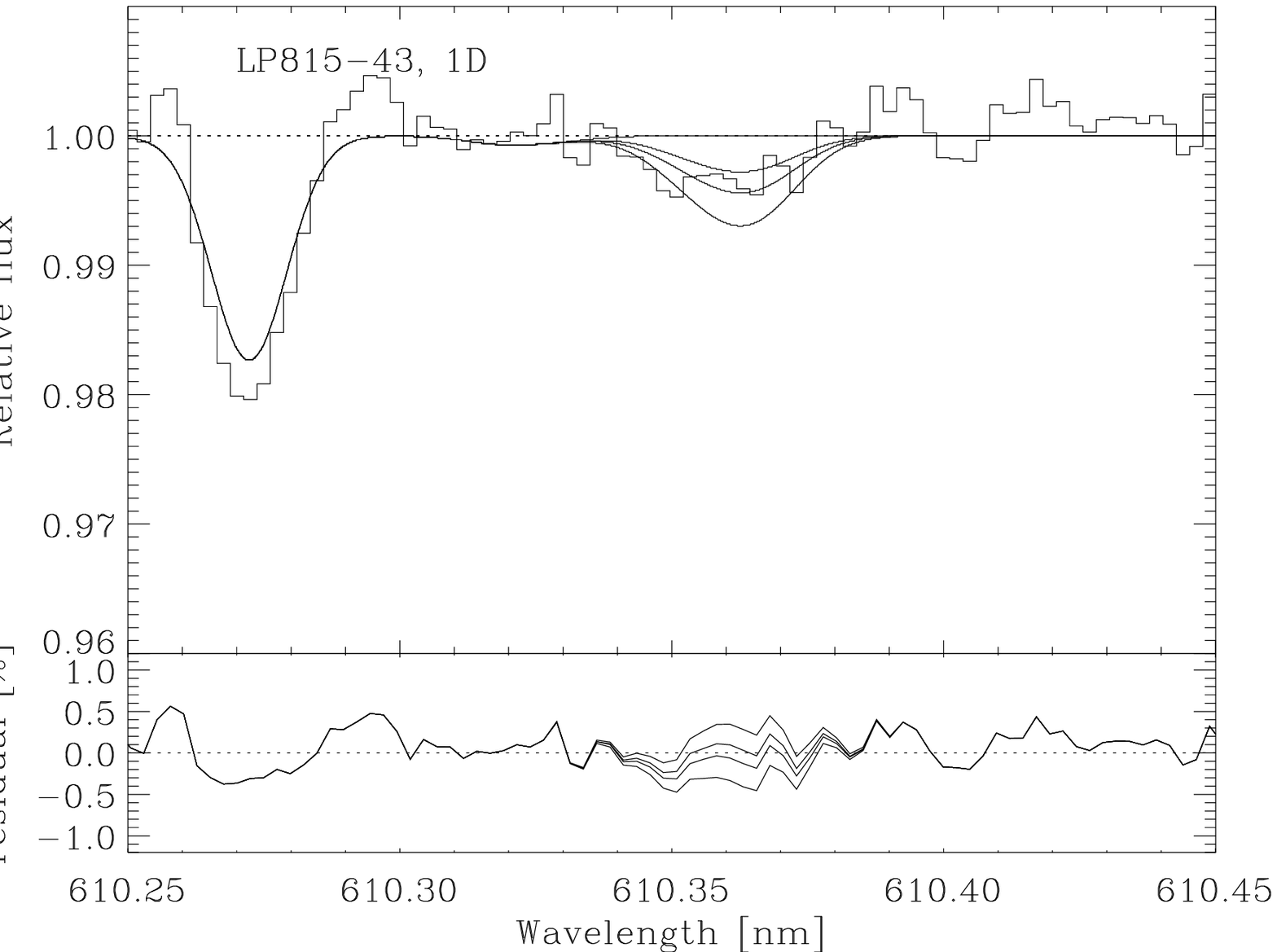}}
\resizebox{\hsize}{!}{\includegraphics{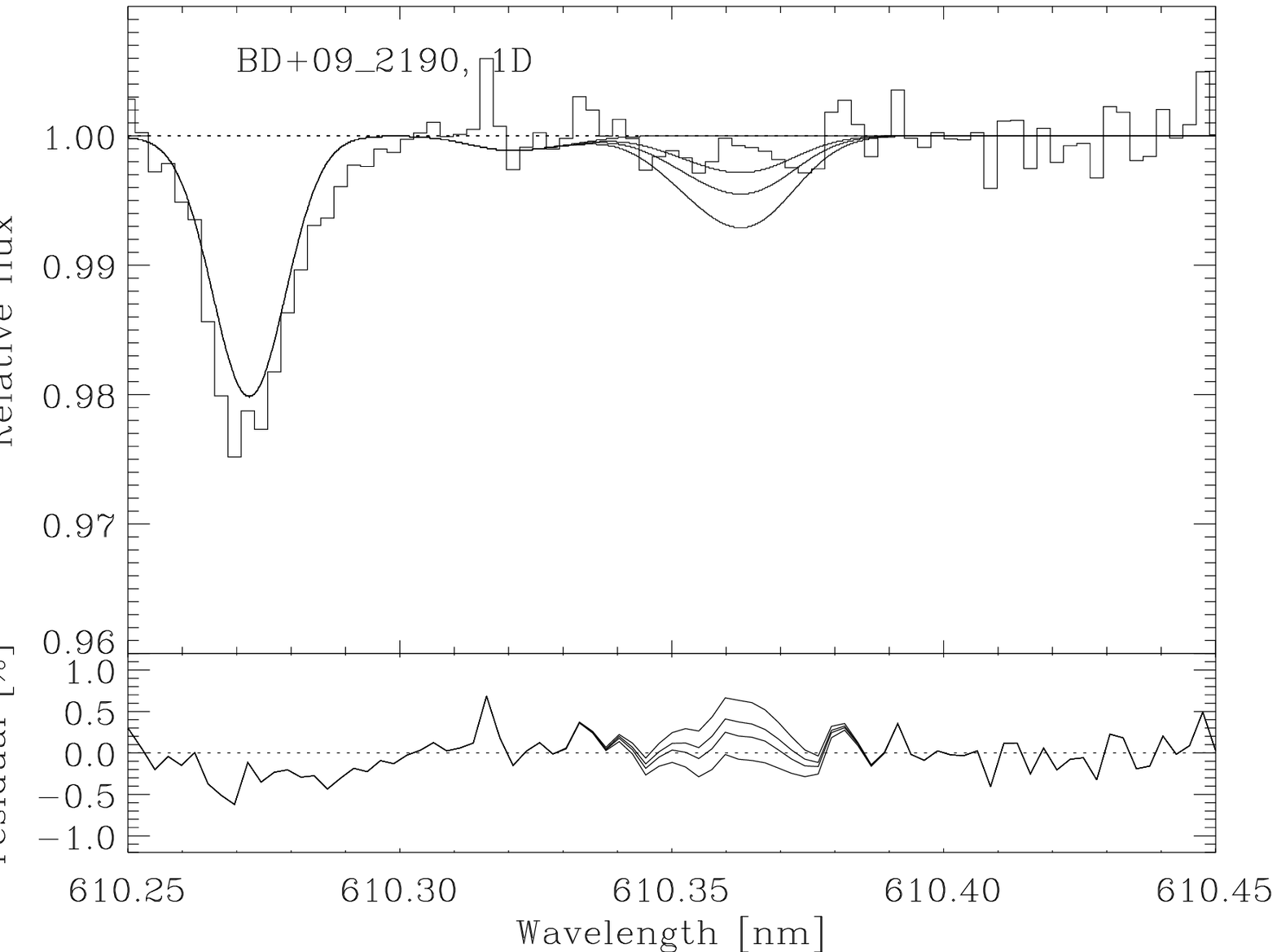}}
\caption{Same as for Fig. \ref{f:li6104_spec1} but for LP\,815-43 (upper panel)
and BD\,$+09\arcdeg 2190$ (lower panel). No significant detection of the 
\lii\ 610.4\,nm line is made in the latter star according to a \chitwo -test, although 
there is a possible feature at the correct wavelength that would correspond to
\logli\,$\approx 2.0$. }
\label{f:li6104_spec2}
\end{figure}

The exceptionally high quality of our UVES spectra  allows 
Li abundances to be derived from the subordinate \lii\ 610.36\,nm line,
which in these halo stars has an equivalent width of just
 $0.1-0.2$\,pm ($1-2$\,m\AA ).
We have performed a similar \chitwo -analysis of a wavelength region
around 610.4\,nm, taking into account the hyperfine and isotope
splitting for the Li line and neighboring lines of \fei , \feii\ 
and \cai . The adopted Fe and Ca abundances were those 
implied by the other considered Fe and Ca lines.
While no attempt was undertaken to optimize the agreement between
observations and predictions for these Fe and Ca lines by adjusting the
$gf$-values or elemental abundances, we emphasize that this had
no impact on the derived Li abundances as the neighboring lines
are quite well described  and  sufficiently far away
from the Li feature not to cause any problems. 
Fig. \ref{f:li6104_spec1} and Fig. \ref{f:li6104_spec2} 
compare the synthesized profiles for
different Li abundances with the observed spectra for a few 
stars.

The derived Li abundances from the subordinate \lii\ 610.4\,nm line 
are listed in Table \ref{t:li_abundances}.
The line is unambiguously detected in
22 of the 24 program stars with the two exceptions being the
very metal-poor stars BD\,$+03\arcdeg 0740$ and BD\,$+09\arcdeg 2190$.
Also for those two stars
there is a tantalizing suggestion
of a feature at the correct wavelength, although the \chitwo -analysis
does not indicate a significant detection. 
As a consequence, we only give $3\sigma$ upper limits to the 
Li abundance in these two stars. 

The derived
1D LTE Li abundances from the \lii\ 610.4\,nm line have been corrected for
departures from LTE using the detailed 1D statistical
equilibrium calculations of Carlsson et al. (1994).
The non-LTE abundance corrections for the subordinate line are 
positive and slightly larger than for the resonance lines:
typically $\Delta$\logli$\simeq +0.05$\,dex. The maximum
non-LTE effect occurs for the metal-poor subgiant HD\,140283 ($+0.09$\,dex).

\subsection{Comparison with Results from the Li Resonance Line}
\label{s:li6104vsli6708}

\begin{figure}
\resizebox{\hsize}{!}{\includegraphics{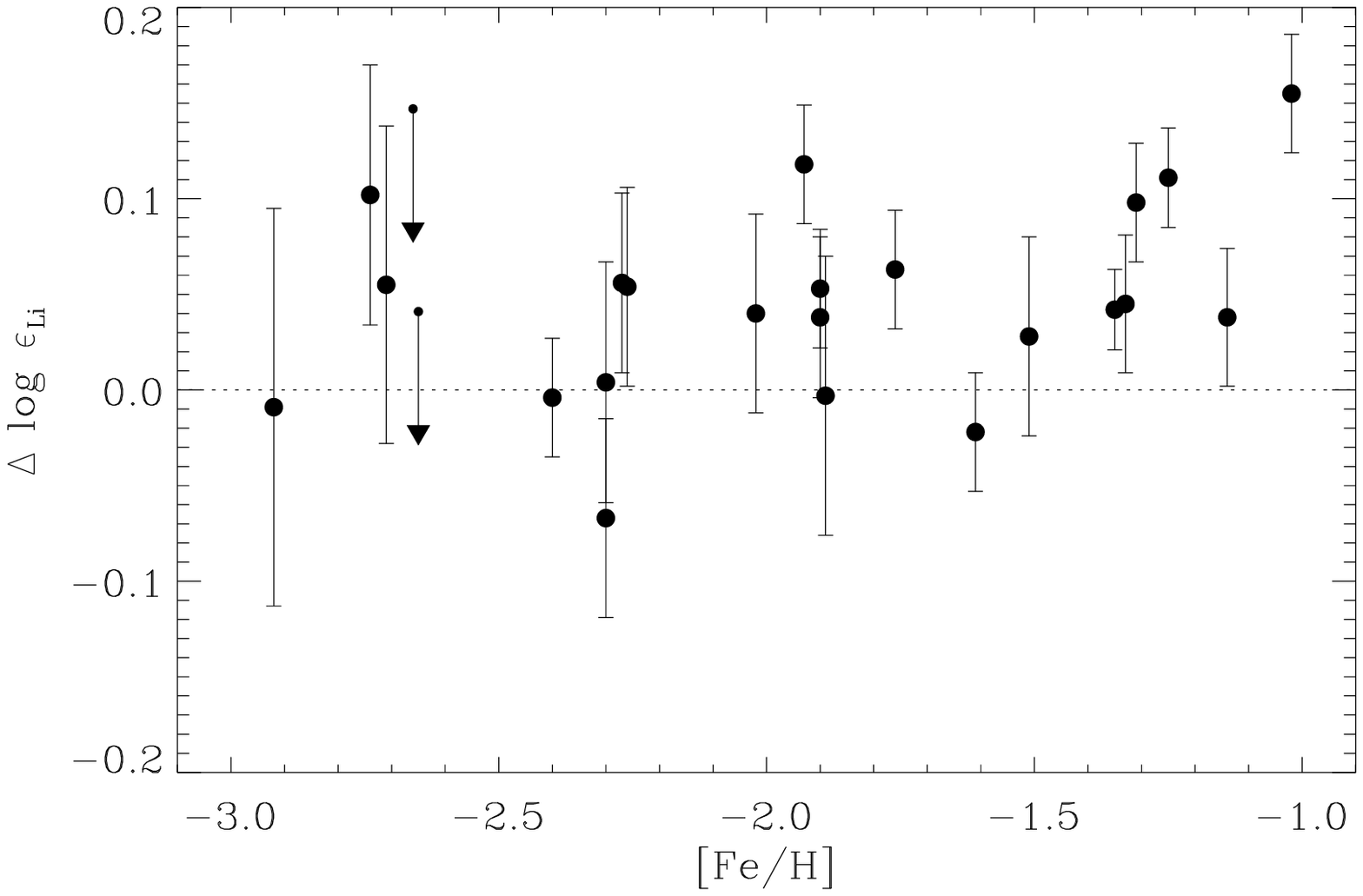}}
\resizebox{\hsize}{!}{\includegraphics{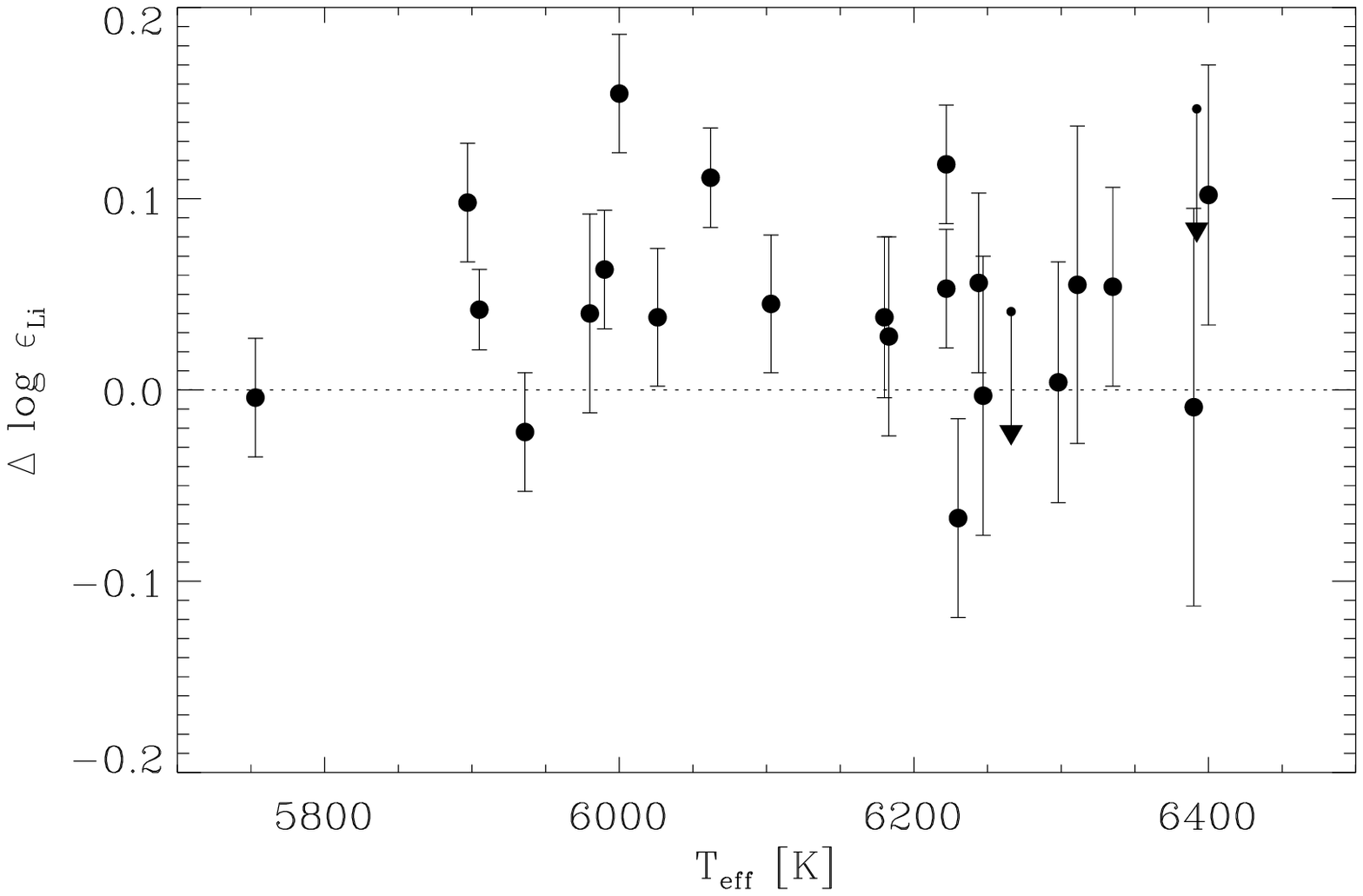}}
\caption{Differences in the 1D non-LTE abundances from the \lii\ 610.4\,nm and 670.8\,nm lines
as a function of  \teff\ ({\em upper panel}) and \feh\ ({\em lower panel}).}
\label{f:li6104vsli6708}
\end{figure}

Figure \ref{f:li6104vsli6708} compares the derived Li abundances from
the weak subordinate \lii\ 610.4\,nm line with those from the stronger \lii\ 670.8\,nm
resonance line. 
The agreement is very encouraging with a mean difference of only
$0.05\pm0.05$\,dex for the 22 stars with measured Li abundances. 
For all practical purposes, the two lines return the same
lithium abundance for a given star.
The differences show no obvious trends with metallicity or effective
temperature.
The upper limits for the Li abundances from the subordinate line in 
BD\,$+03\arcdeg 0740$ and BD\,$+09\arcdeg 2190$ are very close to the 
abundances from the resonance line. 
It is unlikely that the explanation for this minor abundance difference
can be found in erroneous $gf$-values for either of the two lines.
One (remote)
 possibility is that the adopted \teff\  values are underestimated;  a
$\approx 150$\,K higher \teff -scale would erase the mean difference.
Another explanation could be inappropriate non-LTE abundance
corrections for the 1D models. Use of 3D models with non-LTE
line formation might erase the differences
 (Asplund et al. 2003).
A study aimed at investigating the 3D non-LTE line formation
for a wider range of stellar parameters as well as revisiting
the 1D non-LTE abundance corrections in light of recent
improvements in collisional data (Belyaev \& Barklem 2003; Barklem et al. 2003)
would be worthwhile.

The \lii\ 610.4\,nm line was first detected in the halo star HD\,140283 by
Bonifacio \& Molaro (1998). 
They also found very good agreement in the abundances
inferred from the two Li lines. Likewise, Ford et al. (2002)
derived the same abundances from the 610.4 and 670.8\,nm lines 
for this star using a very high quality
spectrum obtained with Subaru/HDS (the same spectrum as used by Aoki et al. (2004)
to derive an upper limit to \liratio\ in this star as discussed above).
However, for the other seven stars for which Ford et al. (2002) 
claimed a detection of the subordinate line, the derived Li abundances are
$0.2-0.5$\,dex higher than suggested by the resonance line. 
This is in sharp contrast with our findings. 
While we are not in a position to unequivocally identify the 
reasons for the  very different conclusions of Ford et al., we suspect
that the key is the quality of the 
observations. The $S/N$ ratios of the spectra of Ford et al. are
between 200 and 300 around the 610.4\,nm line, which may still
not be sufficient to detect this very weak line in metal-poor 
turn-off stars (note that this line is about 50\% stronger in
a subgiant like HD\,140283 with \teff\,$\approx 5700$\,K than
in a turn-off star with \teff\,$\approx 6200$\,K for the same
Li abundance).
Under these circumstances it is easy to overestimate the
line strength due to noise and thus derive too high abundances. 
In comparison, our spectra have $S/N \approx 500$ or better in
the relevant wavelength region. 
Finally, we note that our resolving power is about twice as large 
as obtained by Ford et al. (110\,000 vs 50\,000).

\section{Origin and Evolution of Lithium}
\label{s:evolution}

\subsection{Lithium Abundance Trends and Scatter}
\label{s:trends}

Since our sample is relatively small and does not extend to extremely
low metallicities, our discussion of the lithium
abundances will be  relatively brief.
Furthermore, our results for the total lithium abundance are in broad
agreement with many published analyses. The principal novelty from our
study is an estimate of the isotopic ratio for stars on the Spite plateau.

Our data may be used to quantify the trend of the lithium abundance with
metallicity (here, Fe/H or O/H), to extrapolate the trend to zero
metallicity, and to determine the scatter in lithium abundance about the
mean trend.
In applying our data,
we have excluded G\,271-162 since no accurate H$\alpha$-based \teff\ could
be derived from this star (see Sec. \ref{s:teff}).
IIn addition,  the Li-rich star HD\,106038 has been 
excluded because it
has strong overabundances in Si, Ni, Y and Ba
(Nissen \& Schuster 1997), and  a very high Be abundance 
relative to other halo stars of similar \feh\ (Primas et al., in preparation).
This peculiar abundance pattern
suggests that the star has undergone a highly unusual
nucleosynthesis enrichment, possibly due to mass transfer, which
justifies its exclusion here.
Such Li-rich stars as HD\,106038 appear to be extremely rare. Perhaps the only
known other example is the subgiant BD\,$+23\arcdeg 3912$  with [Fe/H] $=-1.5$
and \logli\ $= 2.6$ (King et al. 1996). 
In addition, our sample deliberately excluded any of the rare Li-deficient
halo dwarfs. The nature of these stars is currently under debate (cf. Ryan et al.
2002; Frebel et al. 2005; Aoki et al. 2006; Charbonnel \& Primas 2005). 
Our assumption is that these stars
have also undergone a very special event, a binary merger perhaps (Ryan \& Elliott 2005).

\begin{figure}[t!]
\resizebox{\hsize}{!}{\includegraphics{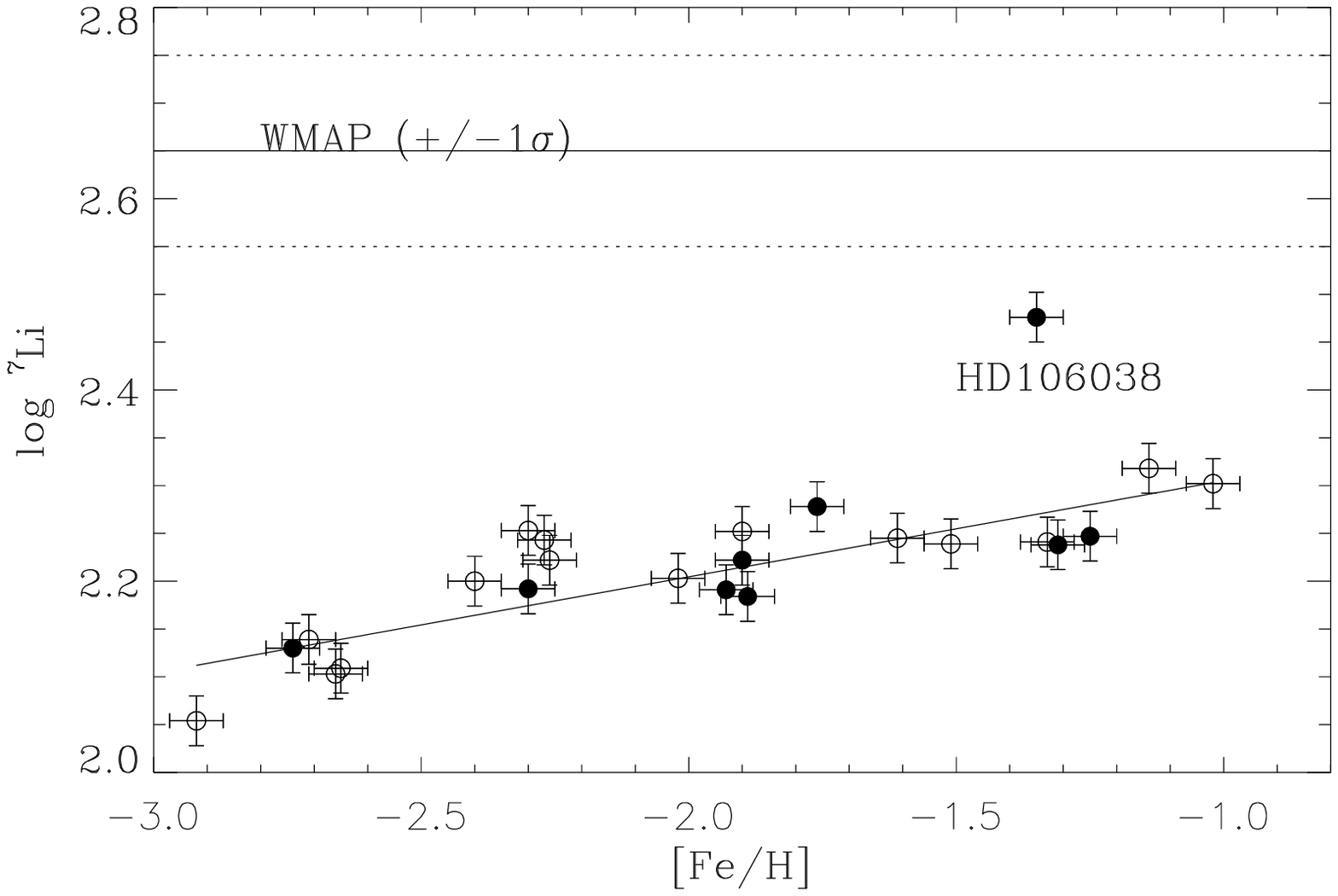}}
\resizebox{\hsize}{!}{\includegraphics{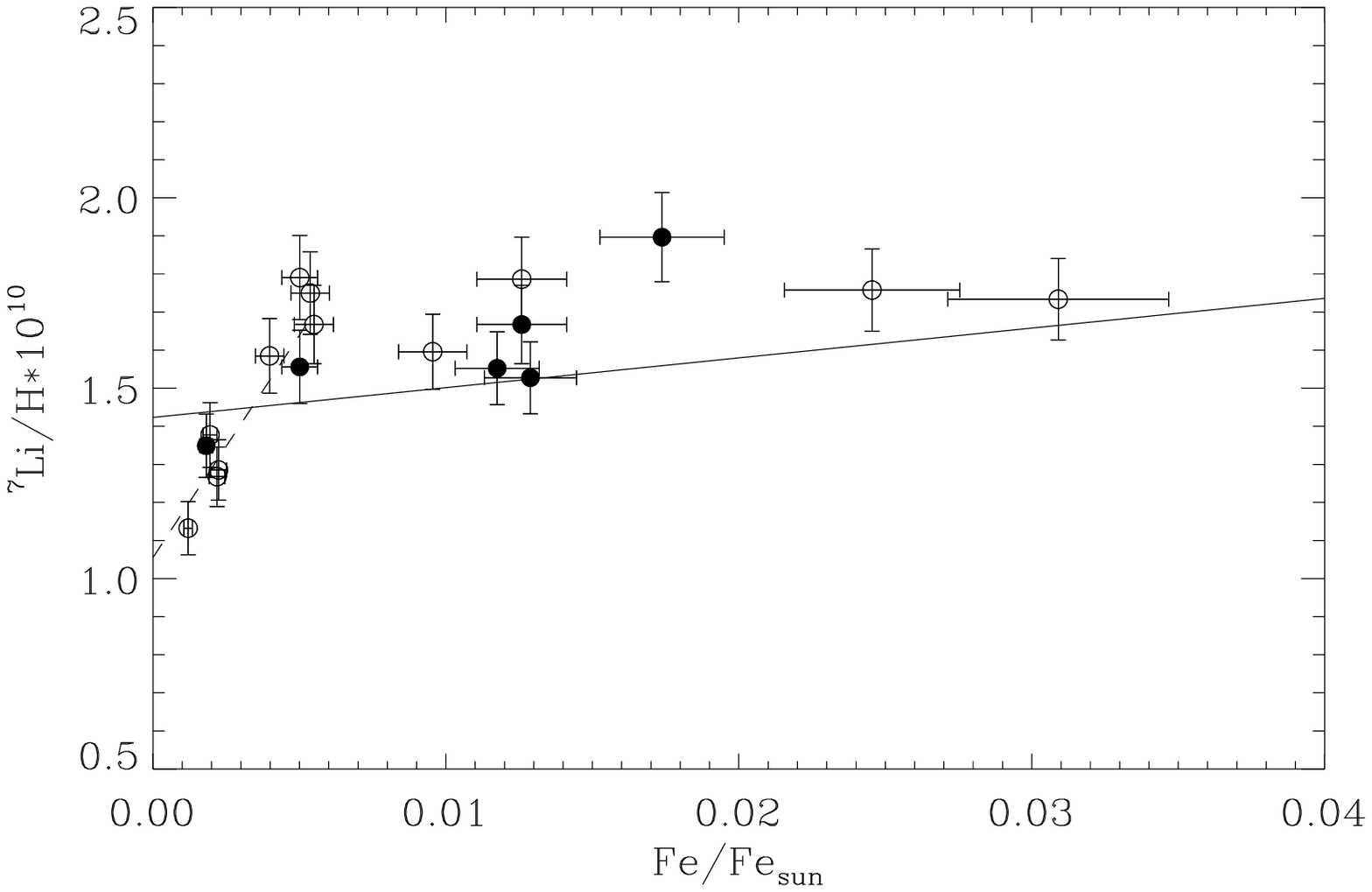}}
\caption{Non-LTE \liseven\ abundances (1D) from the \lii\ 670.8\,nm line as a function
of metallicity; as for Fig. \ref{f:li6708_summary}
solid circles here denote stars for which \lisix\ appears to have been detected
while open circles are non-detections. 
In the upper panel logarithmic abundances are shown (\logli\ vs \feh )
while in the lower panel linear abundances ($^7$Li/H vs \fefesun ) 
are used, which are more appropriate when
extrapolating to the primordial composition. The solid lines denote the best linear fits to
the data when excluding  G\,271-162 (no \ha -based \teff ) and HD\,106038 (abnormally high
Li abundance and other peculiar abundances, see text). 
In the lower panel, the dashed line corresponds to the linear fit when only including the
stars with \feh\,$<-2.2$ (G\,271-162 again excluded).
The horizontal solid line in the upper panel shows the predicted primordial Li abundance
from Big Bang nucleosynthesis and the baryon density inferred from WMAP (Spergel et al. 2003)
together with the corresponding $\pm 1\sigma$ uncertainties as dotted lines. }
\label{f:li_feh}
\end{figure}

From the abundances listed in Table \ref{t:li_abundances}, the
mean \liseven\ abundance of the remaining 22 stars
is \logliseven\,$=2.21 \pm 0.07$. 
As shown in Fig. \ref{f:li_feh}, there is, however, a pronounced
trend with metallicity, which resembles the correlation with \feh\ 
found in some previous Li studies (e.g. Thorburn 1994; Norris et al. 1994;
Ryan et al. 1996, 1999; Bonifacio et al. 2003, 2006). 
A univariate linear least-square-fit taking into account the estimated uncertainties in 
\feh\ and the Li abundances (excluding HD\,106038 and G\,271-162 for the above
reasons) implies

\begin{equation}
\log \epsilon_{\rm ^7Li} = \left( 2.409\pm0.020\right) + \left(0.103\pm0.010\right) \cdot [{\rm Fe/H}].
\end{equation}

\noindent
The observed slope 
is very similar in value to that presented by Ryan et al. (1999):  
$0.103\pm0.010$ vs $0.118\pm0.023$ for Ryan et al.
The scatter around our fit is only $\sigma_{\rm obs} = 0.033$\,dex. 
None of the remaining 22 stars is identified as an outlier by statistical tests.
Repeating the exercise using \oh\ instead of \feh\ as the independent variable yields

\begin{equation}
\log \epsilon_{\rm ^7Li} = \left( 2.372\pm0.017\right) + \left(0.115\pm0.011\right) \cdot [{\rm O/H}].
\end{equation}

\noindent
Here the non-LTE abundances from the \oi\ 777\,nm triplet have been used to estimate [O/H]. Again, the scatter
is very small: $\sigma_{\rm obs} = 0.034$\,dex.

In Sect. \ref{s:teff} we estimated the relative precision in our adopted \teff -values to be about 30\,K,
which would introduce an uncertainty in Li abundance of 0.022\,dex. 
Adding also the minor uncertainties from the $S/N$ and the other stellar parameters gives
a total  expected random error of 0.026\,dex. 
This expectation is only slightly less than the
observed scatter of $\sigma_{\rm obs} = 0.033$\,dex around the linear fit.
The indication is that the star-to-star scatter in lithium
abundances at a given [Fe/H] is very small, say 0.02\,dex or less.
(If the  temperature error were increased to 40\,K, the expected scatter in
the abundances would match the observed scatter.)
The observed scatter around the linear fit to our data is very similar
to that found by Ryan et al., who  achieved similarly high precision in \teff\ 
($\sigma_{\rm T} \approx 30$\,K)
but utilised lower quality spectra ($S/N \approx 100-150$ and $R \approx 50 000$).
They found $\sigma_{\rm obs} = 0.031$\,dex after rejecting one star, 
BD\,$+09\arcdeg$2190, as an outlier;  we note that with the equivalent width measured from our
$S/N \approx 650$ spectrum, it would no longer be
considered  an outlier. 
A different and especially a larger sample of stars may shed new light on the
intrinsic scatter in the lithium abundances.
For example, Nissen et al. (2005b) find that the otherwise almost
identical stars G\,64-12 and G\,64-37
differ in their Li abundances by about 0.16\,dex. Their reported
 observed scatter is 0.041\,dex when
including G\,64-12 but 0.035\,dex when excluding it. 
No doubt, any intrinsic scatter must be very small: $\la 0.04$\,dex.

The near-common belief has been that the primordial lithium abundance
is obtainable by extrapolation of the observed abundances to zero
metallicity. One supposes that extrapolation should be performed
using Fe/H (or an alternative such as O/H) and not the logarithmic
[Fe/H] (or [O/H]). 
As seen in Fig. \ref{f:li_feh}, on such a linear scale
the most metal-poor stars seem to indicate a steeper slope 
than the less metal-poor stars, but there is no known explanation
for the apparent change of slope close to the origin Fe/Fe$_{\rm sun}$ = 0. 
Somewhat different results are, therefore, obtained depending on which
 stars are included. 
When including only the nine stars with \feh\,$<-2.2$ (as usual, excluding G\,271-162),
 the estimated
primordial \liseven\ abundance is $^7{\rm Li/H} \approx 1.1 \cdot 10^{-10}$ or
\logli\ $\approx 2.04$.
Taking all 22 stars into account results in an increase to $^7{\rm Li/H} \approx 1.4 \cdot 10^{-10}$
of \logli\ $\approx 2.15$.
Very similar results are obtained using  extrapolation to
\oosun\,$=0$. 
If  the presence of \lisix\ at the level of that now observed 
 for LP\,815-43 is a general feature among
the most metal-poor stars and it is produced in spallation or $\alpha$-fusion reactions, the
\liseven\ abundances would have a minor non-Big Bang contribution which would reduce
the estimated primordial \liseven\ abundance further by about 0.05\,dex 
(Smith et al. 1998; Ryan et al. 1999; 2000).

Our estimates of the primordial abundance do
 not differ significantly from the majority of published determinations,
although they may be somewhat on the low side of previous estimates
(e.g., Spite \& Spite 1982; Thorburn 1994; 
Ryan et al. 1996, 1999; Bonifacio \& Molaro 1997; Bonifacio et al. 2003, 2006).
Our estimate of the primordial abundance is based on 1D model
atmospheres and includes a correction for non-LTE effects. Presently
available calculations (Asplund et al. 2003; Barklem et al. 2003;
also Sec. \ref{s:li6708_3D}) suggest that only a minor revision will be needed
when the abundance analyses are redone using
3D models  with inclusion of non-LTE effects.

It has  been claimed by Melend\'{e}z \& Ram\'{i}rez (2004) that 
their new calibration of the infrared flux method (IRFM) of effective
temperature determinations 
(Ram\'{\i}rez \& Melend\'{e}z 2005a,b) raises substantially the lithium
abundances derived from metal-poor dwarf stars.
This new \teff -scale is significantly hotter at low metallicity 
than  our \ha -based \teff -scale. It is also hotter than  
the IRFM calibration of Alonso et al. (1996), even
though  the photometric measurements required by the  IRFM  of the two studies agree well for the stars in common. 
For stars with \feh\,$<-3$, such as G\,64-12, G\,64-37 and LP\,831-70, the new temperatures
are 240, 460 and 320\,K, respectively, higher than those estimated using the Alonso et al. \vk\ calibration.
Ram\'{\i}rez \& Melend\'{e}z argue that the strikingly different behaviour for the colour calibrations
at low \feh\ is due to the shortage of low metallicity stars in Alonso et al.'s sample. 

A detailed scrutiny of  Ram\'{\i}rez \& Melend\'{e}z' \teff -scale is
beyond the scope of the present investigation. Nevertheless,
 it is pertinent to estimate the impact of the scale
on our results.
For the most metal-poor stars in our sample (\feh\,$<-2.6$), the \teff -values
of Melend\'{e}z \& Ram\'{i}rez (2004) are on average $182\pm72$\,K higher than ours
based on \ha , which increases the mean 1D non-LTE \liseven\ abundance for those particular stars
from \logliseven\,$=2.10$ to 2.23.
Independently of the new \teff\ scale,
there are reasons why 
Melend\'{e}z \& Ram\'{i}rez (2004) arrive at the significantly higher mean \liseven\ abundance of 
\logliseven\,$=2.37\pm0.06$ compared with our estimate: 
their use of convective overshoot model atmospheres
($\approx +0.1$\,dex), their neglect of non-LTE corrections ($\approx +0.02$\,dex) and their
use of Kurucz models rather than \marcs\ model atmospheres ($\approx +0.02$\,dex).
The remaining difference with our primordial \liseven\ abundance estimate stems from the extrapolation
to Fe/Fe$_{\rm sun}$ = 0 in the presence of our derived slope with metallicity.
Observational evidence does not favor the convective overshoot option as introduced in
the Kurucz (1993) model atmospheres (e.g. Castelli et al. 1997) nor does this approach find any support
in 3D hydrodynamical simulation of surface convection in metal-poor stars (Asplund et al. 1999).
Non-LTE abundance corrections, while small, should be taken into account. 
We suggest therefore that Melend\'{e}z \& Ram\'{i}rez' estimate of
the primordial \liseven\ abundance should be revised downward by at least 0.12\,dex regardless
of whether or not their \teff -scale is preferable to the one employed here. 
We note that our \ha -based temperatures are in excellent agreement on average with 
the well-established IRFM \vk\ and \by\ calibrations of Alonso et al. (1996), which seems to
give little room for such large increases in \teff\ as advocated by Melend\'{e}z \& Ram\'{i}rez (2004),
In addition, accounting for 3D effects should  push the \ha -based \teff -values downwards.
More work is  needed to  settle the \teff -scale of metal-poor stars, 

Recently Bonifacio et al. (2006) have attempted to push the abundance determinations to
yet lower \feh . In general their results agree well with ours. Their mean 1D non-LTE 
Li abundance for 17 stars (excluding a subgiant and a spectroscopic binary) is 
\logli\,$=2.17 \pm 0.09$; they argue that the expected uncertainty is about 0.08\,dex 
from the \teff -errors and  quality of the spectra (median $S/N =180$), leaving little or no room for intrinsic scatter,
in particular when considering the possibility of a trend in their Li abundances with metallicity.
Due to the small range in \feh\ of their sample, this correlation may or may not be statistically
significant by itself but when  including also our results a consistent picture emerges.
We note that their \ha-based \teff-scale should be very similar to ours but that their \feh\ are 
likely underestimated by $0.1-0.2$\,dex due to their use of \fei\ lines (Sect. \ref{s:feh});
unfortunately the two samples have no stars in common, which prevents a direct comparison
of the parameters and Li abundances.
Interestingly, the combination of the two samples suggest a steeper slope for \feh\,$\la -2.3$ than
for higher metallicity.
Whether this effect is real or spurious due to the analysis methods (1D model atmospheres, \teff-scale etc)
is an open question for the future. 
An alternative interpretation is that the intrinsic scatter increases significantly for \feh\,$\la -2.3$.

Our major new result concerns the measurements of the lithium isotopic ratio
and the apparent appearance of the analogue for \lisix\ of the Spite plateau 
(Fig. \ref{f:li6708_summary}). It is impossible to determine if the observed
plateau is tilted, as the Spite plateau is; as shown in Sect. \ref{s:depletion},
when accounting for \lisix\ depletion during the pre-main sequence evolution
the observed \lisix\ abundances apparently increase towards higher metallicity. 
There may be a real scatter
associated with this \lisix\ plateau. The average observed abundance from
eight stars with a positive \lisix\ detection (HD\,106038 is not included)
is \loglisix\,$ = 0.82$, i.e. without accounting for any \lisix\ depletion.
Taking the mean \liratio\ of all stars but HD\,106038 (unusually high Li abundance) and 
HD\,19445 (expected to have no \lisix ) together with
the mean \liseven\ abundance of the same stars would instead imply \loglisix\,$=0.54$.
To both of these estimates should be added the expected \lisix\ depletion which
is $\ge 0.3$\,dex during the pre-main sequence during which \liseven\ remains
largely intact, as discussed in Sect. \ref{s:depletion}.
If, in addition, some \liseven\ has been destroyed as well this would increase 
the \lisix\ depletion further.

\subsection{The Lithium Problems}
\label{s:liproblems}

Since Spite \& Spite's (1982) seminal paper, the lithium abundance of
metal-poor warm dwarfs has been widely equated with the abundance
provided by  the Big Bang. The predicted abundance from a standard Big
Bang is dependent on
one free parameter, namely, the baryon density. Now, the free
parameter is determined independently of any inferred primordial abundances
from the WMAP-observations of the anisotropies of
the cosmic microwave background: $^7{\rm Li/H} = 4.4\pm1.0 \cdot 10^{-10}$
or \logliseven\,$=2.65\pm0.1$
(Coc et al. 2004; Cuoco et al. 2004; Cyburt 2004).
As different calculations yield slightly different results even when adopting exactly
the same baryon density, we have  taken an average of the cited results. 
These same calculations show that very little \lisix\ is
synthesised, say \liratio\ $\sim 10^{-5}$.

While the predictions from standard Big Bang nucleosynthesis are in good
agreement with the mean deuterium abundance from the six available damped Ly-$\alpha$
systems (e.g. Steigman 2005 and references therein), 
they are in substantial conflict with the observations for Li.
The \liseven\ problem is now well established: the
inferred primordial \liseven\ abundance is about 0.5 dex less than the
prediction. As will become clear from the discussion in Sect. \ref{s:spallation},
our survey has in addition identified a \lisix\ problem: the observed high 
\lisix\ abundance at very low \feh\ can not be explained by neither
standard Big Bang nucleosynthesis nor apparently by Galactic cosmic
ray spallation and $\alpha$-fusion reactions.  
Possible resolutions of these problems were 
reviewed by Lambert (2004). 
It should also be noted that the problem of reconciling the inferred and predicted \liseven\
abundance is not entirely decoupled from the goal of achieving 
an understanding of the synthesis of \lisix . Possible modes of
\lisix\ synthesis imply simultaneous synthesis of \liseven\ but
this aggravates the severity of the \liseven\ problem.
In turn, the presence of \lisix\ constrains the possible explanations invoked
for the \liseven\ problem. 

The solution to the \liseven\ problem may lie in one or more of the following
areas:

 \smallskip
\noindent
{\em Systematic errors in the abundance analysis:}
The introduction of an abundance analysis 
based on 3D model atmospheres and non-LTE line formation
for the lithium resonance and subordinate lines has not resulted
in a significant upward revision of the lithium abundance
obtained from 1D atmospheres and LTE or non-LTE line
formation. The derived lithium abundance is dependent on the adopted
\teff -scale but this  cannot  accommodate
the higher temperatures necessary to bring inferred and predicted
abundance into agreement (see Sec. \ref{s:teff} and \ref{s:trends}).
Although the last word has not been said on the representation of the
physics of the stellar atmospheres and the formation of the lithium
lines, it would be extremely surprising were systematic errors in these
areas be shown to resolve the lithium problem.

\smallskip
 \noindent
 {\em Present-day abundances do not reflect the primordial Li abundance:} 
 Dilution and destruction of lithium in metal-poor stars are discussed in the
following section (Sec. \ref{s:depletion}). 

\smallskip
\noindent
{\em Erroneous nuclear reaction rates:}
Nuclear reactions included in the  customary networks have been
well studied and uncertainties in the reaction rates at the relevant
temperatures for Big Bang nucleosynthesis cannot be held responsible for
the lithium problem. 
Coc et al.'s (2004) suggestion that addition to the network
of a previously neglected reaction
 -- $^7$Be($^2$D, p)2 $^4$He --  might possibly reduce the predicted
\liseven\ abundance is not substantiated by a cross-section measurement
(Angulo et al. 2005). ($^7$Be decays to $^7$Li following 
 Big Bang nucleosynthesis.)
 Based on the measured solar neutrino flux, Cyburt et al. (2004) have been
 able to show that employed nuclear data for the $^3$He($\alpha,\gamma$)$^7$Be 
 reaction can not be in error by a large factor as to modify the predicted
 $^7$Li abundances significantly. 

\smallskip
\noindent
{\em Non-standard Big Bang nucleosynthesis:}
Comments on the synthesis of the lithium isotopes
by  non-standard Big Bang models are included in our discussion of
the synthesis of \lisix\ in Sec. \ref{s:spallation}.

\subsection{Lithium Destruction and Dilution in Halo Stars}
\label{s:depletion}

Identification of the primordial lithium abundance from the Spite
plateau with the abundance provided by the Big Bang rests on
the  assumption that the lithium abundance of
the atmosphere of a star on the plateau
has been unchanged since the star was formed from gas having the
lithium abundance provided by the Big Bang. Perhaps, the most
promising solution to the lithium problem represents a
failure of this assumption. Solutions have consequences for the
correction of observed \lisix\ abundances to the \lisix\ abundance
of the gas from which a star formed.

In standard stellar models, the expected depletion of \liseven\ for metal-poor
turn-off stars is negligible, i.e., less than 0.02\,dex (Deliyannis et al. 1990; 
Pinsonneault et al. 1992). A larger depletion, $\ge 0.3$\,dex, is predicted for
\lisix\ with most of this occuring during the pre-main
sequence phase when \lisix\ is destroyed by protons at the base of the
star's deep convective envelope (e.g. Richard et al. 2005). 
The cross-section for destruction of \lisix\ is about 60 times
greater than for \liseven .
Destruction of \lisix\ is predicted to
increase sharply with decreasing stellar mass for main sequence
stars below the turn-off. Evidently, depletion (destruction) of \liseven\
is  according to standard models
not the answer to the lithium problem but is a factor to
be considered when we turn to examine synthesis of \lisix .

The search for a solution to the lithium problem then turns to investigation of
 non-standard
models of stellar evolution (Lambert 2004).
 Physical processes neglected by standard models but
 incorporated into non-standard
models 
in connection with halo stars in recent years include
rotationally-induced mixing (e.g. Vauclair 1988; Chaboyer \& Demarque 1994; Pinsonneault et al. 1999, 2002; Piau 2005), 
diffusion\footnote{The term diffusion normally includes the effects of gravitational settling, thermal diffusion and
radiative acceleration.} (e.g. Michaud et al. 1984; Salaris \& Weiss 2001; Richard et al. 2002; 2005) 
and internal gravity waves 
(e.g. Talon \& Charbonnel 2004; Charbonnel \& Talon 2005), 
sometimes working in tandem with other mechanisms
such as mass loss (e.g. Vauclair \& Charbonnel 1995) and turbulent mixing.
In most cases, the predictive power of these processes is restricted by their
need to introduce one or more free parameters.

We comment briefly on two non-standard models drawn from the suite of proposals: 
rotationally-induced mixing and diffusive separation. 
An obvious requirement of 
successful non-standard  models is that they provide a depletion of surface lithium
by about 0.5\,dex over the 10-13 Gyr life of the observed stars
 (we take it as given that the stars began life with
the WMAP-inferred predicted lithium abundance.) Surely, a more demanding requirement
(and a certain clue to the dominant process)
is that this depletion be achieved uniformly over the observed sample
of stars spanning a range in mass, metallicity, age, and rotational angular momentum:
the depletion sought is about 0.5\,dex with a star-to-star scatter of not more
than about 0.03\,dex.
Our present interest is in identifying non-standard models that may solve the
lithium problem and in obtaining an estimate of the correction to be applied to a
\lisix\ abundance to obtain the star's original \lisix\ abundance.  

A description of lithium destruction from rotationally-induced mixing depends on
processes and initial conditions that are as yet poorly known, including the
initial distribution of angular momentum in the star, the transport of angular momentum
through the star and its loss from the surface, and attendant mixing of the
interior material. To make progress requires  calibrations  using observations
of rotational velocities and lithium abundances for stars, e.g., the Sun
and stars belonging to open clusters. Calibrations have not involved metal-poor dwarfs.
A probable result of rotational effects on lithium depletion is a spread in
the final lithium abundance  for a sample of stars differing only in
their rotational characteristics. One expects lithium depletion to be most severe
for the stars initially rotating rapidly and least severe to vanishingly
small for the slowest rotators.  Then, the observed (small) spread in
lithium abundances on the Spite plateau  is, in principle, an indicator of
rotationally-induced depletion of lithium.
Conversion of the observed spread to an indicator of depletion depends on
the description of rotational-induced mixing and the attendant calibrations.

The recipe and calibrations (Sun, open clusters) invoked by Pinsonneault et al. (1999)
show  that the depletion of  \liseven\ ($D_7$ in dex)
was related to the dispersion ($\sigma$ in dex) where
$\sigma/D_7 \simeq 0.4$ and, hence, the required depletion by 0.5 implies
$\sigma$ = 0.2, a value far in excess of the observed upper limit.
The depletions of \liseven\ and \lisix\ were predicted to be correlated:
$D_7/D_6 \simeq 0.4$ yielding $D_6 \approx 1.2$ dex. Given that the observed
\loglisix\ $\simeq 0.8$ for the stars with detected \lisix , the initial
abundance would have been \loglisix\ $\simeq 2.0$, the value that has
historically been identified with the Spite plateau for \liseven\, and,
more importantly, a value that implies most or all of the \liseven\
came from synthesis by cosmic rays and not from the Big Bang!
Pinsonneault et al. (2002) argue that the Li data of Ryan et al. (1999), who
like us measured $\sigma_{\rm obs} = 0.03$\,dex, imply a most-likely  \liseven\
depletion of $0.13$\,dex although destruction factors as high as 0.3\,dex 
were not ruled out. The corresponding \lisix\ 
destruction is then $\ge 0.3$\,dex, and
\loglisix\ $\ge 1.1$ for our stars with \lisix\ detections.
In summary, rotationally-induced mixing as presented in the models
of Pinsonneault et al. appears not to be the solution to
the \liseven\ problem. 

Stars on the Spite plateau have a thin convective envelope of uniform
composition. Atomic diffusion occurs in the radiative zone below the
convective envelope. The base of the latter mixes with and attains the
composition of the top of the radiative zone. Diffusion is a slow process
but is predicted to be effective in the halo dwarfs because of their great
age and their thin convective envelopes.
 The diffusive velocity decreases with increasing density. Since the
stars at the turn-off have the thinnest envelopes, the effects of diffusion on
surface composition are likely to be more severe for these stars than for
stars further down the main sequence. Lithium is predicted to diffuse inwards in the
radiative zone and the surface abundance to decrease. Inward diffusion of lithium
can result in its destruction by protons.

Model calculations  including diffusion suggest 
that the lithium abundance for stars on the Spite plateau can 
be reduced to the observed level from a starting value corresponding to the
WMAP-based prediction (Salaris \& Weiss 2001; Richard et al. 2002; Richard et al. 2005). 
The earliest calculations (Michaud et al. 1984), 
as well as recent ones (Salaris \& Weiss 2001), predicted
lower lithium abundances for stars at the hot end of the plateau, in
contradiction to the observations. Richard and colleagues
invoke `turbulent diffusion' with a recipe that restores uniform lithium
abundance to the plateau; this extra turbulence could for example 
be induced by stellar rotation, which would limit the diffusion but not
necessarily mix Li to sufficiently high temperatures where it is burnt by
nuclear reactions as in the models by Pinsonneault et al. (1999, 2002)
and others described above. 
Richard et al. (2002, 2005 and private communication) have presented detailed calculations
of the combined effects of diffusion and a parameterized recipe for
turbulent diffusion. 
The predicted \liseven\ and \lisix\ depletion factors for our program stars
based on their calculations are given in Table \ref{t:depletion}.
We note that the pre-main sequence depletion factors are likely more
uncertain than the main sequence and post-main sequence counterparts
(Proffitt \& Michaud 1989).
Furthermore, we note that the predictions are presently available
only for a very limited number of metallicities, and thus many
depletion estimates in Table \ref{t:depletion} are based on 
extrapolations, which likely introduce further uncertainties, in particular
towards higher \feh . For the disk stars HD\,68284 and HD\,130551, in which
\lisix\ has been detected (Nissen et al. 1999), we estimate the
pre-main sequence \lisix\ depletion to be about 1 dex; these two stars
have nearly identical mass and \feh , which explains their similarity 
in depletion in spite of their \teff\ differences.

The following results are pertinent to our discussion. One, there is
a recipe for the turbulent diffusion (the so-called T6.25 model) that
reduces the Big Bang lithium abundance by $0.4-0.5$\,dex for
turn-off stars to the observed value and
preserves a largely flat plateau (Richard et al. 2005). Two,  simulations suggest that stars of different
ages and metallicities may have lithium quite uniformly depleted, but a
quantitative comparison with the small scatter reported by us and
others was not made. Third, \lisix\ is partly destroyed during the pre-main sequence 
evolution even without any accompanying \liseven\ destruction; 
due to the lower \teff\ in general for our higher \feh\ stars, the observed \lisix\ plateau
becomes tilted when accounting for this depletion (Fig. \ref{f:li6vsfeh}).
Fourth, any substantial \liseven\ destruction is accompanied by an even
larger \lisix\ depletion; for example, the T6.25 model which provides a 0.4 dex depletion
of \liseven\ predicts 1.6 dex depletion for \lisix\ in LP\,815-43, in addition to
about $0.3$\,dex \lisix\ destruction during 
the pre-main sequence phase. It then follows that the stars with
\lisix\ detections began with \logliseven\ $= 2.6$ and \loglisix\ $ \approx 2.6$ too
-- not a combination that can be said to reconcile observations with the Big Bang
predictions. The choice of a model that provides a smaller \liseven\ depletion 
also gives a smaller \lisix\ depletion; for example, the T6.09 model reduces
\liseven\ by about 0.25 dex (lithium problem is lessened but not eliminated)
and \lisix\ by about 0.6 dex including the pre-main sequence destruction
(the initial  abundance was \loglisix\ $\approx 1.4$). 
The question is whether alternative recipes for turbulent diffusion can
resolve the \liseven\ problem without
requiring very high initial \lisix\ abundances.

Although no quantitative predictions yet exist for Population II stars, a further
promising candidate for Li depletion is internal gravity waves, which
lead to angular momentum transport and thus mixing
(e.g. Garc\'{\i}a Lop\'{e}z \& Spruit 1991; Talon \& Charbonnel 2004; 
Charbonnel \& Talon 2005).
Recent calculations including these effects successfully reproduce both
the quasi-flat solar rotation profile and the degree of Li depletion in disk open clusters
as a function of age (Charbonnel \& Talon 2005). We are eagerly awaiting
similar modelling for metal-poor stars with predictions for the \liseven\ and \lisix\
depletions, in particular whether large Li depletions can be achieved while still
retaining the small observed abundance scatter. 

Our commentary on rotationally-induced mixing and diffusion shows that
plausible (rotationally-induced mixing) and seemingly inevitable (diffusion)
processes may solve the \liseven\ problem, but the observers' challenge
to fit the Spite plateau's shallow slope with respect to \teff\ and \feh\
and its smoothness  are as yet unmet.  These potential solutions to the
\liseven\ problem indicate that the observed \lisix\ abundance is almost
certainly a lower bound to the initial \lisix\ abundance. Some solutions
lead to very high initial \lisix\ abundances that, if correct, imply that
our understanding of the pre-galactic and Big Bang nucleosynthesis is
poor indeed.

\subsection{Lithium Production in the Early Universe}
\label{s:spallation}

\begin{figure}
\resizebox{\hsize}{!}{\includegraphics{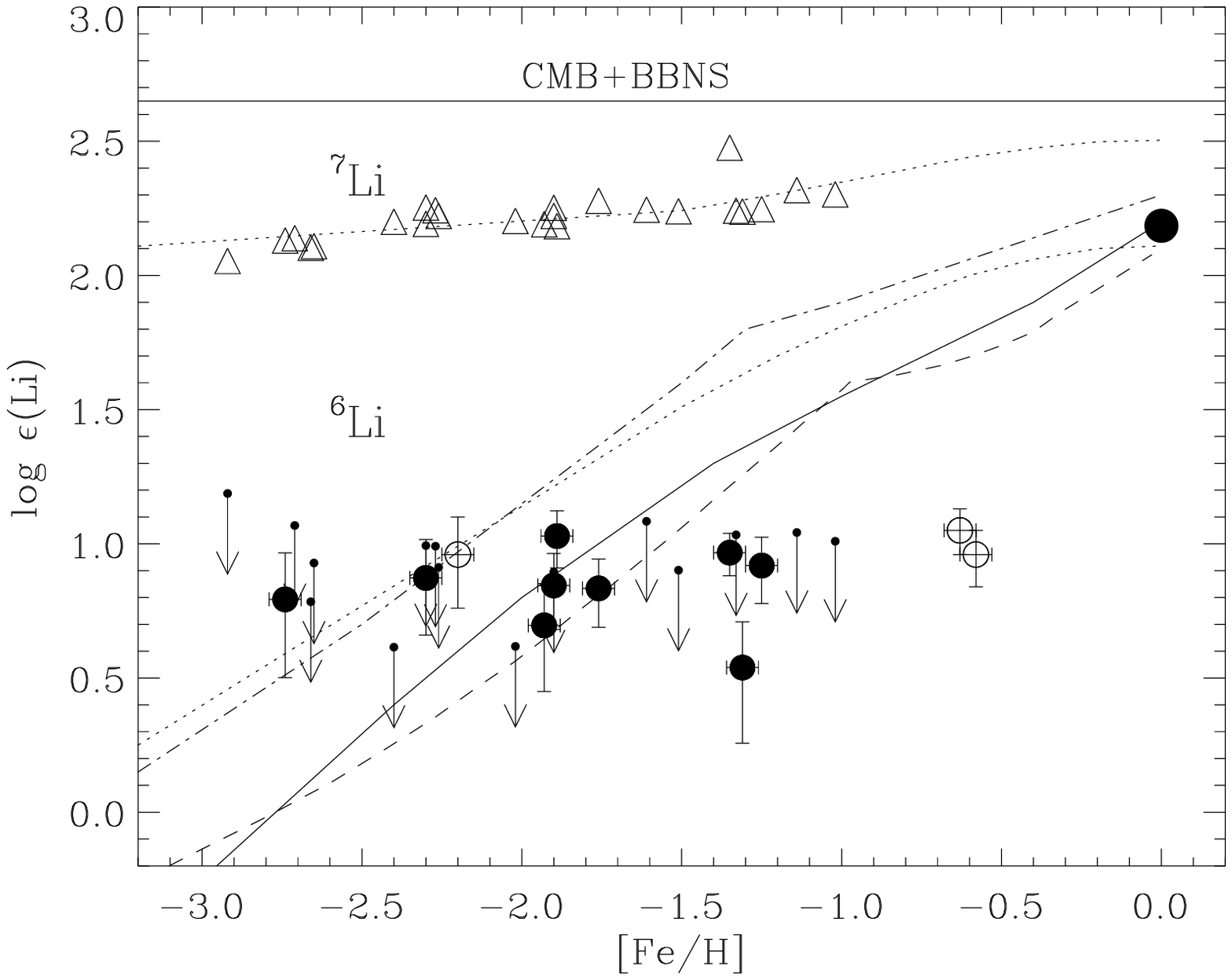}}
\resizebox{\hsize}{!}{\includegraphics{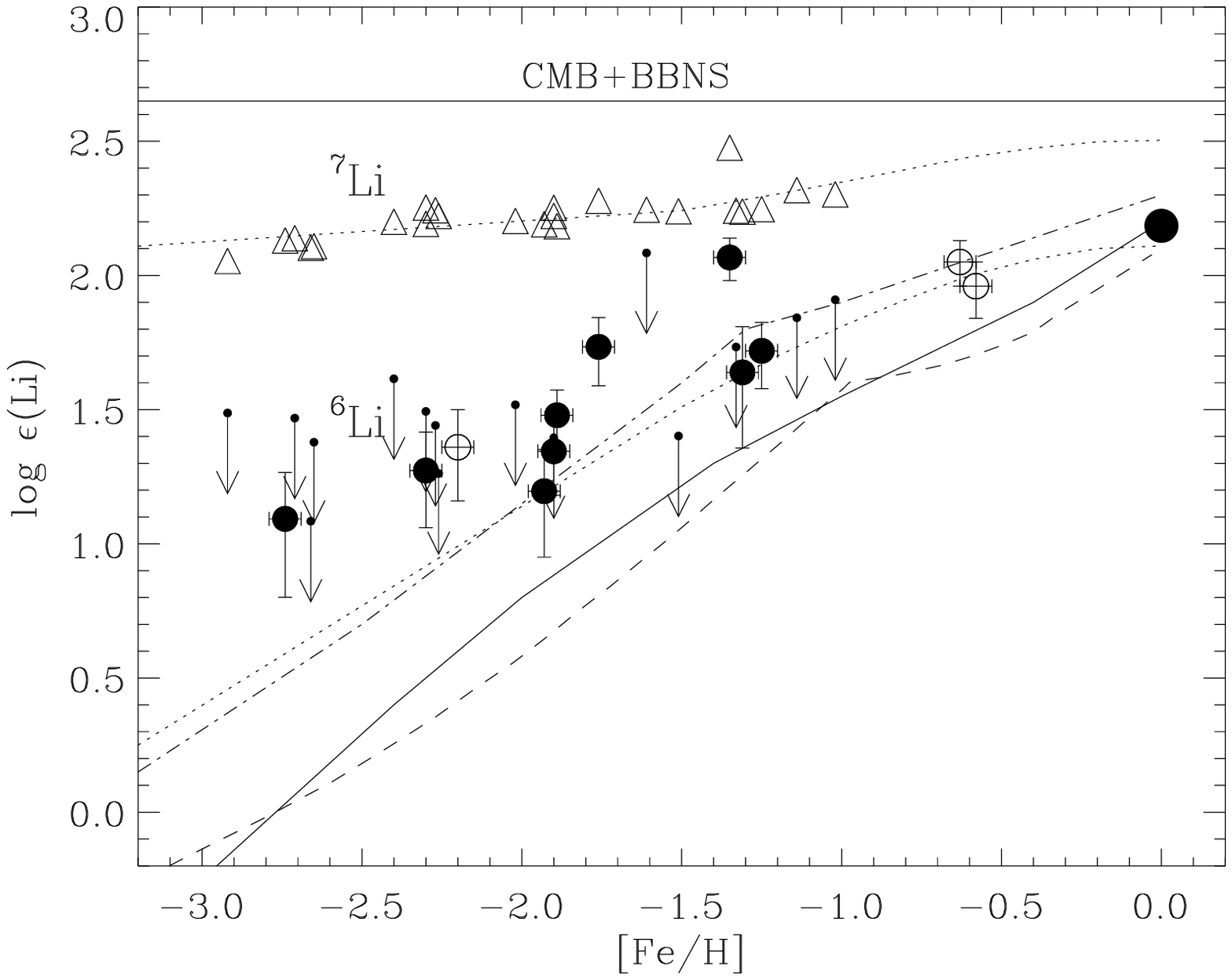}}
\caption{{\em Upper panel:} 
Observed logarithmic abundances of \liseven\ (open triangles) and \lisix\ (solid circles) 
as a function of \feh\ for our program stars; $3\sigma$ upper limits to the \lisix\ abundances are denoted 
with arrows. Also shown as open circles are the \lisix\ detections in the halo turn-off star 
HD\,84937 (Smith et al. 1993, 1998; Cayrel et al. 1999) and in the two Galactic disk stars
HD\,68284 and HD\,130551 (Nissen et al. 1999). The large circle correspond to the 
solar system meteoritic \lisix\ abundance (Asplund et al. 2005), while the horizontal solid line
is the predicted \liseven\ abundance from Big Bang nucleosynthesis and the baryon density
as determined by WMAP (Spergel et al. 2003; Coc et al. 2004; Cuoco et al. 2004; Cyburt 2004).
The (lower) solid, dashed, (lower) dotted and dash-dotted lines correspond 
to the models for cosmic ray \lisix\ production
by Prantzos (2006), Ramaty et al. (2000), Fields \& Olive (1999) and Vangioni-Flam et al. (2000), respectively;
see text for details. The higher dotted line shows the \liseven\ abundance assuming the
Fields \& Olive model with a \liratioinverse\ production ratio of 1:1.5 and a primordial abundance
of \logliseven\,$=2.1$.
{\em Lower panel:} Same as in the upper panel but taking into account the predicted
pre-main sequence \lisix\ depletion from Richard et al. (2002, 2005; and 2005, private
communication) to the observed \lisix\ abundances; these are the minimum
values corresponding to negligible \liseven\ depletion (see Sect. \ref{s:depletion} for details).}
\label{f:li6vsfeh}
\end{figure}

The observed \lisix\ abundance is several orders of magnitude larger than
that predicted from the standard Big Bang. On the assumption that the
standard Big Bang sequence is the correct representation of the
primordial fireball, synthesis of \lisix\ is attributed to collisions in 
the intergalactic or interstellar medium between high-energy particles
(cosmic rays) and ambient nuclei.
The two leading processes are
the $\alpha +\alpha$ fusion reactions and spallation reactions involving
protons (also, $\alpha$s) and $^{16}$O (also $^{12}$C and $^{14}$N)
nuclei. Both processes also produce \liseven\ with \liratioinverse\,$\sim 1$ to 2
(Mercer et al. 2001).
Only the spallation reactions produce $^9$Be, $^{10}$B, and $^{11}$B.
In the classical, or direct, Galactic cosmic ray spallation scenario as formulated by 
Reeves et al. (1970) and Meneguzzi et al. (1971), the accelerated particles are 
the protons etc hitting ambient interstellar
CNO nuclei. As a consequence, this reaction is secondary in nature: \beh\,$\propto$\oh\,$^2$.
In inverse spallation the collisions instead occur through fast CNO nuclei and interstellar protons
and $\alpha$-particles. Since the CNO nuclei are both produced and accelerated by 
supernovae, the resulting spallation yields will be essentially independent of the metallicity,
a primary process: \beh\,$\propto$\oh .
One possible way of achieving this is through superbubbles (Higdon et al. 1998; Parizot \& Drury 1999
but see Prantzos 2005, 2006 for a critique of this scenario): 
a sequence of supernovae in rapid concession in a cluster or OB association accelerates 
nuclear-process enriched material. 
Establishing the behaviour of [O/Fe] with metallicity is clearly important for estimating
the spallation production rate but unfortunately no unanimous verdict have appeared to date in this regard
(e.g. Israelian et al. 1998, 2001; Boesgaard et al. 1999; Asplund \& Garc\'{\i}a P\'{e}rez 2001;
Nissen et al. 2002; Fulbright \& Johnson 2003; Garc\'{\i}a P\'{e}rez et al. 2006).

Since spallation reactions are thought to be the sole process by
which Be is synthesised and the Be abundance is obtainable for
stars on the Spite plateau (Boesgaard et al. 1999; Primas et al., in preparation),
 the efficiency of a proposed
scenario invoking spallation may be rather well calibrated
using observed Be abundances. The simultaneous production from
the $\alpha + \alpha$ fusion reactions is not simply calibratable.
The observed \lisixbe\ ratio shows that
the \lisix\ must be produced primarily from the fusion
reactions at low metallicities with Be necessarily from spallation; 
the meteoritic \lisixbe\ ratio ($\approx 6$) 
is a representative measure of the production
ratios for spallation but the ratio is $\approx 40$ for HD\,84937
(Smith et al. 1998) and may reach $\approx 150$ for LP\,815-43
(Primas et al., in preparation). The initial \lisixbe\ ratios were likely
even greater due to \lisix\ depletion, underscoring the need for
$\alpha + \alpha$ fusion reactions.

A large number of Galactic chemical evolution models including spallation and $\alpha$-fusion reactions
have been presented in recent years and compared with observed Li, Be and B 
isotopic abundances (e.g. Yoshii et al. 1997; Fields \& Olive 1999; Parizot 2000;
Vangioni-Flam et al. 2000; Ramaty et al. 2000; Suzuki \& Yoshii 2001; Alibes et al. 2002; Prantzos 2005, 2006).
Four representative predictions are presented in Fig. \ref{f:li6vsfeh}: 
the model by Fields \& Olive (1999) is based on direct cosmic ray spallation but with a very steep [O/Fe] trend
(Israelian et al. 1998; Boesgaard et al. 1999); 
the model by Ramaty et al. (2000) is for inverse spallation with special consideration of the energetics
of the light element production; the Vangioni-Flam et al. (2000) model is for inverse spallation 
in superbubbles with an
assumed low energy ($\la 30$\,MeV/nucleon) cosmic ray component (Meneguzzi \& Reeves 1975);
and the calculations by Prantzos (2006) assume that the cosmic ray composition does not vary with time
and that the light isotopes are produced as primaries, i.e. like in the inverse spallation models. 
These calculations fit about equally well the observed run of the Be
abundance with the Fe abundance, although Ramaty et al. (2000) argue that the Fields \& Olive
Be production is significantly overestimated through the adoption of unrealistic Fe yields . 
A fair fit to the observed \lisix\ abundance in 
HD\,84937 (previously the sole data point for this isotope) when ignoring the inevitable
pre-main sequence destruction is  achieved for the Fields \& Olive 
and Vangioni-Flam et al. models but
it requires some assumptions to maximise \lisix\ production.
There are other problems and difficulties with the calculations: for example,
the efficiency assumed for cosmic ray acceleration may be unrealistic
given the available kinetic energy from supernovae, which 
may rule out the low energy cosmic ray scenario
(Ramaty et al. 2000; Suzuki \& Yoshii 2001; Prantzos 2005, 2006). 
Furthermore, several of the spallation models presented here invoke
a [O/Fe] trend that is significantly higher than observed by Nissen et al. (2002) and us (Sect. \ref{s:oh});
note that in addition our adopted solar O abundance is $0.1-0.2$\,dex lower than assumed 
by Fields \& Olive and Ramaty et al. As a consequence, their predicted \lisix\ abundances should
be shifted downwards by $\approx 0.3$\,dex for [Fe/H]\,$\le-2$. 
Finally, the subsequent downward revision by a factor of two of the measured cross-section for
$^{6}$Li production by $\alpha$-fusion (Mercer et al. 2001) will further push the predictions to lower yields 
for all but the Prantzos model.

Now that additional \lisix\ detections are available with the high \lisix\
abundance in LP\,815-43 in particular,
it is clear that the model calculations do not reproduce the \lisix\
measurements and the apparent observed \lisix\ plateau.
Of the here presented models, the one by Prantzos (2006, shown as a solid line in Fig. \ref{f:li6vsfeh})
is likely the most realistic as it requires the least number of assumptions in terms of maximising
the production efficiency and is based on the most recent reaction cross-sections. 
Still we note that it makes the as yet unproven assumption that the cosmic ray composition stays 
constant with time and thus may well overestimate the \lisix\ production.
The discrepancy between models and observations at low \feh\ is further aggravated   
when accounting for the pre-main sequence \lisix\ depletion of $\ge 0.3$\,dex
(Table \ref{t:depletion}),
as seen in Fig. \ref{f:li6vsfeh}; invoking any significant amount of \liseven\ depletion would
make the situation worse still (Sect. \ref{s:depletion}). 
The inclusion of \lisix\ pre-main sequence depletion tilts the \lisix\ observations to follow
roughly the metallicity-dependence of the models, albeit at a higher \lisix\ abundance; 
we recall here that the pre-main sequence depletion calculations are rather uncertain and
that most of the \lisix\ depletion may in fact take place during the main sequence. 
At face value, it seems like a remarkable coincidence that the \lisix\ production from
cosmic ray spallation and Li-depletion can result in a nearly flat \lisix\ abundance
plateau as observed. However, we do note that we are only able to detect \liratio\ ratios
of $\ga 0.02$ for observational reasons and thus would not be able to identify the
stars in which the depletion is much larger than the corresponding production.

A means to boost the light element production over direct and indirect cosmic ray 
spallation in the interstellar medium has recently been put forward by Nakamura et al. (2006, 
see also Fields et al. 1996 and Nakamura \& Shigeyama 2004). 
They speculate that the fast CNO-rich ejecta of very energetic ($\approx 10^{52}~{\rm erg}=10^{45}~{\rm J}$)
supernovae having lost all but a small fraction of their H- and He-rich envelopes
can interact directly with the circumstellar material without shock
acceleration of cosmic rays. Such SNe Ic events may produce significant amounts of primary \lisix\ and \benine\
under suitable conditions. 
Nakamura et al. (2006) have identified a parameter space that seem to explain the high 
\lisix /O ratio we observe in LP\,815-43 but it requires some fine-tuning in regards 
the explosion energy and the amount of available He in the ejecta. It remains to be seen
whether real supernovae fulfill these requirements. 

An additional and alternative scenario places lithium synthesis by
$\alpha+\alpha$ fusion reactions during the formation of
the Galaxy within the framework
of hierarchical structure formation (Suzuki \& Inoue 2002, 2004). Infalling and merging 
subgalactic clumps would have produced shocks that could have accelerated 
the $\alpha$-particles at early epochs. 
Due to our incomplete understanding of the details of galaxy formation, models
for the accompanying \lisix\ production by necessity include several free parameters. 
To achieve the observed high \lisix\ abundances at \feh\,$\le -2.3$ requires 
strong structure formation shocks to have been in place very early on ($\approx 0.1$\,Gyr after
Big Bang) but vanished relatively shortly thereafter ($\approx 0.1$\,Gyr). 
An argument against this scenario is that  mergers in $\Lambda$CDM  continue to
the present day with only a relatively small dark matter halo present at early times, which
would appear to make it difficult for the necessary strong shocks to have formed
during the required time interval (Prantzos 2006). 
Rollinde et al (2005, see also Montmerle 1977) have presented a similar scenario 
of pre-Galactic production of \lisix\ (and \liseven )
due to an ad-hoc early burst of cosmological cosmic rays, possibly due
to Population III stars. The energetics of such a production were not investigated.
Prantzos (2006) also speculate that accretion to a supermassive black hole such as
the one believed to exist currently in the Galactic center could generate
sufficient energy to accelerate the necessary cosmic rays. This scenario requires some extreme 
assumptions about the time of its formation and its efficiency in converting accretion energy
to particle acceleration to produce sufficient \lisix\ at \feh\,$\approx -2.7$. 

The final potential site for \lisix\ synthesis is the Big Bang.
In a standard Big Bang  much too little \lisix\ is produced
to explain our observations. 
An extension of the standard model for particle physics such as supersymmetry predicts the 
existence of various exotic particles, including the gravitino, axion and the neutralino;
the first two of these particles are predicted to be unstable but the neutralino is likely stable. 
The decay and the annihilation
during or shortly after the era of Big Bang nucleosynthesis
can alter the resulting light element abundances, provided the masses and life-times of
these putative particles are right (e.g. Dimopoulos et al. 1988; Jedamzik 2000, 2004a,b, 2005;
Kawasaki et al. 2004; Ichikawa et al. 2004).
The annihilation of the neutralino can release sufficient energy to produce \lisix\ by
non-thermal reactions like $^3$H($\alpha$,n)\lisix\ and $^3$He($\alpha$,p)\lisix\ (Jedamzik 2004a).
The amount of \lisix\ produced depends on the mass of the neutralino and the exact annihilation channels but
can reach the levels observed in some of our very metal-poor stars; this scenario does not, however, resolve
the \liseven\ dilemma described above. 
Another option to produce \lisix\ is through the decay of particles like the gravitino and axion.
While the electromagnetic decay routes of such particles result in
$^3$He/D ratios inconsistent with observations (Ellis et al. 2005), the injection of energetic nucleons
through the hadronic decay about $10^3$\,s after Big Bang can lead to substantial 
\lisix\ production without spoiling the agreement with D and the He isotopes.
Indeed, for the right combination of particle properties, a simultaneous 
production of \lisix\ and destruction of \liseven\ appears achievable
and may explain both the observed \lisix\ plateau and the low \liseven\
abundances in comparison with standard Big Bang nucleosynthesis 
(Jedamzik 2004b, 2005). 
Thus, both of the Li problems can conceivably be solved at the same time.
While these idea are very attractive, they rests on as yet unproven and speculative physics.

\section{Concluding Remarks}
\label{s:conclusions}

Remarkable as it may have seemed in 1982 following the discovery of the Spite plateau and the
near-universal association of the plateau's lithium abundance with primordial nucleosynthesis, 
the debate over that association has not been stilled. Two key recent
factors heating the debate have been the  measurement of the
Big Bang's photon-to-baryon ratio ($\eta$) using the WMAP measurements of the
anisotropies in the cosmic microwave background and  the detections, here and previously, of the 
lighter isotope $^6$Li. A standard model of the Big Bang with the WMAP estimate
of $\eta$ predicts a lithium ($^7$Li) abundance greater
by about 0.5 dex than the value generally attributed  to the Spite plateau,
and a $^6$Li abundance several orders of magnitude below the value reported here.

These two discrepancies resist simple resolution, as should be clear from the discussion in Section 7.
One supposes that resolution of the  discrepancies may be found in one or more of the following
three areas:

\smallskip
\noindent
{\em 1. Composition of the gas from which stars on the Spite plateau  formed}

\noindent
Ignoring the possibility that the $^6$Li was synthesised after the formation of the star,
a conjecture is that $^6$Li was synthesised either in a non-standard Big Bang (e.g., hadronic
decays of exotic particles resulting in post-processing the $^7$Li from a standard Big Bang) or 
through $\alpha - \alpha$ fusion reactions in the proto-Galaxy (e.g. hierarchical structure
formation shocks and/or supermassive black holes).
Introduction of a non-standard
Big Bang  may lower the predicted $^7$Li abundance and
increase the predicted $^6$Li abundance. Qualitatively, these changes resolve both
discrepancies provided that points 2 and 3 below do not come into play.  Addition of lithium
synthesis from fusion reactions in pre-Galactic gas, as the sole additional site over
a standard Big Bang, may remove the $^6$Li discrepancy but may aggravate the
discrepancy between the predicted   $^7$Li abundance and the
observed abundance, unless one adopts the heresy that the WMAP estimate
of $\eta$ should be replaced by the lower value that results in the minimum
synthesis of $^7$Li. (The nuclear reaction network appears to be free of significant
uncertainties.) 

\smallskip
\noindent
{\em 2. Relationship between the stars' initial composition set 10 -- 13 Gyr ago and the present composition
of their stellar atmospheres}

\noindent
There is tension on this point between those who stress the significance of the observed very small scatter
of the lithium abundances along the Spite plateau and those who identify physical processes that
ought to change the atmospheric compositions of plateau residents (Section 7.4).
The very small scatter would seem to imply that the change in surface lithium abundance has 
been itself slight. 
 Yet, some argue
that reduction of the lithium abundance through
 diffusion, perhaps supplemented by rotationally-induced mixing and other processes, must
occur but can result in a small star-to-star scatter of lithium abundances.
 The presence of
$^6$Li, as we noted,  complicates interpretations that rely on depletion/destruction
of $^7$Li to reduce the WMAP-based Big Bang $^7$Li abundance to the observed value for the
Spite plateau; very high to extraordinarily high initial $^6$Li abundances may be implied.
There are clearly open questions for pursuit by observers.

Diffusion does not exclusively act on lithium.  Relative abundances of abundant elements
are changed but these elements are not subject to destruction as lithium is.
 Alterations of the abundances are lessened and then removed as stars
evolve from the main sequence to subgiants and then to giants. Careful comparisons
of stars along this sequence might confirm that diffusion was operable in their
main sequence progenitors. At present, abundance data from dwarfs and giants have
given seemingly consistent results for composition as a function of metallicity, i.e.,
the same run of  [X/Fe] versus [Fe/H] (e.g. Gratton et al. 2001). 

\smallskip
\noindent
{\em 3. Measurement of the present $^6$Li and $^7$Li abundances}

\noindent
Systematic errors are now unlikely to be uncovered that will lift the Spite plateau's
lithium abundance by the 0.5 dex to match the Big Bang WMAP-based abundance
(Section 7.1). The adjustments (e.g, the necessary increase of effective temperatures
by about 700\,K) unless specific to lithium would result in
large changes to abundances of other elements. Presently, there is generally
good agreement between the compositions of metal-poor dwarfs and giants of the
same metallicity and this agreement would be compromised unless adjustments
were also made to the abundance analyses of giants with their very
different atmospheric structures. The simplest conceivable adjustment --
raising the effective temperatures by 700 K -- introduces more problems
than the one it solves.

Absence of systematic errors affecting the determinaton of the $^6$Li/$^7$Li ratio
cannot be so strongly dismissed: the presence of $^6$Li is deduced from a slight
additional asymmetry to the intrinsically asymmetric $^7$Li line. The various tests 
outlined in Sect. \ref{s:li6708} give confidence that $^6$Li has indeed
been detected. This confidence should be tested by observations of additional
stars like
HD\,19445 for which pre-main sequence destruction of $^6$Li is so great that
this isotope should be undetectable. Obviously, additional \lisix\ observations 
at very low \feh\ would be extremely valuable, including a confirmation of our
detection in LP\,815-43.

The confidence commonly and frequently expressed after 1982 that primordial lithium had been
discovered and measured and could be matched with a prediction from a standard Big Bang
is today weakened. There remains a belief that much of the lithium is traceable to
primordial nucleosynthesis but quantitative agreement between prediction and observation
of the $^6$Li and $^7$Li abundances in stars on the Spite plateau is proving elusive.
Observers will suppose that well designed additional spectroscopic analyses
of metal-poor stars will provide some of the clues necessary to determining the
primordial abundance of the two lithium isotopes. Theoreticians may combine these
clues with other evidence and probe deeper into primordial nucleosynthesis.
Interest in the lithium of the Spite plateau is greater than in 1982, even as the
lack of concordance between predictions of the standard Big Bang and measured lithium
abundances has been demonstrated.

\acknowledgements
This paper has benefitted from helpful assistance,
illuminating discussions and penetrating questions from
numerous colleagues, including
Paul Barklem, Lars Bergstr\"om, Corinne Charbonnel, Bengt Gustafsson, Susume Inoue, 
Karsten Jedamzik, Georges Michaud, John Norris, Keith Olive, 
Olivier Richard, Gary Steigman, Elisabeth Vangioni and Achim Weiss. 
We thank Wako Aoki for kindly providing the Subaru spectrum of HD\,140283 for
an independent analysis and Piercarlo Bonifacio for sharing the results of the
First Stars program prior to publication.
This work has been supported by the Australian Research Council through the
grants DP0342613 and DP0558836. DLL acknowledges the support of the Robert A. Welch
Foundation of Houston, Texas.

\newpage


\begin{deluxetable}{lrccccr}
\tablecaption{Observational details, including date of 
observation, Universal Time for the middle of the exposure, total
exposure time, signal-to-noise around the \lii\ 670.8\,nm line in
the reduced spectrum, and the measured heliocentric radial velocity
of the star.\label{t:observations}}
\tablehead{\colhead{Star} & \colhead{$V$} & \colhead{Date}  & \colhead{UTC} & 
\colhead{$t_{\rm exp}$} & \colhead{$S/N$} & \colhead{$v_{\rm rad}$} \\
\colhead{}     & \colhead{[mag]}&  \colhead{}     &    \colhead{}  & 
\colhead{[min]}     &   \colhead{}    & \colhead{[\kms]}  \\    }
\startdata
   HD\,3567 &  9.26 & 2000-07-24 & 08:43 &  45 & 560 & $-$47.3 \\
  HD\,19445 &  8.05 & 2000-08-04 & 10:13 &  20 & 790 & $-$139.7 \\
  HD\,59392 &  9.73 & 2002-02-05 & 02:53 &  60 & 710 &    267.9 \\
 HD\,102200 &  8.74 & 2002-02-05 & 07:36 &  40 & 610 &    161.1 \\
 HD\,106038 & 10.18 & 2000-07-23 & 23:57 &  60 & 450 &     99.6 \\
           &     & 2000-07-24 & 23:56 &  40 & 350 &     99.9 \\
 HD\,140283 &  7.21 & 2000-07-24 & 23:15 &  13 & 1020 & $-$170.4 \\
           &     & 2002-02-05 & 08:52 &  15 & 770 & $-$170.4 \\
           &     & 2004-08-28 & 00:38 &  19 & 820 & $-$170.6 \\
 HD\,160617 &  8.73 & 2000-07-25 & 01:34 &  25 & 650 &    100.6 \\
 HD\,213657 &  9.66 & 2000-08-04 & 07:23 &  45 & 640 &     50.8 \\
 HD\,298986 & 10.05 & 2002-02-05 & 06:24 &  80 & 470 &    198.3 \\
 HD\,338529 &  9.37 & 2000-07-25 & 04:12 &  60 & 680 & $-$128.3 \\
 G\,013-009 & 10.00 & 2000-07-25 & 00:43 &  60 & 530 &     58.4 \\
 G\,020-024 & 11.13 & 2000-07-24 & 02:19 &  90 & 430 &     34.2 \\
           &     & 2000-07-25 & 02:47 &  90 & 420 &     34.5 \\
 G\,075-031 & 10.52 & 2000-07-24 & 09:46 &  60 & 500 &     57.8 \\
           &     & 2000-08-04 & 09:26 &  60 & 500 &     57.8 \\
 G\,126-062 &  9.48 & 2000-08-04 & 06:30 &  45 & 700 & $-$288.6 \\
 BD\,$+09\arcdeg2190$ & 11.15 & 2002-02-05 & 04:25 &  90 & 400 &  266.6 \\
                    &     & 2002-02-06 & 05:16 &  90 & 400 &  266.0 \\
 BD\,$+03\arcdeg0740$ &  9.80 & 2002-02-05 & 01:03 &  60 & 740 &  174.2 \\
 BD\,$-13\arcdeg3442$ & 10.29 & 2002-02-06 & 06:27 &  90 & 400 &  116.1 \\
 CD\,$-30\arcdeg18140$&  9.95 & 2000-07-24 & 05:51 &  60 & 450 &   17.9 \\
                     &     & 2000-07-25 & 06:03 &  25 & 350 &   17.9 \\
 CD\,$-33\arcdeg1173$ & 10.94 & 2002-02-06 & 01:13 &  90 & 610 &   47.6 \\
 CD\,$-33\arcdeg3337$ &  9.08 & 2002-02-05 & 01:57 &  30 & 740 &   73.7 \\
 CD\,$-35\arcdeg14849$& 10.57 & 2000-07-24 & 07:27 &  90 & 550 &  108.0 \\
                     &     & 2000-07-25 & 06:33 &  25 & 350 &  108.1 \\
 CD\,$-48\arcdeg2445$ & 10.55 & 2002-02-06 & 03:01 & 100 & 640 &  319.2 \\
   LP\,815-43 & 10.91 & 2000-07-24 & 04:12 &  90 & 400 &  $-$3.8 \\
            &     & 2000-07-25 & 05:35 &  40 & 300 &  $-$3.5 \\
            &     & 2004-08-30 & 04:52 & 150 & 420 &  $-$3.6 \\
            &     & 2004-08-31 & 01:35 & 150 & 490 &  $-$3.9 \\
\enddata
\end{deluxetable}

\begin{deluxetable}{lcccccc}
\tablecaption{Derived values of effective temperature,
surface gravity, metallicity (as measured by \feii\ lines),
oxygen abundance (non-LTE values from the \oi\ triplet),
microturbulence and absolute magnitude. \label{t:parameters}}
\tablehead{
\colhead{Star} & \colhead{\teff} & \colhead{\logg} & \colhead{\feh} & 
\colhead{\oh} & \colhead{\micro} & \colhead{\mv}  \\
\colhead{}     & \colhead{[K]}    & \colhead{[cgs] }  &  \colhead{}  & 
\colhead{}  & \colhead{[\kms]} &  \colhead{}    \\}
\startdata
    HD\,3567 &    6026 &   4.08 &  $-$1.14 & $-$0.79 &   1.5 &    4.01  \\
   HD\,19445 &    5980 &   4.42 &  $-$2.02 & $-$1.38 &   1.3 &    5.13  \\
   HD\,59392 &    5936 &   3.99 &  $-$1.61 & $-$1.04 &   1.5 &    3.97  \\
  HD\,102200 &    6062 &   4.15 &  $-$1.25 & $-$0.85 &   1.3 &    4.27  \\
  HD\,106038 &    5905 &   4.30 &  $-$1.35 & $-$0.76 &   1.2 &    4.84  \\
  HD\,140283 &    5753 &   3.70 &  $-$2.40 & $-$1.74 &   1.5 &    3.35  \\
  HD\,160617 &    5990 &   3.79 &  $-$1.76 & $-$1.43 &   1.5 &    3.35  \\
  HD\,213657 &    6180 &   3.92 &  $-$1.90 & $-$1.34 &   1.5 &    3.60  \\
  HD\,298986 &    6103 &   4.22 &  $-$1.33 & $-$0.98 &   1.3 &    4.41  \\
  HD\,338529 &    6335 &   4.04 &  $-$2.26 & $-$1.68 &   1.5 &    3.73  \\
  G\,013-009 &    6298 &   3.99 &  $-$2.30 & $-$1.71 &   1.5 &    3.70  \\
  G\,020-024 &    6247 &   3.98 &  $-$1.89 & $-$1.35 &   1.5 &    3.72  \\
  G\,075-031 &    6000 &   4.08 &  $-$1.02 & $-$0.68 &   1.3 &    3.96  \\
  G\,126-062 &    6183 &   4.11 &  $-$1.51 & $-$1.05 &   1.5 &    4.05  \\
  G\,271-162 &    6230 &   3.93 &  $-$2.30 & $-$1.79 &   1.5 &    3.61  \\
 BD\,$+09\arcdeg2190$ &    6392 &   4.09 &  $-$2.66 & $-$2.22 &   1.5 &    3.89  \\
 BD\,$+03\arcdeg0740$ &    6266 &   4.04 &  $-$2.65 & $-$2.06 &   1.5 &    3.93  \\
 BD\,$-13\arcdeg3442$ &    6311 &   3.86 &  $-$2.71 & $-$2.09 &   1.5 &    3.43  \\
CD\,$-30\arcdeg18140$ &    6222 &   4.11 &  $-$1.90 & $-$1.28 &   1.5 &    4.06  \\
 CD\,$-33\arcdeg1173$ &    6390 &   4.28 &  $-$2.92 & $-$2.33 &   1.5 &    4.36  \\
 CD\,$-33\arcdeg3337$ &    5897 &   4.01 &  $-$1.31 & $-$0.76 &   1.3 &    3.92  \\
CD\,$-35\arcdeg14849$ &    6244 &   4.30 &  $-$2.27 & $-$1.70 &   1.5 &    4.56  \\
 CD\,$-48\arcdeg2445$ &    6222 &   4.25 &  $-$1.93 & $-$1.40 &   1.5 &    4.42  \\
           LP\,815-43 &    6400 &   4.17 &  $-$2.74 & $-$2.20 &   1.5 &    4.10  \\
\enddata
\end{deluxetable}

\begin{deluxetable}{lcccccccccc}
\tabletypesize{\scriptsize}
\tablecaption{Li isotopic abundances derived from the \lii\ 670.8\,nm resonance line together with the
Li abundances estimated from the \lii\ 610.4\,nm subordinate line. Both the 1D LTE and non-LTE
Li abundances are listed for the two lines. The quoted uncertainties in the LTE abundances are
those only due to the finite $S/N$ and do not for example take into account errors in the stellar
parameters. The second and third column
give the projected rotational velocity \vsini\ (assumed to be $0.5$\,\kms\ for all stars) and the
macroturbulence \macro ; note that instrumental broadening has also been applied. 
The listed equivalent widths for
the 670.8 and 610.4\,nm lines are those coming from the best-fitting theoretical line profiles
to the observed profiles.\label{t:li_abundances}}
\tablehead{
\colhead{}& \colhead{}& \colhead{}& \multicolumn{4}{c}{\lii\ 670.8\,nm} &\colhead{}& \multicolumn{3}{c}{\lii\ 610.4\,nm}  \\
\cline{4-7} \cline{9-11}
\\
\colhead{Star} & \colhead{\vsini} & \colhead{\macro}  & \colhead{$W_\lambda$} & \colhead{\logliseven} & \colhead{\logliseven} & \colhead{\liratio} &
        & \colhead{$W_\lambda$} & \colhead{\logli} & \colhead{\logli} \\
        & \colhead{[\kms]} & \colhead{[\kms]} & \colhead{[pm]}  &  \colhead{(LTE)}       & \colhead{(non-LTE)}     &      \colhead{}    &
\colhead{}        & \colhead{[pm]} &  \colhead{(LTE)}        & \colhead{(non-LTE)}        \\}
\startdata
            HD\,3567                & 0.50 & $4.71\pm0.19$ & 3.83 & $  2.32 \pm 0.01$ &   2.32 & $0.017\pm0.012$ && 0.19 & $  2.30 \pm   0.04$ &   2.36\\
            HD19445                & 0.50 & $3.83\pm0.13$ & 3.51 & $  2.21 \pm 0.01$ &   2.20 & $0.002\pm0.008$ && 0.16 & $  2.19 \pm   0.05$ &   2.24\\
            HD\,59392              & 0.50 & $4.37\pm0.35$ & 3.92 & $  2.24 \pm 0.01$ &   2.24 & $0.021\pm0.016$ && 0.15 & $  2.16 \pm   0.03$ &   2.22\\   
            HD\,102200            & 0.50 & $4.44\pm0.24$ & 3.30 & $  2.25 \pm 0.01$ &   2.25 & $0.047\pm0.013$ && 0.18 & $  2.30 \pm   0.03$ &   2.36\\
            HD\,106038            & 0.50 & $3.88\pm0.12$ & 6.31 & $  2.49 \pm 0.01$ &   2.48 & $0.031\pm0.006$ && 0.30 & $  2.45 \pm   0.02$ &   2.52\\
            HD\,140283            & 0.50 & $4.87\pm0.07$ & 4.78 & $  2.17 \pm 0.01$ &   2.20 & $0.008\pm0.006$ && 0.18 & $  2.11 \pm   0.03$ &   2.20\\
            HD\,160617            & 0.50 & $4.54\pm0.18$ & 4.05 & $  2.28 \pm 0.01$ &   2.28 & $0.036\pm0.010$ && 0.19 & $  2.28 \pm   0.03$ &   2.34\\
            HD\,213657            & 0.50 & $5.02\pm0.08$ & 3.00 & $  2.27 \pm 0.01$ &   2.25 & $0.011\pm0.011$ && 0.15 & $  2.24 \pm   0.04$ &   2.29\\
            HD\,298986            & 0.50 & $4.70\pm0.20$ & 3.04 & $  2.25 \pm 0.01$ &   2.24 & $0.017\pm0.015$ && 0.15 & $  2.24 \pm   0.04$ &   2.29\\
            HD\,338529            & 0.50 & $5.65\pm0.09$ & 2.37 & $  2.25 \pm 0.01$ &   2.22 & $0.010\pm0.013$ && 0.13 & $  2.24 \pm   0.05$ &   2.28\\
            G\,013-009              & 0.50 & $5.85\pm0.07$ & 2.39 & $  2.22 \pm 0.01$ &   2.19 & $0.048\pm0.019$ && 0.11 & $  2.15 \pm   0.06$ &   2.20\\
            G\,020-024              & 0.50 & $5.12\pm0.22$ & 2.49 & $  2.21 \pm 0.01$ &   2.18 & $0.070\pm0.017$ && 0.11 & $  2.14 \pm   0.07$ &   2.18\\
            G\,075-031              & 0.50 & $4.95\pm0.22$ & 3.78 & $  2.30 \pm 0.01$ &   2.30 & $0.015\pm0.012$ && 0.24 & $  2.40 \pm   0.03$ &   2.46\\
            G\,126-062              & 0.50 & $4.76\pm0.30$ & 2.76 & $  2.26 \pm 0.01$ &   2.24 & $0.001\pm0.015$ && 0.14 & $  2.22 \pm   0.04$ &   2.27\\
            G\,271-162              & 0.50 & $5.78\pm0.13$ & 2.91 & $  2.27 \pm 0.01$ &   2.25 & $0.019\pm0.012$ && 0.11 & $  2.14 \pm   0.05$ &   2.19\\
  BD\,$+09\arcdeg$2190 & 0.50 & $6.27\pm0.35$ & 1.66 & $  2.13 \pm 0.01$ &   2.10 & $-0.033\pm0.027$ && $<0.11$ & $  <2.20$           &   $<2.25$\\
  BD\,$+03\arcdeg$0740 & 0.50 & $6.71\pm0.15$ & 2.08 & $  2.12 \pm 0.01$ &   2.11 & $0.015\pm0.017$ && $<0.10$ & $  <2.10$           &   $<2.15$ \\
  BD\,$-13\arcdeg$3442  & 0.50 & $6.38\pm0.20$ & 2.04 & $  2.16 \pm 0.01$ &   2.14 & $0.001\pm0.028$ && 0.11 & $  2.15 \pm   0.08$ &   2.19\\
 CD\,$-30\arcdeg$18140 & 0.50 & $5.07\pm0.11$ & 2.76 & $  2.25 \pm 0.01$ &   2.22 & $0.042\pm0.013$ && 0.14 & $  2.23 \pm   0.03$ &   2.27\\
  CD\,$-33\arcdeg$1173  & 0.50 & $5.82\pm0.61$ & 1.60 & $  2.08 \pm 0.01$ &   2.05 & $0.013\pm0.041$ && 0.07 & $  2.01 \pm   0.10$ &   2.04\\
  CD\,$-33\arcdeg$3337  & 0.50 & $4.91\pm0.14$ & 3.90 & $  2.23 \pm 0.01$ &   2.24 & $0.020\pm0.010$ && 0.20 & $  2.27 \pm   0.03$ &   2.34\\
 CD\,$-35\arcdeg$14849 & 0.50 & $4.68\pm0.12$ & 2.84 & $  2.27 \pm 0.01$ &   2.24 & $0.020\pm0.012$ && 0.15 & $  2.26 \pm   0.05$ &   2.30\\
  CD\,$-48\arcdeg$2445  & 0.50 & $5.18\pm0.10$ & 2.56 & $  2.22 \pm 0.01$ &   2.19 & $0.032\pm0.014$ && 0.15 & $  2.27 \pm   0.03$ &   2.31\\
            LP\,815-43              & 0.50 & $6.20\pm0.19$ & 1.89 & $  2.16 \pm 0.01$ &   2.13 & $0.046\pm0.022$ && 0.11 & $  2.17 \pm   0.06$ &   2.21\\
\enddata
\end{deluxetable}

\begin{deluxetable}{lcccccc}
\tabletypesize{\small}
\tablecaption{Effects of the derived \liratio\ ratio to the
stellar parameters and suite of 1D model atmospheres
\label{t:li_parameters}}
\tablehead{
\colhead{Star} & \multicolumn{6}{c}{\liratio} \\
\cline{2-7}
\colhead{}     & \colhead{normal} & \colhead{$\Delta$\teff} & 
\colhead{$\Delta$\logg} & \colhead{$\Delta$\feh} & 
\colhead{$\Delta$\micro} & \colhead{Kurucz\tablenotemark{a}}
}
\startdata
              HD3567 &  0.017 &  0.016 &  0.015 &  0.017 &  0.015 &  0.018  \\
             HD19445 &  0.002 &  0.000 &  0.002 &  0.002 &  0.001 &  0.003  \\
             HD59392 &  0.021 &  0.020 &  0.021 &  0.021 &  0.020 &  0.021  \\
            HD102200 &  0.047 &  0.043 &  0.045 &  0.047 &  0.046 &  0.044  \\
            HD106038 &  0.031 &  0.028 &  0.030 &  0.032 &  0.029 &  0.028  \\
            HD140283 &  0.008 &  0.008 &  0.008 &  0.008 &  0.007 &  0.006  \\
            HD160617 &  0.036 &  0.033 &  0.034 &  0.035 &  0.035 &  0.035  \\
            HD213657 &  0.011 &  0.008 &  0.011 &  0.011 &  0.010 &  0.011  \\
            HD298986 &  0.017 &  0.014 &  0.016 &  0.017 &  0.017 &  0.018  \\
            HD338529 &  0.010 &  0.008 &  0.011 &  0.010 &  0.011 &  0.011  \\
            G013-009 &  0.048 &  0.047 &  0.049 &  0.048 &  0.049 &  0.049  \\
            G020-024 &  0.070 &  0.070 &  0.069 &  0.070 &  0.069 &  0.070  \\
            G075-031 &  0.015 &  0.013 &  0.014 &  0.015 &  0.013 &  0.016  \\
            G126-062 &  0.001 &  $-0.001$ &  0.000 &  0.000 &  $-0.001$ &  0.003  \\
            G271-162 &  0.019 &  0.018 &  0.019 &  0.018 &  0.018 &  0.019  \\
  BD\,$+09\arcdeg$2190 & $-0.033$ & $-0.035$ & $-0.033$ & $-0.033$ & $-0.032$ & $-0.033$  \\
  BD\,$+03\arcdeg$0740 &  0.015 &  0.013 &  0.015 &  0.014 &  0.013 &  0.015  \\
  BD\,$-13\arcdeg$3442 & 0.001 & $-0.001$ & 0.001 & 0.001 & 0.001 & 0.001  \\
 CD\,$-30\arcdeg$18140 &  0.042 &  0.039 &  0.043 &  0.041 &  0.041 &  0.044  \\
  CD\,$-33\arcdeg$1173 &  0.013 &  0.010 &  0.012 &  0.013 &  0.013 &  0.011  \\
  CD\,$-33\arcdeg$3337 &  0.020 &  0.019 &  0.019 &  0.021 &  0.017 &  0.021  \\
   CD\,$-35\arcdeg$14849 &  0.020 &  0.017 &  0.020 &  0.020 &  0.020 &  0.020  \\
  CD\,$-48\arcdeg$2445 &  0.032 &  0.031 &  0.033 &  0.032 &  0.033 &  0.034  \\
            LP815-43 &  0.046 &  0.045 &  0.046 &  0.048 &  0.046 &  0.046  \\
\enddata
\tablenotetext{a}{Derived \liratio\
when using Kurucz model atmospheres (without convective overshoot as computed by
Castelli et al. 1997)
instead of the standard \marcs\ models
otherwise used
}
\end{deluxetable}

\begin{deluxetable}{lrrr}
\tablecaption{Estimated projected stellar rotational velocities and \liratio\ based on a 3D analysis
\label{t:li6_3D}}
\tablehead{
\colhead{Star} & \colhead{\vsini\ }& \colhead{\liratio(3D)} & \colhead{\liratio(1D)}  \\ 
\colhead{}        & \colhead{[\kms ]}  &  \colhead{}   &  \colhead{} 
}
\startdata
              HD3567 &  $0.35\pm0.60$ & $0.000\pm0.012$ &  $0.017\pm0.012$ \\
             HD19445 &  $2.55\pm0.14$  & $0.030\pm0.008$ &  $0.002\pm0.008$ \\
            HD140283 &   $2.01\pm0.14$ & $0.023\pm0.006$ &  $0.008\pm0.006$ \\     
            HD160617 &   $1.19\pm0.61$ & $0.052\pm0.012$ &  $0.036\pm0.010$ \\
            HD213657 &    $1.93\pm0.23$ & $0.030\pm0.012$ &  $0.011\pm0.011$ \\
            HD338529 &   $2.50\pm0.18$ & $0.017\pm0.013$ &  $0.010\pm0.013$ \\
            G013-009 &   $2.77\pm0.24$ & $0.056\pm0.021$ &  $0.048\pm0.019$ \\
            G020-024 &   $2.30\pm0.17$ & $0.081\pm0.016$ &  $0.070\pm0.017$ \\
            G075-031 &   $0.67\pm0.82$ & $0.000\pm0.012$ &  $0.015\pm0.012$ \\
            G271-162 &    $2.65\pm0.20$ & $0.026\pm0.012$ &  $0.019\pm0.012$ \\
  BD\,$+09\arcdeg$2190 &  $2.84\pm0.58$  & $0.000\pm0.027$ &  $-0.033\pm0.027$ \\
  BD\,$+03\arcdeg$0740 &   $3.80\pm0.27$  & $0.026\pm0.019$ &  $0.015\pm0.017$ \\
  BD\,$-13\arcdeg$3442 &   $3.02\pm0.39$ & $0.000\pm0.028$ &  $0.001\pm0.028$ \\
 CD\,$-30\arcdeg$18140 &    $2.21\pm0.26$ & $0.058\pm0.014$ &  $0.042\pm0.013$ \\
  CD\,$-33\arcdeg$1173 &   $1.81\pm1.45$ & $0.012\pm0.027$ &  $0.013\pm0.041$ \\
   CD\,$-35\arcdeg$14849 &   $0.10\pm0.19$ & $0.026\pm0.011$ &  $0.020\pm0.012$ \\
  CD\,$-48\arcdeg$2445 &  $2.34\pm0.19$  & $0.046\pm0.014$ &  $0.032\pm0.014$ \\
            LP815-43 &   $2.82\pm0.25$ & $0.036\pm0.021$ & $0.046\pm0.022$  \\
\enddata
\end{deluxetable}

\begin{deluxetable}{lccccccccccc}
\tablecaption{Predicted \liseven\ and \lisix\ depletion factors (in dex) for our program
stars interpolated from predictions by Richard et al. (2002, 2005; and 2005, private
communication). The  second and third columns give the expected depletion during 
the pre-main sequence (PMS) while the others are for the evolution thereafter to 
which the PMS depletion factors should be added. Diffusion denotes the case of
pure diffusion while T6.09 and T6.25 are for two different models of
turbulent diffusion (see text for details). 
\label{t:depletion}}
\tablehead{
\colhead{Star} & \multicolumn{2}{c}{PMS} && \multicolumn{2}{c}{Diffusion} 
&& \multicolumn{2}{c}{T6.09} && \multicolumn{2}{c}{T6.25}  \\
\noalign{\smallskip}
\cline{2-3} \cline{5-6} \cline{8-9} \cline{11-12}  
\noalign{\smallskip}
\colhead{}& \colhead{\liseven} & \colhead{\lisix} &\colhead{}& 
\colhead{\liseven} & \colhead{\lisix} &\colhead{}& \colhead{\liseven} & 
\colhead{\lisix} &\colhead{}& \colhead{\liseven} & \colhead{\lisix}
}
\startdata
    HD\,3567 &      0.01 & 0.8 &&  0.3 & 0.3 &&  0.2 & 0.6 &&  0.5 & 2.1 \\
   HD\,19445 &     0.01 & 0.9 &&  0.2 & 0.2 &&  0.2 & 0.6 &&  0.5 & 2.2   \\
   HD\,59392 &     0.02 & 1.0 &&  0.2 & 0.2 &&  0.2 & 0.6 &&  0.5 & 2.3 \\
  HD\,102200 &    0.01 & 0.8 &&  0.3 & 0.3 &&  0.2 & 0.6 &&  0.5 & 2.0  \\
  HD\,106038 &    0.02 & 1.1 &&  0.2 & 0.2 &&  0.2 & 0.7 &&  0.6 & 2.5  \\
  HD\,140283 &    0.01 & 0.4 &&  0.2 & 0.3 &&  0.2 & 0.4 &&  0.5 & 1.9  \\
  HD\,160617 &    0.01 & 0.6 &&  0.4 & 0.3 &&  0.2 & 0.3 &&  0.5 & 1.8  \\
  HD\,213657 &    0.01 & 0.5 &&  0.4 & 0.4 &&  0.2 & 0.4 &&  0.5 & 1.8   \\
  HD\,298986 &    0.01 & 0.7 &&  0.3 & 0.3 &&  0.2 & 0.5 &&  0.5 & 1.9   \\
  HD\,338529 &    0.00 & 0.3 &&  0.7 & 0.7 &&  0.2 & 0.4 &&  0.4 & 1.7  \\
  G\,013-009 &      0.00 & 0.4 &&  0.6 & 0.6 &&  0.2 & 0.4 &&  0.4 & 1.7 \\
  G\,020-024 &      0.00 & 0.4 &&  0.5 & 0.5 &&  0.2 & 0.4 &&  0.4 & 1.8 \\
  G\,075-031 &      0.01 & 0.9 &&  0.3 & 0.3 &&  0.2 & 0.6 &&  0.5 & 2.1  \\
  G\,126-062 &      0.01 & 0.5 &&  0.4 & 0.4 &&  0.2 & 0.4 &&  0.4 & 1.8  \\
  G\,271-162 &      0.00 & 0.5 &&  0.5 & 0.5 &&  0.2 & 0.4 &&  0.4 & 1.8  \\
 BD\,$+09\arcdeg2190$ &     0.00 & 0.3 &&  0.9 & 0.8 &&  0.2 & 0.3 &&  0.4 & 1.6  \\
 BD\,$+03\arcdeg0740$ &     0.00 & 0.4 &&  0.5 & 0.5 &&  0.2 & 0.4 &&  0.4 & 1.7  \\
 BD\,$-13\arcdeg3442$ &      0.00 & 0.4 &&  0.6 & 0.6 &&  0.2 & 0.4 &&  0.4 & 1.7 \\
CD\,$-30\arcdeg18140$ &    0.00 & 0.5 &&  0.5 & 0.5 &&  0.2 & 0.4 &&  0.4 & 1.8  \\
 CD\,$-33\arcdeg1173$ &     0.00 & 0.3 &&  0.8 & 0.9 &&  0.2 & 0.3 &&  0.4 & 1.6 \\
 CD\,$-33\arcdeg3337$ &     0.02 & 1.1 &&  0.2 & 0.2 &&  0.2 & 0.7 &&  0.6 & 2.5 \\
CD\,$-35\arcdeg14849$ &    0.00 & 0.4 &&  0.5 & 0.5 &&  0.2 & 0.4 &&  0.4 & 1.7   \\
 CD\,$-48\arcdeg2445$ &     0.00 & 0.5 &&  0.5 & 0.5 &&  0.2 & 0.4 &&  0.4 & 1.8  \\
           LP\,815-43 &                 0.00 & 0.3 &&  0.9 & 0.8 &&  0.2 & 0.3 &&  0.4 & 1.6  \\    
\enddata
\end{deluxetable}

\end{document}